\documentclass[10pt,english,notitlepage,a4paper,english,superscriptaddress,floatfix,longbibliography,pre]{revtex4-2}
\usepackage[utf8]{inputenc}
\usepackage[T1]{fontenc}
\usepackage{xcolor}
\usepackage{ulem}
\usepackage{graphicx,epsfig}
\usepackage{epstopdf}
\usepackage{bbm, amsmath, amssymb, mathrsfs, float, mathtools, cases, color, verbatim}
\usepackage{dsfont}
\usepackage{xcolor}
\usepackage{printlen}
\usepackage[percent]{overpic}

\usepackage[unicode=true,pdfusetitle,
bookmarks=true,bookmarksnumbered=false,bookmarksopen=false,
 breaklinks=false,pdfborder={0 0 1},backref=false,colorlinks=true]
 {hyperref}

\newcommand{\figpanel}[1]{(\textbf{\lowercase{#1}})}

\newcommand{\poli}{Dipartimento di Scienza Applicata e Tecnologia, Politecnico di Torino, Corso Duca degli Abruzzi 24, 10129, Torino, Italy}
\newcommand{\polimi}{Dipartimento di Elettronica, Informazione e Bioingegneria, Politecnico di Milano, Via Ponzio 34/5, 20133 Milano, Italy}
\newcommand{\iigm}{Statistical inference and computational biology, Italian Institute for Genomic Medicine, c/o IRCSS, 10060 Candiolo, Torino, Italy}
\newcommand{\carloalberto}{Collegio Carlo Alberto, P.za Arbarello 8, 10122, Torino, Italy}
\newcommand{\ucm}{Departamento de Física Teórica, Universidad Complutense, 28040 Madrid, Spain}

\begin{document}

\title{Effectiveness of probabilistic contact tracing in epidemic containment: the role of super-spreaders and transmission path reconstruction \\
}

\author{Anna Paola Muntoni}
\email{anna.muntoni@polito.it}
\affiliation{\poli}
\affiliation{\iigm}
\author{Fabio Mazza}
\affiliation{\poli}
\affiliation{\polimi}
\author{Alfredo Braunstein}
\affiliation{\poli}
\affiliation{\iigm}
\author{Giovanni Catania}
\affiliation{\ucm}
\author{Luca Dall'Asta}
\affiliation{\poli}
\affiliation{\iigm}
\affiliation{\carloalberto}

\begin{abstract}
The recent COVID-19 pandemic underscores the significance of early-stage non-pharmacological intervention strategies.
The widespread use of masks and the systematic implementation of contact tracing strategies provide a potentially equally effective and socially less impactful alternative to more conventional approaches, such as large-scale mobility restrictions.
However, manual contact tracing faces strong limitations in accessing the network of contacts, and the scalability of currently implemented protocols for smartphone-based digital contact tracing becomes impractical during the rapid expansion phases of the outbreaks, due to the surge in exposure notifications and associated tests.
A substantial improvement in digital contact tracing can be obtained through the integration of probabilistic techniques for risk assessment that can more effectively guide the allocation of new diagnostic tests. 
In this study, we first quantitatively analyze the diagnostic and social costs associated with these containment measures based on contact tracing, employing three state-of-the-art models of SARS-CoV-2 spreading. Our results suggest that probabilistic techniques allow for more effective mitigation at a lower cost.
Secondly, our findings reveal a remarkable efficacy of probabilistic contact-tracing techniques in performing backward and multi-step tracing and capturing super-spreading events.

\end{abstract}

\maketitle
\section{Introduction} 

The recent experience of the COVID-19 pandemic has shown that mobility restrictions and lockdowns can have severe social and economic consequences \cite{bonaccorsi2020economic}. 
In light of the potential unavailability of vaccines, particularly in the early stages of a pandemic, it is then imperative to develop and implement non-pharmacological intervention measures capable of ensuring the containment or gradual slowing down of epidemic outbreaks while concurrently preserving economic and social activities \cite{perra2021non,kretzschmar2022challenges}. Together with increased attention to hygiene and the use of masks, contact tracing represents the most promising non-pharmacological measure for this purpose \cite{hellewell2020feasibility}, and has been successfully employed to identify and eradicate small outbreaks of COVID-19 \cite{lavezzo2020suppression,bi2020epidemiology}.
Manual contact tracing (MCT) becomes impractical for large epidemic outbreaks, implying high costs and temporal delays \cite{keeling2020efficacy,hellewell2020feasibility,firth2020using}. Moreover, MCT is unlikely to discover contacts outside of immediate family or close relationships \cite{smieszek2016contact,mastrandrea2015contact}. Building on previous studies related to the Ebola virus disease \cite{sacks2015introduction,danquah2019use}, it has been argued that such limitations could be overcome with the systematic use of automated contact tracing procedures, which could scale up to the case of large outbreaks and favor the discovery of potentially infectious contacts even among occasional ones \cite{ferretti2020quantifying,kucharski2020effectiveness} (see also \cite{braithwaite2020automated}).
Indeed, aggressive containment policies based on digital contact tracing (DCT) technologies, such as smartphone apps and GPS beacons, proved effective during the first wave of COVID-19 in countries like Taiwan \cite{chien2022taiwan}, South Korea \cite{oh2020national}, China \cite{aslam2020fighting}, and Singapore \cite{huang2020performance}. These techniques sparked debates in Western countries on the threat of individual privacy \cite{mello2020ethics,bengio2020need,amann2021digital} and the need for voluntary adoption of contact-tracing apps by a large portion of the population \cite{ferretti2020quantifying,jacob2021adoption,munzert2021tracking}. 
Privacy-preserving protocols for digital contact tracing (DCT) have been introduced, using either centralized \cite{bay2020bluetrace,NHSapp,Aarogyasetu} or distributed \cite{troncoso2020decentralized,chan2020pact,AppleGoogle} approaches, primarily relying on Bluetooth low-energy (BLE) communication to detect physical proximity without geolocation. The analysis of data obtained from early implementations of DCT apps indicates a tangible contribution to epidemic containment, providing an additional quantitative and qualitative advantage over MCT \cite{kendall2020epidemiological,salathe2020early,wymant2021epidemiological,rodriguez2021population}.

In most DCT apps, exposure notifications are triggered for every contact with individuals who have tested positive, irrespective of a variety of factors determining the risk associated with the contact. As a consequence, the proliferation of exposure notifications and quarantines, responsible for the reduction in the number of infected individuals,
can lead to very high social costs (e.g., the number of isolated individuals) and economic costs (e.g., the number of diagnostic tests used) \cite{barrat2021effect,cencetti2021digital,contreras2021challenges}. 
A crucial step towards improving the efficacy of DCT and reducing notification redundancy is represented by probabilistic contact tracing methods, which could naturally account for multiple exposures \cite{alsdurf2020covi,fenton2020privacy,baker2021epidemic,murphy2021risk}. Using a Bayesian framework that incorporates all available data on individuals who tested positive (or negative), Baker et al. \cite{baker2021epidemic} proposed an efficient distributed method, based on Belief Propagation \citep{braunstein_inference_2016, altarelli2014bayesian}, to compute the individual probabilities of infection. This information can be leveraged by the contact tracing app to determine individual risk levels presented to the users, favoring self-isolation and more efficient testing strategies.
In the present study, following the approach of Baker et al. \cite{baker2021epidemic}, we demonstrate the superiority of probabilistic contact tracing methods over standard ones, both in terms of higher containment capacity and lower cost-to-benefit ratio. This is done through a comparative analysis using different epidemic simulators \cite{hinch2021openabm,kerr2021covasim,lorch2022quantifying}, obtaining results that are robust across various disease transmission models and parameter ranges. 

This study also offers the opportunity to delve deeper into the mechanisms and causal relationships that control automated contact tracing, investigating the reasons behind the claimed superiority of probabilistic methods. As positive tested individuals are more likely to come from contagion clusters than to generate them \cite{kojaku2021effectiveness}, it is believed that the detection of sources of individual infections (backward tracing) and super-spreading events can significantly improve containment strategies, especially in the presence of overdispersed secondary infections~\cite{tufekci_this_2020,bradshaw2021bidirectional}, a common trait of modern diseases such as COVID-19 \cite{stein2011super,endo2020estimating,lau2020characterizing,althouse2020superspreading,sun2021transmission,lemieux2021phylogenetic} and Mpox \cite{smith2022yes,paredes2024underdetected,ward2024understanding}. Countries like Japan~\cite{oshitani2020cluster}, South Korea~\cite{lee2020nationwide}, and Uruguay \cite{taylor2020uruguay} are credited with successfully implementing backward tracing in their contact tracing campaigns. 
However, current app-based digital contact tracing implementations predominantly engage in simple forward tracing \cite{endo2020implication}, where tracked individuals are primarily those who could have been exposed to someone who has tested positive.
Innovative digital contact tracing methods based on statistical inference~\cite{baker2021epidemic}, which ground their predictive power on reconstructing causal relationships in transmission paths \cite{altarelli2014bayesian}, are instead expected to more efficiently discover multi-step forward and backward traces and capture super-spreading events. This is here quantitatively demonstrated by analyzing these features for various contact tracing strategies across different epidemic models in the early-containment phase, providing a possible explanation of the superior containment ability of probabilistic contact tracing.

\section{Results}

Mathematical models of epidemic spreading are largely used to forecast the evolution of outbreaks at different spatial and temporal scales, to evaluate the effects of public health interventions, and ultimately to guide governments' decisions \cite{flaxman2020estimating,chang2021mobility,pullano2021underdetection}. In this respect, agent-based models provide stylized but sufficiently reliable representations of the actual contact networks on which contagion between individuals could take place, thus becoming a natural and necessary tool for analyzing the consequences of non-pharmaceutical intervention strategies based on contact tracing.
Among the abundance of agent-based models proposed during the first waves of the COVID-19 pandemic~\cite{aleta2020modelling,gupta2020covi,hinch2021openabm,kerr2021covasim,lorch2022quantifying,lasser2022assessing}, some of them can be considered exemplary for formulating a critical analysis of the containment capabilities of the different contact tracing methods and evaluate their cost-to-benefit ratio.
The agent-based models analyzed in the present work, namely the OpenABM model by Hinch et al. \cite{hinch2021openabm}, Covasim by Kerr et al. \cite{kerr2021covasim} and the Spatiotemporal Epidemic Model (StEM) by Lorch et al. \cite{lorch2022quantifying}, can be considered rather simple generalizations of the Susceptible-Exposed-Infected-Recovered (SEIR) model, in which additional states are included to account for different levels of symptomaticity and disease severity. Agent populations are endowed with realistic features, including demographic data and different layers of social interactions, also obtained from simulated mobility (see Methods and the Supplementary Information for details). As a consequence, such models are capable of reproducing the empirically observed non-Poissonian statistics and overdispersion in contact patterns and individual viral loads.

These three agent-based models, each characterized by their unique attributes, serve as an ideal platform to assess the efficacy of contact tracing methods based on statistical inference, demonstrating their superiority in comparison to conventional test-trace-quarantine approaches. The probabilistic methods under study are those appearing in Baker et al. \cite{baker2021epidemic}, namely Simple Mean Field (SMF) and Belief Propagation (BP). For comparison, other contact tracing methods are considered: a basic form of digital contact tracing (DCT), and a more advanced ``informed'' contact tracing (ICT) approach that leverages all available information from medical test results. Additionally, for Covasim, we employed Test-Trace-Quarantine (TTQ), the integrated containment method presented in the work by Kerr et al. \cite{kerr2021controlling}. This method employs information about the symptomatic status of the tested individuals; even though encoding this data into BP is always possible, we do not use this information while running BP, SMF, DCT, and ICT to allow for a fair comparison among the four methods.
The Methods section provides a brief overview of the contact tracing algorithms (see  Supplementary Information for further details). The containment effectiveness of these different contact tracing methods is evaluated by a quantitative study across various intervention scenarios generated using these three agent-based models. Our analysis demonstrates that contact tracing based on statistical inference techniques facilitates effective mitigation at low medical costs, measured in terms of diagnostic tests, and social costs, quantified by the fraction of the population subjected to quarantine.
Finally, tracing techniques based on statistical inference are shown to outperform other approaches in effectively tracing both backward and forward transmissions and therefore in identifying superspreading events associated with the overdispersion of secondary infections.

\begin{figure}[t!]
	\centering
	\includegraphics[width=1.0\linewidth]{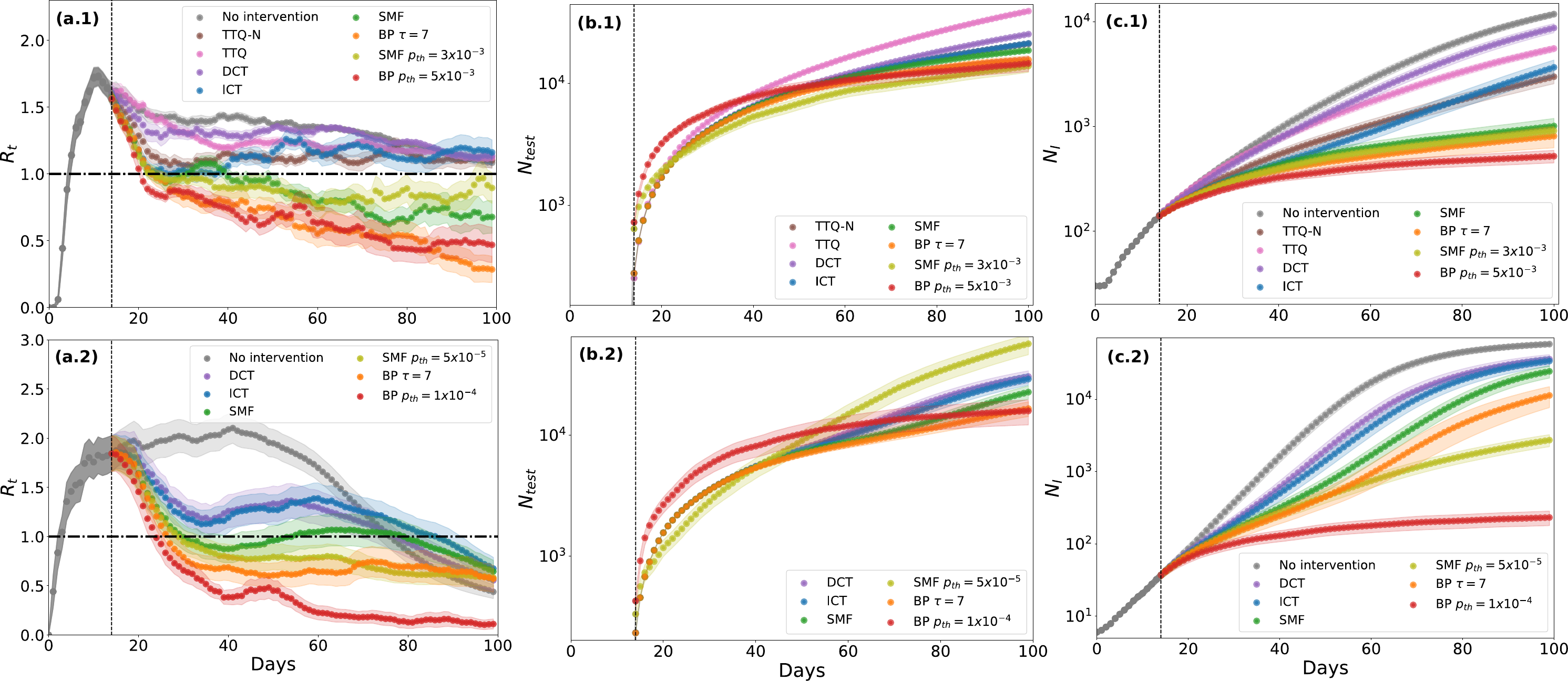}
	\caption{\textbf{Effective epidemic mitigation.} Columns labeled (a), (b), and (c) show, respectively, the behavior in time of the effective reproduction number $R_{t}$ (see the Supplementary Information for a detailed description), the cumulative number of diagnostic tests, and the cumulative number of infected individuals.  
		For the Covasim model (first row), simulations are run on a population of $70\,000$ people, for $T=100$ days. Each simulation starts with $N_{pz}=30$ patients zero, all in the exposed state, and each day half of the unidentified symptomatic individuals are observed ($p_{sym} = 50 \%$), while tracing-based interventions start after $t_{i} = 14$ days.
		For the StEM model (second row), simulations are performed on the urban area of T{\"u}bingen for $T = 100$ days, and the number of initial cases $N_{pz}$ is 6 (1 in the exposed state, 2 in the asymptomatic state, and 3 of them are pre-symptomatic individuals). The same fraction of the symptomatic individuals is observed ($p_{sym} = 50 \%$), with interventions starting at $t_{i} = 14$. In the StEM model, households are confined whenever a member is tested positive.
		Lines reflect the average behavior of the metrics computed from $20$ realizations of the Covasim population model and $30$ realizations of the StEM mobility model. The shaded regions indicate the associated standard error.
		\label{fig:fig1}}
\end{figure}

\subsection{Epidemic Containment\label{subsec:results_containment}}
Digital contact tracing-based strategies possess a remarkable capability to contain the spread of epidemics by reducing their impact. This was recently demonstrated within the realistic framework provided by OpenABM \cite{baker2021epidemic}. A similar analysis is carried out here on several instances of epidemic spreading generated using Covasim and StEM from a small initial number of infected individuals (patient zeros). The contact tracing protocol involves daily testing of a fixed fraction of symptomatic individuals. Different contact-tracing methods exploit the initial phase to gather information and update a ranking of potentially infected individuals. Starting from the first day of intervention $t_{i}$, an additional number of individuals is tested daily according to the risk predictions provided by the different methods. Those who test positive are subsequently confined. To formulate the ranking, each contact tracing algorithm incorporates the diagnostic test results and the contacts collected by the underlying contact tracing app over a predefined period. 
We assume that the app gathers the same information for all contact tracing methods, contingent on the app's adoption fraction ($AF$) within the population (assumed to be $AF=1.0$ here). The effects of lower adoption fractions ($AF<1$) were investigated in \cite{baker2021epidemic} for the case of OpenABM, but we expect similar behaviors for the two other models studied here. Note, however, that even for $AF=1$, the transmission network may be significantly different than the contact network detected by the app. In StEM, some exogenous transmissions are added within the simulation, and in Covasim the relative transmission among individuals, i.e. the weights of the transmission network, is highly heterogeneous and inaccessible to the contact tracing app and inference method (see Methods).
The test results are subject to error due to a non-zero false-negative rate ($f_N$). In our simulations, we set $f_N = 0.285$, an estimated value derived from published data \cite{dinnes2021rapid}, representing a relatively high false-negative rate associated with rapid COVID-19 tests that provide quick and affordable, but less accurate contagion assessment.

A standard testing strategy, applicable to all contact tracing methods, entails performing a fixed number of tests per day. In the strategies labeled as $\text{DCT}$, $\text{ICT}$, $\text{SMF}$, and $\text{BP}\,\tau = 7$ the number of tests is fixed to $N_{\text{test}} = 220$ for StEM and $N_{\text{test}} = 230$ for Covasim.
However, it is worth noting that probabilistic contact tracing methods like BP and SMF allow for an alternative testing strategy. This approach involves observing individuals whose estimated probability of being infected exceeds a threshold value ($p_{\text{th}}$). In this case, the number of tests based on the ranking changes adaptively over time. For StEM, we set $p_{\text{th}} = 1 \times 10^{-4}$, and $p_{\text{th}} = 5 \times 10^{-5}$ for BP and SMF respectively, while for Covasim, we set $p_{\text{th}} = 3 \times 10^{-3}$ for SMF and $p_{\text{th}} = 5 \times 10^{-3}$ for BP (see Figure S4 and S5 of the Supplementary Information for the performances of the two algorithms under varying thresholds).  
The significant advantage inherent in this testing strategy is that each test is performed based on an estimate of the individual's medical status. This has a twofold impact. First, when no individual is eligible for testing, no diagnostic test is administered, leading to a more parsimonious use of medical resources compared to the fixed $N_{\text{test}}$ setting. Secondly, this approach addresses ethical considerations by encouraging testing only for individuals with a high likelihood of being infected.

To quantify the effectiveness of each containment policy and to set the stage for the analysis carried out in the next sections, Figure~\ref{fig:fig1} shows the effective reproduction number $R_{t}$ (refer to the Methods section for a definition), the cumulative number $N_I$ of the infected individuals and the cumulative number $N_{\rm test}$ of performed tests (included those administered to symptomatic individuals) over time. In both models, all non-probabilistic methods face challenges in sustaining $R_t$ below one, even in the long run, whereas BP, and to a lesser extent SMF, prove to be more adept at achieving this goal swiftly.

\subsection{Cost-Benefit Analysis}
\begin{figure}[t!]
	\centering
	\includegraphics[width=\linewidth]{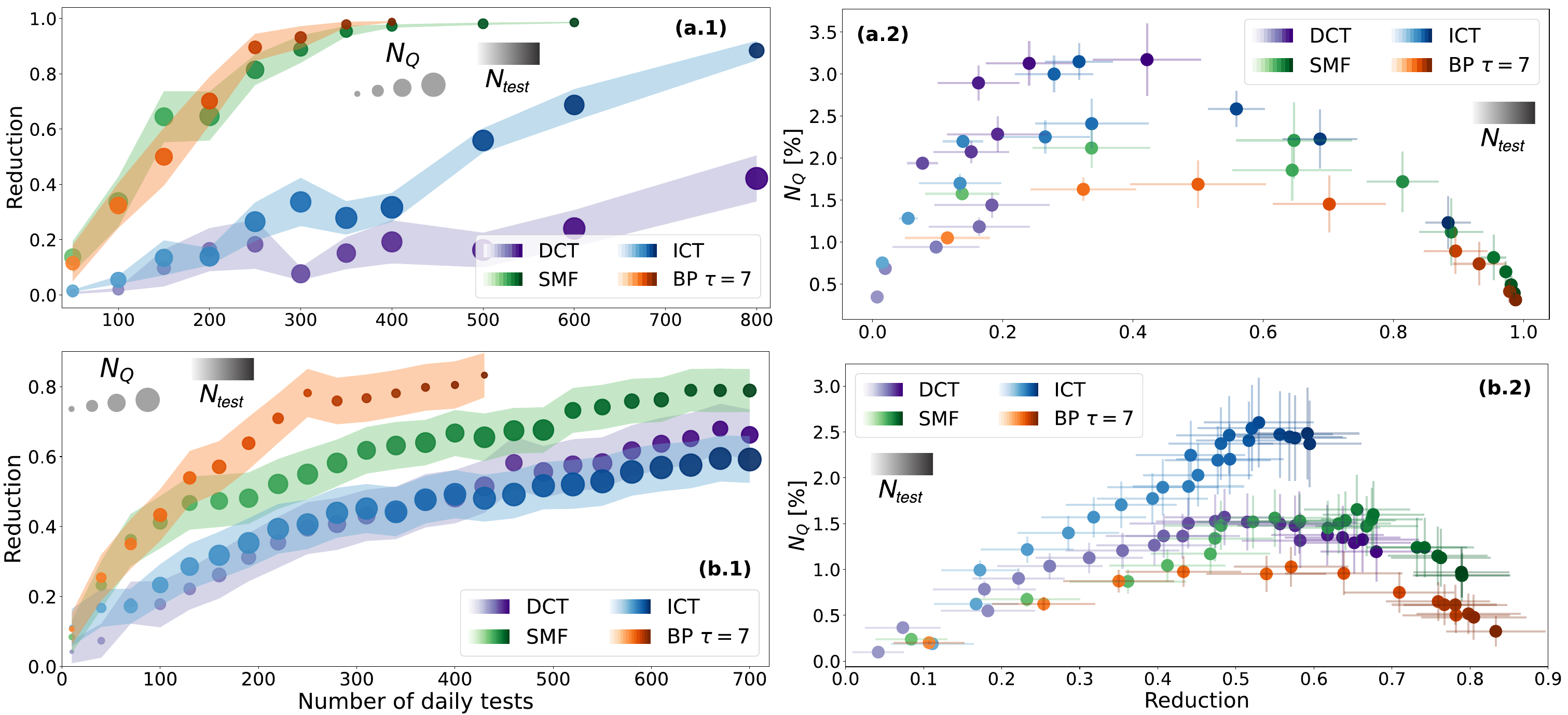}
	\caption{\textbf{Spreading reduction, social, and diagnostic cost.}
		Panels (a.1) and (b.1) show the reduction measure of the epidemic spreading as a function of the number of medical tests performed daily during the simulations; panels (a.2) and (b.2) display $N_{Q}$, the percentage of the confined individuals due to the different confinement strategies as a function of the reduction (see Ref. \cite{barrat2021effect}). These quantities are computed for $T= 100$ and $T = 150$ for StEM and Covasim respectively, when the number of infected individuals reaches a plateau in the corresponding uncontrolled simulations. For StEM (Covasim), the population has a size of $90,546$ ($70,000$) individuals (see Methods).
		The panels on the top display the two measures associated with the Covasim model, while the panels on the bottom show the results while running StEM dynamics. 
		The reduction measure is formally defined as the difference between the cumulative number of infected in an unconstrained propagation (where only the fixed percentage of symptomatic is confined) and the mitigated one, normalized by the cumulative number of infected in unconstrained dynamics. The higher the reduction, the more effective the containment measure. 
		The size of the markers in panels (a.1) and (b.1) is proportional to the number of quarantines (the quantity plotted in the y-axis of the (a.2) and (b.2) panels), the larger the dots, the larger the number of confined individuals. The color code used in all the panels mirrors the number of tests performed on a daily basis: the darker the color, the larger this number.
		\label{fig:fig2}}
\end{figure}

In addition to the economic costs associated with medical tests, non-pharmacological epidemic containment policies also impose a social cost due to mobility restrictions. This cost can be quantified by measuring the cumulative number, or percentage, $N_Q$ of individuals in quarantine as a result of different contact tracing strategies. This quantity is then compared to the effective reduction in epidemic spread, defined as one minus the ratio between the infected individuals in a mitigated scenario and that in an uncontrolled regime, where only a fixed percentage of symptomatic individuals are tested and quarantined. The values of reduction are computed when the number of infected individuals in uncontrolled simulations reaches a plateau (which happens roughly at $T=100$ for the StEM and at $T=150$ for Covasim). Higher values of reduction indicate better containment performance. 
This cost-to-benefit analysis was first introduced in Ref.~\cite{barrat2021effect}, where the authors investigated a theoretical expectation of the number of required quarantines to achieve a specific reduction in the final epidemic size when manual and digital contact tracing is applied.
For the comparison, the settings described in Figure \ref{fig:fig1} for both the Covasim model and StEM are adopted. Figures \ref{fig:fig2} (a.1) and (b.1) show the reduction measure defined above as a function of the number of tests performed daily, for Covasim and StEM respectively. The size of the markers reflects the cumulative number of quarantined individuals resulting from the employed contact tracing strategy (larger dots correspond to larger numbers). The color gradient represents the number of daily tests conducted during the simulation, with darker colors indicating a larger number of observations. 
As clearly shown by these results, the two probabilistic methods (i.e. SMF and BP) always reach higher performances in terms of reduction at a fixed number of medical tests. 

Similarly, panels (a.2) and (b.2) display the percentage of individuals in quarantine generated by the intervention strategy (excluding isolation associated with symptomatic individuals) as a function of the reduction (see Figure S6 of the Supplementary Information for the plot of the number of confined individuals as a function of the number of daily tests). Regardless of the number of available rapid tests, the number of confined individuals is significantly smaller for probabilistic contact tracing techniques (BP and SMF) than for the others (DCT and ICT).
This suggests that not only the two techniques are preferable in terms of effectiveness, but they also incur a lower social cost as fewer individuals need to be isolated.
Our numerical estimates appear qualitatively similar to the results in Ref.~\cite{barrat2021effect} where the authors predicted a behavior similar to a downward opening parabola for the number of quarantines as a function of the reduction. In our case, we stress that BP-based curves are always associated with lower values of the isolated cases $N_{Q}$ at fixed reduction values. The color gradient in panels (a.2), and (b.2) also reveals that this result is achieved at a lower diagnostic cost as the number of necessary tests to reach the same performance in terms of reduction, is lower than that used by the other methods. This behavior is particularly pronounced in StEM: BP obtains a reduction greater than $0.8$ using about $400$ daily tests while SMF needs at least $700$ observations, and DCT and ICT never reach this value with the number of tests considered for this experiment (see panel (b.1)).

\subsection{Overdispersion and superspreaders \label{subsec:results_supersperaders}}

\begin{figure}[t!]
	\centering
	\includegraphics[width=\textwidth]{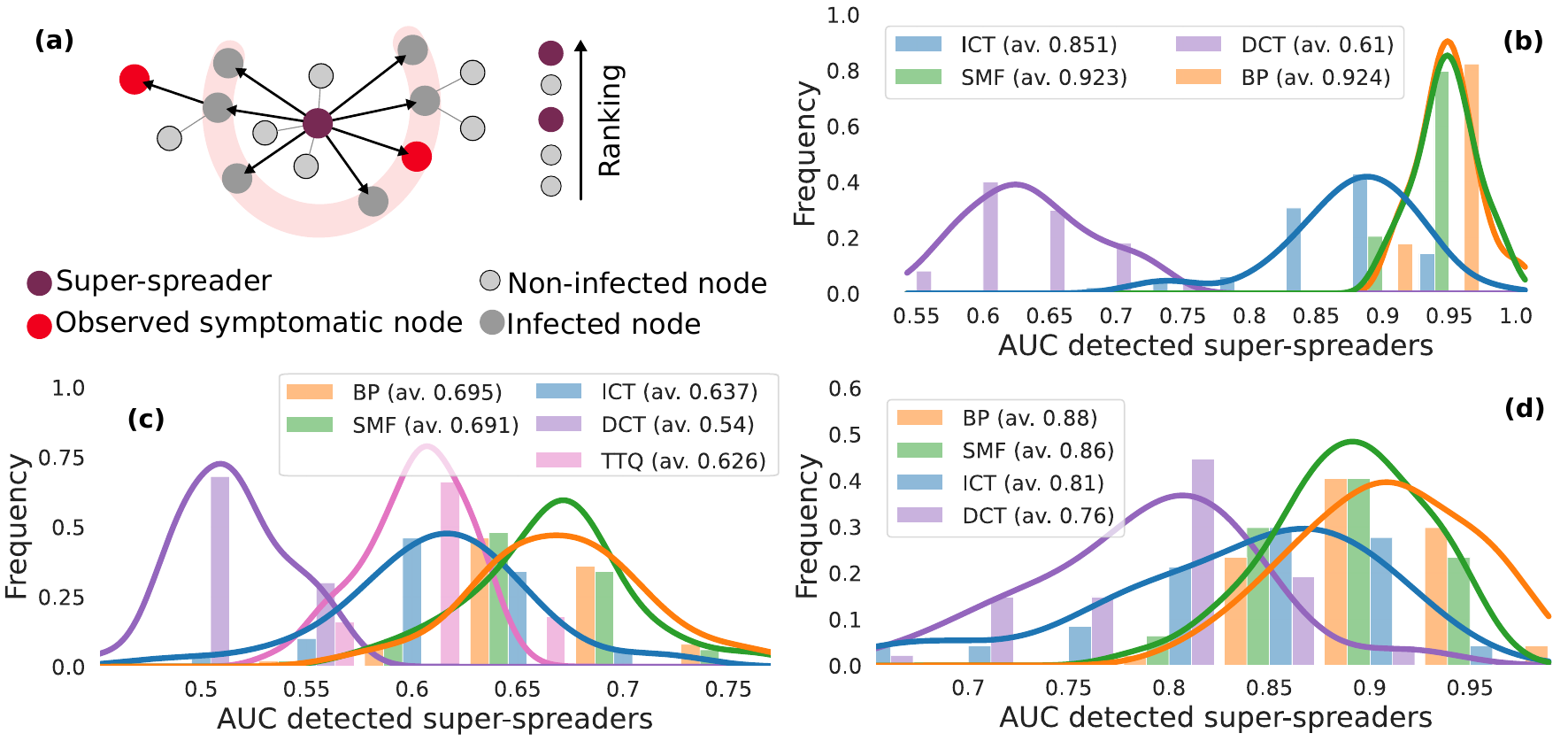}
	\caption{\textbf{Detection of the super-spreaders.} 
		(a) Schematic representation of the experimental setup. A posteriori, the superspreader individuals (the purple nodes) are identified as those responsible for over-dispersed transmissions (see the main text for a proper definition for the three models), here marked as the nodes within the pink shadow. To fairly evaluate the ability of each contact tracing method to detect superspreaders the ROC curves are built only for a subset of the individuals composed of the true superspreaders and susceptible individuals (small light grey nodes). Information about the epidemic dynamics entirely comes from the contact network and the daily observation of a fixed fraction of symptomatic (red nodes). The methods employed to compute the ROC curves are Belief Propagation (BP, orange), Simple Mean Field (SMF, green), Informed Contact Tracing (ICT, blue), Digital Contact Tracing (DCT, purple), Trace-Test-Quarantine (TTQ, pink). The statistics of the AUC associated with the ROC curves obtained by different methods are shown for (b) OpenABM, (c) Covasim, and (d) StEM. Lines are kernel density estimation plots used as guides for the eyes, while mean AUC values are reported in the legend. All parameters used in these simulations are the same as used in the epidemic containment results, except for the time $T$, the number of patients zero $N_{\rm{pz}}$, and the probability of self-testing. These numbers have been tuned to ensure that the maximum number of true positives in the ROC curves is at least a few tens. In particular, the duration of the free epidemic propagation before estimation is set to $T = 15$ for StEM, $T = 30$ for Covasim, and $T = 20$ for OpenABM. The number of initially infected individuals is set to $N_{\rm{pz}} = 200$ for StEM, $N_{\rm{pz}} = 90$ for Covasim and $N_{\rm{pz}} = 100$ for OpenABM. The fraction of observed symptomatic individuals is set to $p_{\rm{sym}} = 0.1$ for StEM and for Covasim, while for OpenABM all severe symptomatic individuals are observed ($p_{\rm{ssym}} = 1.0$) together with a fraction $p_{\rm{msym}} = 0.3$ of mild ones. 
		\label{fig:fig3} }

\end{figure}

Probabilistic-based tracing methods also exhibit a remarkable ability to effectively detect super-spreaders. 
Super-spreading transmission can have distinct origins, contingent on the properties of both the viral disease and the underlying population. This diversity is represented and exemplified by the three agent-based models under study. In OpenABM~\cite{ferretti2020quantifying}, superspreading events occur due to an overdispersed distribution of contacts in one of the three network layers used to model the population structure. Similarly, in StEM~\cite{lorch2022quantifying}, overdispersion arises naturally from the contact graph, as a result of realistic mobility simulations based on geolocalized data within an urban area. In both cases, the empirical distribution of the number of infections exhibits significant non-Poissonian statistics, characterized by a variance-to-mean ratio (VMR) larger than one (refer to the Supplementary Information for further details).
For these two models, individuals who infect at least seven contacts within their infectious time window are identified as superspreaders, following the definition provided in Wong et al. \cite{wong2020evidence}. In contrast, in Covasim \cite{kerr2021covasim}, the overdispersion of infections directly arises from the properties of individual viral load, which is drawn from a fat-tailed distribution (see Supplementary Information): superspreaders can therefore be identified by looking at the individual relative transmission intensity $T_{\rm rel}$, a quenched parameter not accessible to the tracing methods. In particular, in each simulation, individuals displaying $T_{\rm rel} \geq 5$ are classified as superspreaders.

The ability of the different contact tracing methods to detect superspreaders among the infected individuals is evaluated through numerical experiment employing the following procedure: in each epidemic realization, the propagation is allowed to evolve freely without intervention up to a time $T$, whereupon the contact tracing methods are applied once, and the corresponding ranking of potentially infected individuals is collected. The value of $T$ is here chosen to be of the order of a few weeks, representing the typical time window for which contact information can be retained in digital contact tracing applications \cite{baker2021epidemic}. 
To mimic a realistic setting, we assume that individuals showing symptoms spontaneously take tests and their results are collected by the contact tracing app. This is encoded in our simulations by observing a fixed fraction of the symptomatic individuals daily (see caption of Figure~\ref{fig:fig3} for additional details).
Individuals identified by means of the different contact tracing methods, and ranked based on their epidemic risk, are then classified according to their true infection status, obtaining corresponding ROC curves.  
To specifically study the detection of superspreaders (and not other infected individuals), only the subset consisting of (a posteriori determined and non-observed) superspreaders and susceptible individuals at time $T$ was considered (refer to Figure \ref{fig:fig3}a for a schematic representation of the setup). Superspreaders who recovered before time $T$ were not taken into account, as their number is negligible after $T$ days. 
Figures \ref{fig:fig3}b-\ref{fig:fig3}d illustrate the empirical distributions of the area under the curve (AUC) obtained from different contact tracing methods across multiple epidemic realizations for OpenABM, Covasim, and StEM. In all three models, probabilistic methods (SMF and BP) turn out to better differentiate between non-infected and superspreaders, as indicated by both the distribution of the AUC (it is significantly shifted towards larger values for SMF and BP) and the average value of the AUC shown in the legend. Conversely, the distributions associated with ICT, DCT (and TTQ for Covasim) predictions are concentrated at lower values, confirming that non-probabilistic algorithms are less effective in tracing superspreader exposures.

\subsection{Backward and forward tracing\label{subsec:fwdbwd_infections}}
\begin{figure}[t!]
	\centering
	\includegraphics[width=\linewidth]{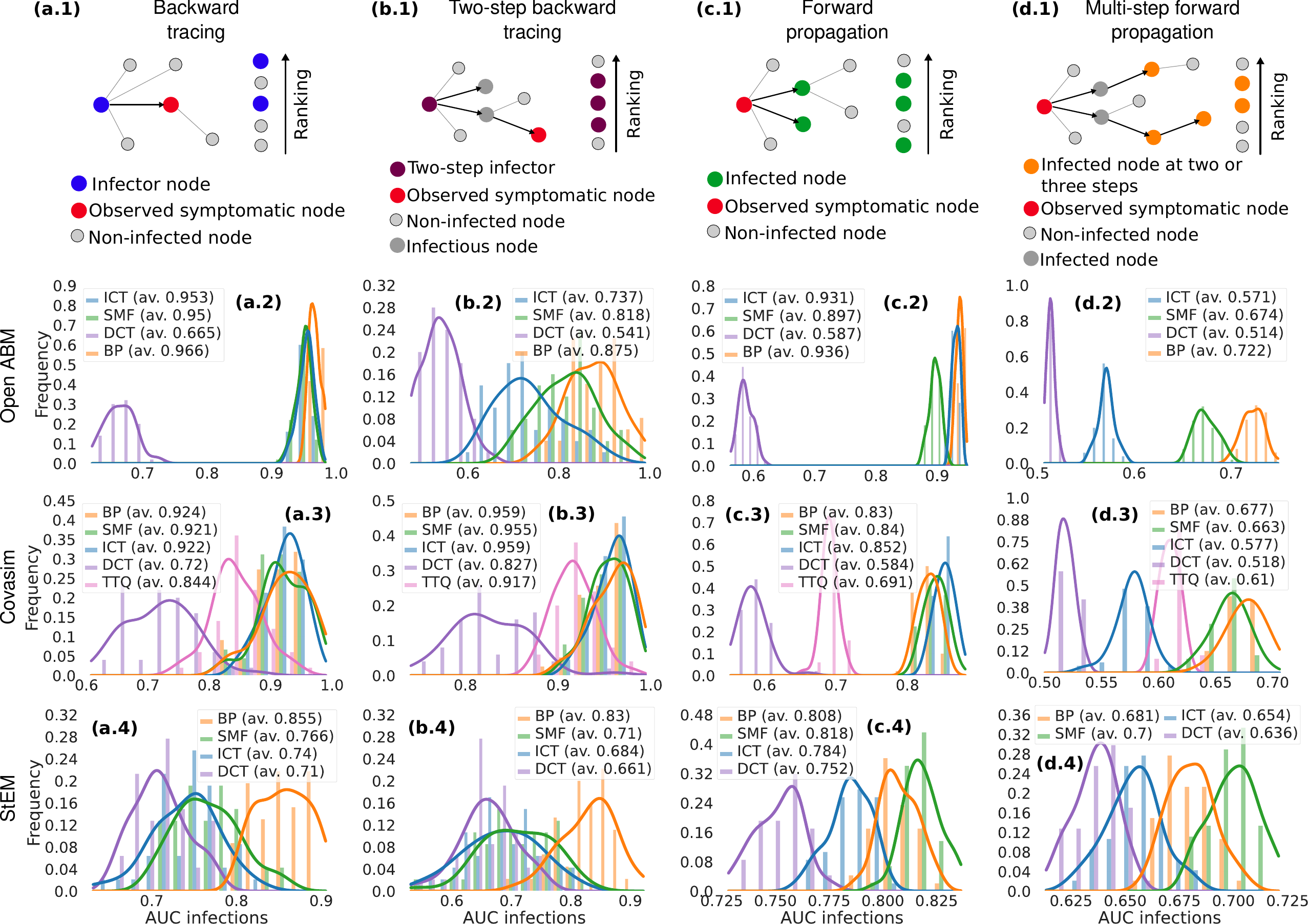}
	\caption{\textbf{Detection of the one-step, two-step backward, and one-step, multi-step forward tracing.} Panels (a.1), (b.1), (c.1), and (d.1) show a schematic representation of the one-step, two-step backward, one-step, and multi-step forward transmissions respectively. See the main text for a formal definition. The second, third, and fourth rows show the histogram of the AUC associated with the detection of the four types of infected individuals, for OpenABM, Covasim, and StEM respectively. The methods used to obtain the ROC curves are Belief Propagation (BP, orange), Simple Mean Field (SMF), Informed Contact Tracing (ICT, blue), Digital Contact Tracing (DCT, purple), and Trace-Test-Quarantine (TTQ, pink). The simulation set-up used for these results is the same exploited for the detection of the super-spreaders illustrated in Figure \ref{fig:fig3}. The average AUC is reported in the legend for the methods, while the lines report kernel density estimates to guide the visualization of the histograms. 
		\label{fig:fig4}}
	
\end{figure}

One of the inherent difficulties in contact tracing is determining the direction of infection among confirmed cases. While tracing new infections (forward tracing) is relatively easier, a more complex task is to trace the source of the observed infections (backward tracing). The ability to identify transmissions backward is crucial for detecting superspreaders and effectively mitigating the spread of an outbreak \cite{tufekci_this_2020,bradshaw2021bidirectional}. To further emphasize the advantages of probabilistic contact tracing methods like SMF and BP, it is valuable to assess their ability to identify secondary and tertiary infections, i.e., new infections that occur two or three steps away from the observed individuals in the transmission history. 
The experimental setup employed in Figure \ref{fig:fig4} consists of the following: for each epidemic realization, the propagation is allowed to evolve without intervention until a time $T$, and a small fraction of symptomatic individuals is observed daily. The backward propagators are defined as the sources of infection for the observed symptomatic individuals (depicted as blue dots in Figure \ref{fig:fig4}a.1); their infectors instead identify the two-step backward propagators (see Figure \ref{fig:fig4}b.1). Forward propagators are defined as the secondary infections of observed individuals (represented by green nodes in Figure \ref{fig:fig4}c.1). New infections occurring at two and three steps from the observed individuals are shown as orange nodes in the example presented in Figure \ref{fig:fig4}d.1. 
To quantify the performances of the ranking methods, a comparison is made using the AUC associated with the classification of the infected individuals in a restricted set, where the false positive set comprises all non-infected individuals (light grey nodes in Figure~\ref{fig:fig4}a.1--\ref{fig:fig4}d.1) while the true positive set consists of the unobserved one-step and two-step backward infectors, forward infections, or new infections occurring at steps two and three, respectively. Other infected individuals not belonging to these three categories (e.g., tested-positive individuals, represented by red nodes in Figure \ref{fig:fig4}a.1--\ref{fig:fig4}d.1) are not considered. Although the performances vary across the three epidemic models, the results in Figure \ref{fig:fig4} demonstrate that probabilistic models such as BP and SMF are highly effective in identifying transmissions forward and backward. For OpenABM (panels a.2--d.2) and StEM (panels a.4--d.4) probabilistic contact tracing methods outperform the others, particularly when detecting one-step, two-step backward. 
In the case of Covasim (panels a.3--d.3), probabilistic methods appear to play a crucial role mainly in detecting multi-step forward transmissions, while their performances are similar to ICT in detecting backward and one-step forward transmissions. In these last three scenarios, simpler and less computationally expensive non-probabilistic contact tracing methods (DCT and TTQ) do not reach the same AUC values achieved by ICT. We stress that although TTQ includes additional information about the symptomatic status of the individuals, it still does not attain the accuracy of probabilistic methods.

\section{Discussion}

Contact tracing stands out as a compelling strategy to support and improve the effectiveness of common non-pharmaceutical mitigation measures, such as social distancing, the use of masks, and other hygiene practices, in order to contain the spread of emerging viral diseases. This approach has the potential to prevent the need for measures with significant socioeconomic impacts, such as lockdowns.
In particular, digital contact tracing overcomes the limitation of manual contact tracing by encompassing the ability to detect pre-symptomatic and asymptomatic individuals outside of close and known relationships with tested individuals, a key aspect in the prevention of highly contagious diseases, such as COVID-19. 
The primary drawback of current implementations of digital contact tracing is that the volume of exposure notifications delivered drastically grows with the outbreak size. Consequently, the number of individuals flagged for testing grows substantially, rendering the overall procedure impractical. A potential solution to this challenge involves enhancing individual-based epidemic risk assessment and using it to guide selective test-trace-isolate/quarantine strategies. This can be accomplished by integrating contact tracing with distributed statistical inference methods, capable of reconstructing contagion channels from locally collected information and providing a more accurate estimate of individual risk \cite{baker2021epidemic}. These algorithms can be implemented in a privacy-preserving distributed way through smartphone apps based on current technology and without the need for centralized calculations. It was estimated that when implementing BP or SMF the amount of information sent and received between two users could be approximately 1 megabyte (MB) or 2 kilobytes (KB) per day, respectively \cite{baker2021epidemic}.

The present work builds on this direction providing a quantitative comparative analysis of the performance of different contact tracing methods across various epidemic regimes using three distinct epidemic models recently developed for COVID-19. 
In all scenarios under study, probabilistic contact-tracing methods effectively curb ongoing outbreaks, as indicated by the rapid reduction of the effective reproduction number below the critical value of one. This is achieved with a substantially lower cumulative number of infected individuals compared to other methods, all while incurring a similar or significantly reduced deployment of testing resources. The cost (number of quarantines) versus benefit (outbreak reduction) analysis clearly shows a more favorable ratio for probabilistic contact-tracing methods, in particular for BP.  
Note that experiments conducted in this work utilize imperfect information about the underlying contact network (specifically, not all exposure events are assumed in StEM to be traceable, and contact strength is highly variable in Covasim but this information is not available to the inference algorithms). Other regimes with more uncertainty on the contact network will be investigated in future developments.

The numerical experiments also revealed that probabilistic methods are better suited than others to detect super-spreading events, whether stemming from an innate variety of transmissibility or the heterogeneity of the contact network. This capability is crucial for the containment of emerging viral diseases characterized by overdispersion in secondary infections. Due to the presence of superspreaders, it becomes essential to work backward to identify the sources of infection for observed cases, as many individuals are likely infected by someone who also transmitted the virus to other people. In this respect, probabilistic contact tracing methods were found to outperform other methods in correctly reconstructing infection channels by one-step and multi-step backward and forward tracing. 
On one hand, these results provide valuable insights into the mechanisms and causal patterns that govern the detailed functioning of contact tracing. On the other hand, the emerged superiority of probabilistic methods demonstrates that greater effectiveness in detecting super-spreading events and backward and multi-step causal relationships is crucial for successful epidemic containment strategies. 

Our findings also have several practical implications. As pointed out by the threshold-based probabilistic methods, early interventions using a large number of tests appear to be always advantageous. This approach ensures a better assessment of the population-wide epidemic risk during the initial phase of the outbreak and prompt employment of a possibly large number of quarantines, if necessary. This strategy is particularly effective when a possibly large number of cheap, low-sensitivity rapid tests is available \cite{kennedy2021perfect}, as the prior information about the sensitivity of the tests can be included in the Bayesian probabilistic approach \cite{baker2021epidemic}.
Notice that a timely intervention ensures better containment in the long run, but also a lower time-integrated social cost (e.g. lower total number of isolated individuals). In this regard, a more in-depth study of probabilistic contact tracing strategies with intervention thresholds appears compelling.

In the numerical experiments, all contact tracing methods are either model-free (DCT, ICT, TTQ) or assume much simpler epidemic models (BP, SMF) compared to those used to generate the underlying epidemic traces. It follows that the superior performance shown by probabilistic contact tracing methods is not attributable to a greater knowledge of the real transmission mechanisms of the specific epidemics. This feature makes us believe that the results discussed in this work will remain consistent, at least qualitatively, in a real-world scenario with epidemic data from the necessarily much more complex diffusion in a human population, in the event of new variants or other emerging diseases with similar properties. 
Finally, these probabilistic contact tracing methods are sufficiently flexible to work with other prevention and mitigation measures, especially in the presence of vaccinated individuals and selective mobility restrictions.

\section{Methods}

\paragraph{Contact tracing methods.}
We report a brief description of the overall set of ranking techniques referring to the Supplementary Information for the implementation details.
\begin{itemize}
	\item{Digital contact tracing (DCT). When individuals are tested positive, their recent contacts (within a one-week time window) are considered eligible for testing. When the number of individuals to be reached exceeds the number of available tests, we uniformly sample for testing as many of them as the maximum number of tests. This protocol is similar to the one published in Barrat et al. \cite{barrat2021effect}. }
	\item{Informed contact tracing (ICT). Similarly to the probabilistic contact tracing technique, this method returns a score quantifying how likely each individual may be infected at the observation time. Exploiting both the positive and negative results of the tests, this method counts the number of potentially exposed events that each individual has had within a temporal window of one week. This type of potentially infectious contact occurs if (a) the considered individual has never been tested or has always been negatively tested in the past, and (b) the time of the contact lies in the time interval ranging from the last time the potential infector has been negatively tested (before being positively tested), and the first time it has been negatively tested after the infection (in case there is no such occurrence, this corresponds to the observation time). Here we have assumed that the process is irreversible, or, in other words, the time window we consider is sufficiently small to assume that, after a first infection, the acquired immunity preserves individuals from further infections after recovery. 
	}
	\item{Simple Mean Field (SMF). This method assumes Markovian Susceptible-Infected-Recovered (SIR) dynamics with, when available to the app, heterogeneous infection probabilities mirroring, for instance, a diverse duration of the contacts. When an individual tests positive, SMF assumes that the infection occurred $t_{\rm SMF}$ days before (here and in Baker et al. \cite{baker2021epidemic}  $t_{\rm SMF} = 5$ as it seems to better fit the COVID-19 features).
		Finally, the SMF-based ranker estimates an approximated marginal probability of the state of all individuals at each time step of the dynamics. The values obtained at the observation time for the infected state are considered as a proxy for the individual risks. More details can be found in Ref. \cite{baker2021epidemic} and in the SI.} 
	\item{Belief Propagation (BP). Similarly to SMF, BP assumes that the underlying infection can be modeled as a SIR dynamic. Though, at difference with SMF, some intrinsic COVID-19 features are encoded in time-dependent infection and recovery rates resulting in a Non-Markovian SIR model. Observations of the states of the individuals, i.e. the results of the medical tests, are properly introduced in the model by means of a Bayesian framework. This allows us to deal with imprecise test outcomes, mirroring the false negative and positive rates of the tests. Through the application of BP, the overall epidemic dynamics are reconstructed by inferring the infection and recovery times of all individuals. From this information, one can compute the individual probability of being infected at the observation time and, therefore, an estimate of the risk. The label $\tau = 7$  refers to the implementation used in Ref. \cite{baker2021epidemic}, in which the risk is computed from the aggregate probability of infection and recovery times in a time window of $\tau = 7$ days. When a threshold is set, all individuals with a probability of being in the infected state larger than the threshold are tested. See Ref. \cite{baker2021epidemic} for a detailed description and Supplementary Information for the implementation details used for the StEM, Covasim, and OpenABM.} 
	\item{Test-Trace-Quarantine (TTQ). This containment strategy is integrated into Covasim \cite{kerr2021controlling} and relies on the manual contact tracing process included in the model. This method traces individuals who have come in contact with confirmed infected ones (with a probability $p_{\rm trace}$ for each contact to be traced) and puts them in the so-called pre-emptive quarantine (PQ). In this state, which is unique to the Covasim model, individuals reduce their infectiousness levels. In the TTQ strategy, individuals are tested each day with a probability that depends both on their state (symptomatic or asymptomatic) and the time elapsed since their entrance into PQ (see Supplementary Information for details).
		As implemented in Covasim, this strategy does not limit the number of tests performed each day. To perform a fair comparison with the other containment techniques, in the regime of a limited number of tests, we adopted a modified version of the process, called TTQ-N, where the individuals to be tested are randomly chosen, drawing first from the set of symptomatic individuals and then with a probability proportional to the one used in TTQ. The process stops when the maximum number of tests is reached. Moreover, since individuals in the PQ state are subject to a reduction in transmission probability, in the results on Covasim shown in Figure \ref{fig:fig1}, manual contact tracing is applied together with the other tracing methods in order to produce fair comparisons between TTQ and the other methods.
	}
\end{itemize}
\paragraph{Agent-based models.} This section contains some important implementation details of the agent-based simulations. A brief description of the three models considered is reported in the SI.
\begin{itemize}
	\setlength\itemsep{0.cm}
	\item{OpenABM. 
		The model introduced in Ref. \citep{hinch2021openabm} exploits discrete-time non-Markovian stochastic processes to simulate an epidemic spreading on an age-stratified population interacting on a multi-layer synthetic graph, with demographic data based on the UK census (additional details are given in the Supplementary Information). 
		The efficacy of probabilistic inference using BP and SMF against standard contact-tracing technique in the epidemic containment was already discussed in \citep{baker2021epidemic}.  Here we focus only on quantifying their performance w.r.t. the detection of super spreaders and forward/backward infections. The results presented in Sections \ref{subsec:results_supersperaders} and \ref{subsec:fwdbwd_infections}
		are obtained by simulating a population of $N=10^5$ individuals for $T=20$ days, with an initial number $N_{pz} = 100$ of infected individuals. All the other model parameters are not changed with respect to the default implementation of the simulator discussed in the original work. As the OpenABM model distinguishes between asymptomatic states and different classes of symptomatic ones (mild, severe), observations are performed on a daily basis on the full population of severe symptomatic and on $30\%$ of mild symptomatic individuals, i.e. the same setting used in \citep{baker2021epidemic} for the online containment.}
	\item{Covasim. The work in Ref.\cite{kerr2021covasim} introduces Covasim, an agent-based model that includes country-specific demographic information such as age structure and population size. The contact networks used in Covasim comprise both an individual scale (these contacts are static) and a community scale (these interactions are randomly redrawn over time) to cope with households and social interactions. For this work, we use a population of $70,000$ individuals, with contact features matching those of the Seattle Metropolitan Area (as done in Ref.\cite{kerr2021covasim}).
		The epidemic model underlying the Covasim dynamic is a discrete-time non-Markovian process involving susceptible, exposed, several infectious states (an asymptomatic and pre-symptomatic state and three symptomatic states to account for mild, severe, and critical conditions), as well as a recovered state. All individual transition times between these states are log-normally distributed. Special attention is devoted to the transmission of the disease; when a susceptible and an infectious individual meet, the transmission probability associated with this event depends on both individual viral-load-based transmissibility and susceptibility, and the social layer the contact belongs to. These ingredients favor the occurrence of super-spreading events.}
	\item{StEM. The model proposed in Ref.~\cite{lorch2022quantifying} combines publicly available demographic data and automatic geo-referencing to produce continuous-time individual mobility traces with realistic features. In particular, we run mobility simulations on the urban area of T{\"u}bingen (Germany), having $90,546$ individuals distributed in $47,309$ houses. All the accessible venues fall into five categories (education, social places, public transport, offices, and supermarkets). Each inhabitant can visit a subset (one education venue, ten social places, five public transportation, one office, and two supermarkets) of the $1,487$ available locations assigned with a probability that depends on the house-location distance. The duration of the visits depends on the location ($2$ hours at education, $1.5$ hours at social places, $0.2$ hours for public transport, $2$ hours for working places, and $0.5$ hours supermarket).
		Simulated mobility data is used to compute infectious contacts within a continuous-time non-Markovian stochastic model that includes an exposed state and multiple infected states to cope with asymptomatic, pre-symptomatic, and symptomatic individuals. Exposures depend on the state of the infectors, an exposure rate (set to $0.05$ for all locations), and a kernel term that allows one to accommodate environmental transmissions. All contacts are available to the containment methods except those due to a small but continuous influx of untraceable exogenous exposures (as in the default setting, we set five of such events per $100,000$ inhabitants and per week).}
\end{itemize}
\vspace{-.5cm}
\section*{Acknowledgements}
We are grateful to Lars Lorch for his help in managing the implementation of StEM.
Computational resources were provided by HPC@POLITO, a project of Academic Computing within the Department of Control and Computer Engineering at the Politecnico di Torino (\href{http://www.hpc.polito.it}{www.hpc.polito.it}), and by the SmartData@PoliTO (\href{http://smartdata.polito.it}{smartdata.polito.it}) interdepartmental center on Big Data and Data Science. 
\subsection*{Funding:} G.C. acknowledges support from the Comunidad de Madrid and the Complutense University of Madrid (UCM) through the Atracción de Talento programs (Refs. 2019-T1/TIC-13298). This study was carried out within the FAIR - Future Artificial Intelligence Research project and received funding from the European Union Next-GenerationEU (Piano Nazionale di Ripresa e Resilienza (PNRR) – Missione 4 Componente 2, Investimento 1.3 – D.D. 1555 11/10/2022, PE00000013). This manuscript reflects only the authors’ views and opinions, neither the European Union nor the European Commission can be considered responsible for them.
\subsection*{Author contributions:} L.D., A.B., and A.P.M. designed the research. A.P.M. and F.M. performed simulations about the epidemic containment. A.P.M., F.M., and G.C. performed static analysis on the transmission features. All the authors analyzed the results and wrote the manuscript. 
\subsection*{Competing interests:} The authors declare that they have no competing interests.
\subsection*{Data and materials availability:} All data, codes, and materials used for producing these results are available at \cite{sibGitHub}. All other data are present in the paper and the Supplementary Information.

\bibliographystyle{apsrev4-1}


\begin{thebibliography}{79}%
	\makeatletter
	\providecommand \@ifxundefined [1]{%
		\@ifx{#1\undefined}
	}%
	\providecommand \@ifnum [1]{%
		\ifnum #1\expandafter \@firstoftwo
		\else \expandafter \@secondoftwo
		\fi
	}%
	\providecommand \@ifx [1]{%
		\ifx #1\expandafter \@firstoftwo
		\else \expandafter \@secondoftwo
		\fi
	}%
	\providecommand \natexlab [1]{#1}%
	\providecommand \enquote  [1]{``#1''}%
	\providecommand \bibnamefont  [1]{#1}%
	\providecommand \bibfnamefont [1]{#1}%
	\providecommand \citenamefont [1]{#1}%
	\providecommand \href@noop [0]{\@secondoftwo}%
	\providecommand \href [0]{\begingroup \@sanitize@url \@href}%
	\providecommand \@href[1]{\@@startlink{#1}\@@href}%
	\providecommand \@@href[1]{\endgroup#1\@@endlink}%
	\providecommand \@sanitize@url [0]{\catcode `\\12\catcode `\$12\catcode
		`\&12\catcode `\#12\catcode `\^12\catcode `\_12\catcode `\%12\relax}%
	\providecommand \@@startlink[1]{}%
	\providecommand \@@endlink[0]{}%
	\providecommand \url  [0]{\begingroup\@sanitize@url \@url }%
	\providecommand \@url [1]{\endgroup\@href {#1}{\urlprefix }}%
	\providecommand \urlprefix  [0]{URL }%
	\providecommand \Eprint [0]{\href }%
	\providecommand \doibase [0]{http://dx.doi.org/}%
	\providecommand \selectlanguage [0]{\@gobble}%
	\providecommand \bibinfo  [0]{\@secondoftwo}%
	\providecommand \bibfield  [0]{\@secondoftwo}%
	\providecommand \translation [1]{[#1]}%
	\providecommand \BibitemOpen [0]{}%
	\providecommand \bibitemStop [0]{}%
	\providecommand \bibitemNoStop [0]{.\EOS\space}%
	\providecommand \EOS [0]{\spacefactor3000\relax}%
	\providecommand \BibitemShut  [1]{\csname bibitem#1\endcsname}%
	\let\auto@bib@innerbib\@empty
	\bibitem [{\citenamefont {Bonaccorsi}\ \emph {et~al.}(2020)\citenamefont
		{Bonaccorsi}, \citenamefont {Pierri}, \citenamefont {Cinelli}, \citenamefont
		{Flori}, \citenamefont {Galeazzi}, \citenamefont {Porcelli}, \citenamefont
		{Schmidt}, \citenamefont {Valensise}, \citenamefont {Scala}, \citenamefont
		{Quattrociocchi} \emph {et~al.}}]{bonaccorsi2020economic}%
	\BibitemOpen
	\bibfield  {author} {\bibinfo {author} {\bibfnamefont {G.}~\bibnamefont
			{Bonaccorsi}}, \bibinfo {author} {\bibfnamefont {F.}~\bibnamefont {Pierri}},
		\bibinfo {author} {\bibfnamefont {M.}~\bibnamefont {Cinelli}}, \bibinfo
		{author} {\bibfnamefont {A.}~\bibnamefont {Flori}}, \bibinfo {author}
		{\bibfnamefont {A.}~\bibnamefont {Galeazzi}}, \bibinfo {author}
		{\bibfnamefont {F.}~\bibnamefont {Porcelli}}, \bibinfo {author}
		{\bibfnamefont {A.~L.}\ \bibnamefont {Schmidt}}, \bibinfo {author}
		{\bibfnamefont {C.~M.}\ \bibnamefont {Valensise}}, \bibinfo {author}
		{\bibfnamefont {A.}~\bibnamefont {Scala}}, \bibinfo {author} {\bibfnamefont
			{W.}~\bibnamefont {Quattrociocchi}},  \emph {et~al.},\ }\href@noop {}
	{\bibfield  {journal} {\bibinfo  {journal} {Proceedings of the National
				Academy of Sciences}\ }\textbf {\bibinfo {volume} {117}},\ \bibinfo {pages}
		{15530} (\bibinfo {year} {2020})}\BibitemShut {NoStop}%
	\bibitem [{\citenamefont {Perra}(2021)}]{perra2021non}%
	\BibitemOpen
	\bibfield  {author} {\bibinfo {author} {\bibfnamefont {N.}~\bibnamefont
			{Perra}},\ }\href@noop {} {\bibfield  {journal} {\bibinfo  {journal} {Physics
				Reports}\ }\textbf {\bibinfo {volume} {913}},\ \bibinfo {pages} {1} (\bibinfo
		{year} {2021})}\BibitemShut {NoStop}%
	\bibitem [{\citenamefont {Kretzschmar}\ \emph {et~al.}(2022)\citenamefont
		{Kretzschmar}, \citenamefont {Ashby}, \citenamefont {Fearon}, \citenamefont
		{Overton}, \citenamefont {Panovska-Griffiths}, \citenamefont {Pellis},
		\citenamefont {Quaife}, \citenamefont {Rozhnova}, \citenamefont {Scarabel},
		\citenamefont {Stage} \emph {et~al.}}]{kretzschmar2022challenges}%
	\BibitemOpen
	\bibfield  {author} {\bibinfo {author} {\bibfnamefont {M.~E.}\ \bibnamefont
			{Kretzschmar}}, \bibinfo {author} {\bibfnamefont {B.}~\bibnamefont {Ashby}},
		\bibinfo {author} {\bibfnamefont {E.}~\bibnamefont {Fearon}}, \bibinfo
		{author} {\bibfnamefont {C.~E.}\ \bibnamefont {Overton}}, \bibinfo {author}
		{\bibfnamefont {J.}~\bibnamefont {Panovska-Griffiths}}, \bibinfo {author}
		{\bibfnamefont {L.}~\bibnamefont {Pellis}}, \bibinfo {author} {\bibfnamefont
			{M.}~\bibnamefont {Quaife}}, \bibinfo {author} {\bibfnamefont
			{G.}~\bibnamefont {Rozhnova}}, \bibinfo {author} {\bibfnamefont
			{F.}~\bibnamefont {Scarabel}}, \bibinfo {author} {\bibfnamefont {H.~B.}\
			\bibnamefont {Stage}},  \emph {et~al.},\ }\href@noop {} {\bibfield  {journal}
		{\bibinfo  {journal} {Epidemics}\ }\textbf {\bibinfo {volume} {38}},\
		\bibinfo {pages} {100546} (\bibinfo {year} {2022})}\BibitemShut {NoStop}%
	\bibitem [{\citenamefont {Hellewell}\ \emph {et~al.}(2020)\citenamefont
		{Hellewell}, \citenamefont {Abbott}, \citenamefont {Gimma}, \citenamefont
		{Bosse}, \citenamefont {Jarvis}, \citenamefont {Russell}, \citenamefont
		{Munday}, \citenamefont {Kucharski}, \citenamefont {Edmunds}, \citenamefont
		{Sun} \emph {et~al.}}]{hellewell2020feasibility}%
	\BibitemOpen
	\bibfield  {author} {\bibinfo {author} {\bibfnamefont {J.}~\bibnamefont
			{Hellewell}}, \bibinfo {author} {\bibfnamefont {S.}~\bibnamefont {Abbott}},
		\bibinfo {author} {\bibfnamefont {A.}~\bibnamefont {Gimma}}, \bibinfo
		{author} {\bibfnamefont {N.~I.}\ \bibnamefont {Bosse}}, \bibinfo {author}
		{\bibfnamefont {C.~I.}\ \bibnamefont {Jarvis}}, \bibinfo {author}
		{\bibfnamefont {T.~W.}\ \bibnamefont {Russell}}, \bibinfo {author}
		{\bibfnamefont {J.~D.}\ \bibnamefont {Munday}}, \bibinfo {author}
		{\bibfnamefont {A.~J.}\ \bibnamefont {Kucharski}}, \bibinfo {author}
		{\bibfnamefont {W.~J.}\ \bibnamefont {Edmunds}}, \bibinfo {author}
		{\bibfnamefont {F.}~\bibnamefont {Sun}},  \emph {et~al.},\ }\href@noop {}
	{\bibfield  {journal} {\bibinfo  {journal} {The Lancet Global Health}\
		}\textbf {\bibinfo {volume} {8}},\ \bibinfo {pages} {e488} (\bibinfo {year}
		{2020})}\BibitemShut {NoStop}%
	\bibitem [{\citenamefont {Lavezzo}\ \emph {et~al.}(2020)\citenamefont
		{Lavezzo}, \citenamefont {Franchin}, \citenamefont {Ciavarella},
		\citenamefont {Cuomo-Dannenburg}, \citenamefont {Barzon}, \citenamefont
		{Del~Vecchio}, \citenamefont {Rossi}, \citenamefont {Manganelli},
		\citenamefont {Loregian}, \citenamefont {Navarin} \emph
		{et~al.}}]{lavezzo2020suppression}%
	\BibitemOpen
	\bibfield  {author} {\bibinfo {author} {\bibfnamefont {E.}~\bibnamefont
			{Lavezzo}}, \bibinfo {author} {\bibfnamefont {E.}~\bibnamefont {Franchin}},
		\bibinfo {author} {\bibfnamefont {C.}~\bibnamefont {Ciavarella}}, \bibinfo
		{author} {\bibfnamefont {G.}~\bibnamefont {Cuomo-Dannenburg}}, \bibinfo
		{author} {\bibfnamefont {L.}~\bibnamefont {Barzon}}, \bibinfo {author}
		{\bibfnamefont {C.}~\bibnamefont {Del~Vecchio}}, \bibinfo {author}
		{\bibfnamefont {L.}~\bibnamefont {Rossi}}, \bibinfo {author} {\bibfnamefont
			{R.}~\bibnamefont {Manganelli}}, \bibinfo {author} {\bibfnamefont
			{A.}~\bibnamefont {Loregian}}, \bibinfo {author} {\bibfnamefont
			{N.}~\bibnamefont {Navarin}},  \emph {et~al.},\ }\href@noop {} {\bibfield
		{journal} {\bibinfo  {journal} {Nature}\ }\textbf {\bibinfo {volume} {584}},\
		\bibinfo {pages} {425} (\bibinfo {year} {2020})}\BibitemShut {NoStop}%
	\bibitem [{\citenamefont {Bi}\ \emph {et~al.}(2020)\citenamefont {Bi},
		\citenamefont {Wu}, \citenamefont {Mei}, \citenamefont {Ye}, \citenamefont
		{Zou}, \citenamefont {Zhang}, \citenamefont {Liu}, \citenamefont {Wei},
		\citenamefont {Truelove}, \citenamefont {Zhang} \emph
		{et~al.}}]{bi2020epidemiology}%
	\BibitemOpen
	\bibfield  {author} {\bibinfo {author} {\bibfnamefont {Q.}~\bibnamefont
			{Bi}}, \bibinfo {author} {\bibfnamefont {Y.}~\bibnamefont {Wu}}, \bibinfo
		{author} {\bibfnamefont {S.}~\bibnamefont {Mei}}, \bibinfo {author}
		{\bibfnamefont {C.}~\bibnamefont {Ye}}, \bibinfo {author} {\bibfnamefont
			{X.}~\bibnamefont {Zou}}, \bibinfo {author} {\bibfnamefont {Z.}~\bibnamefont
			{Zhang}}, \bibinfo {author} {\bibfnamefont {X.}~\bibnamefont {Liu}}, \bibinfo
		{author} {\bibfnamefont {L.}~\bibnamefont {Wei}}, \bibinfo {author}
		{\bibfnamefont {S.~A.}\ \bibnamefont {Truelove}}, \bibinfo {author}
		{\bibfnamefont {T.}~\bibnamefont {Zhang}},  \emph {et~al.},\ }\href@noop {}
	{\bibfield  {journal} {\bibinfo  {journal} {The Lancet infectious diseases}\
		}\textbf {\bibinfo {volume} {20}},\ \bibinfo {pages} {911} (\bibinfo {year}
		{2020})}\BibitemShut {NoStop}%
	\bibitem [{\citenamefont {Keeling}\ \emph {et~al.}(2020)\citenamefont
		{Keeling}, \citenamefont {Hollingsworth},\ and\ \citenamefont
		{Read}}]{keeling2020efficacy}%
	\BibitemOpen
	\bibfield  {author} {\bibinfo {author} {\bibfnamefont {M.~J.}\ \bibnamefont
			{Keeling}}, \bibinfo {author} {\bibfnamefont {T.~D.}\ \bibnamefont
			{Hollingsworth}}, \ and\ \bibinfo {author} {\bibfnamefont {J.~M.}\
			\bibnamefont {Read}},\ }\href@noop {} {\bibfield  {journal} {\bibinfo
			{journal} {J Epidemiol Community Health}\ }\textbf {\bibinfo {volume} {74}},\
		\bibinfo {pages} {861} (\bibinfo {year} {2020})}\BibitemShut {NoStop}%
	\bibitem [{\citenamefont {Firth}\ \emph {et~al.}(2020)\citenamefont {Firth},
		\citenamefont {Hellewell}, \citenamefont {Klepac}, \citenamefont {Kissler},
		\citenamefont {Kucharski},\ and\ \citenamefont {Spurgin}}]{firth2020using}%
	\BibitemOpen
	\bibfield  {author} {\bibinfo {author} {\bibfnamefont {J.~A.}\ \bibnamefont
			{Firth}}, \bibinfo {author} {\bibfnamefont {J.}~\bibnamefont {Hellewell}},
		\bibinfo {author} {\bibfnamefont {P.}~\bibnamefont {Klepac}}, \bibinfo
		{author} {\bibfnamefont {S.}~\bibnamefont {Kissler}}, \bibinfo {author}
		{\bibfnamefont {A.~J.}\ \bibnamefont {Kucharski}}, \ and\ \bibinfo {author}
		{\bibfnamefont {L.~G.}\ \bibnamefont {Spurgin}},\ }\href@noop {} {\bibfield
		{journal} {\bibinfo  {journal} {Nature medicine}\ }\textbf {\bibinfo {volume}
			{26}},\ \bibinfo {pages} {1616} (\bibinfo {year} {2020})}\BibitemShut
	{NoStop}%
	\bibitem [{\citenamefont {Smieszek}\ \emph {et~al.}(2016)\citenamefont
		{Smieszek}, \citenamefont {Castell}, \citenamefont {Barrat}, \citenamefont
		{Cattuto}, \citenamefont {White},\ and\ \citenamefont
		{Krause}}]{smieszek2016contact}%
	\BibitemOpen
	\bibfield  {author} {\bibinfo {author} {\bibfnamefont {T.}~\bibnamefont
			{Smieszek}}, \bibinfo {author} {\bibfnamefont {S.}~\bibnamefont {Castell}},
		\bibinfo {author} {\bibfnamefont {A.}~\bibnamefont {Barrat}}, \bibinfo
		{author} {\bibfnamefont {C.}~\bibnamefont {Cattuto}}, \bibinfo {author}
		{\bibfnamefont {P.~J.}\ \bibnamefont {White}}, \ and\ \bibinfo {author}
		{\bibfnamefont {G.}~\bibnamefont {Krause}},\ }\href@noop {} {\bibfield
		{journal} {\bibinfo  {journal} {BMC infectious diseases}\ }\textbf {\bibinfo
			{volume} {16}},\ \bibinfo {pages} {1} (\bibinfo {year} {2016})}\BibitemShut
	{NoStop}%
	\bibitem [{\citenamefont {Mastrandrea}\ \emph {et~al.}(2015)\citenamefont
		{Mastrandrea}, \citenamefont {Fournet},\ and\ \citenamefont
		{Barrat}}]{mastrandrea2015contact}%
	\BibitemOpen
	\bibfield  {author} {\bibinfo {author} {\bibfnamefont {R.}~\bibnamefont
			{Mastrandrea}}, \bibinfo {author} {\bibfnamefont {J.}~\bibnamefont
			{Fournet}}, \ and\ \bibinfo {author} {\bibfnamefont {A.}~\bibnamefont
			{Barrat}},\ }\href@noop {} {\bibfield  {journal} {\bibinfo  {journal} {PloS
				one}\ }\textbf {\bibinfo {volume} {10}},\ \bibinfo {pages} {e0136497}
		(\bibinfo {year} {2015})}\BibitemShut {NoStop}%
	\bibitem [{\citenamefont {Sacks}\ \emph {et~al.}(2015)\citenamefont {Sacks},
		\citenamefont {Zehe}, \citenamefont {Redick}, \citenamefont {Bah},
		\citenamefont {Cowger}, \citenamefont {Camara}, \citenamefont {Diallo},
		\citenamefont {Gigo}, \citenamefont {Dhillon},\ and\ \citenamefont
		{Liu}}]{sacks2015introduction}%
	\BibitemOpen
	\bibfield  {author} {\bibinfo {author} {\bibfnamefont {J.~A.}\ \bibnamefont
			{Sacks}}, \bibinfo {author} {\bibfnamefont {E.}~\bibnamefont {Zehe}},
		\bibinfo {author} {\bibfnamefont {C.}~\bibnamefont {Redick}}, \bibinfo
		{author} {\bibfnamefont {A.}~\bibnamefont {Bah}}, \bibinfo {author}
		{\bibfnamefont {K.}~\bibnamefont {Cowger}}, \bibinfo {author} {\bibfnamefont
			{M.}~\bibnamefont {Camara}}, \bibinfo {author} {\bibfnamefont
			{A.}~\bibnamefont {Diallo}}, \bibinfo {author} {\bibfnamefont {A.~N.~I.}\
			\bibnamefont {Gigo}}, \bibinfo {author} {\bibfnamefont {R.~S.}\ \bibnamefont
			{Dhillon}}, \ and\ \bibinfo {author} {\bibfnamefont {A.}~\bibnamefont
			{Liu}},\ }\href@noop {} {\bibfield  {journal} {\bibinfo  {journal} {Global
				Health: Science and Practice}\ }\textbf {\bibinfo {volume} {3}},\ \bibinfo
		{pages} {646} (\bibinfo {year} {2015})}\BibitemShut {NoStop}%
	\bibitem [{\citenamefont {Danquah}\ \emph {et~al.}(2019)\citenamefont
		{Danquah}, \citenamefont {Hasham}, \citenamefont {MacFarlane}, \citenamefont
		{Conteh}, \citenamefont {Momoh}, \citenamefont {Tedesco}, \citenamefont
		{Jambai}, \citenamefont {Ross},\ and\ \citenamefont
		{Weiss}}]{danquah2019use}%
	\BibitemOpen
	\bibfield  {author} {\bibinfo {author} {\bibfnamefont {L.~O.}\ \bibnamefont
			{Danquah}}, \bibinfo {author} {\bibfnamefont {N.}~\bibnamefont {Hasham}},
		\bibinfo {author} {\bibfnamefont {M.}~\bibnamefont {MacFarlane}}, \bibinfo
		{author} {\bibfnamefont {F.~E.}\ \bibnamefont {Conteh}}, \bibinfo {author}
		{\bibfnamefont {F.}~\bibnamefont {Momoh}}, \bibinfo {author} {\bibfnamefont
			{A.~A.}\ \bibnamefont {Tedesco}}, \bibinfo {author} {\bibfnamefont
			{A.}~\bibnamefont {Jambai}}, \bibinfo {author} {\bibfnamefont {D.~A.}\
			\bibnamefont {Ross}}, \ and\ \bibinfo {author} {\bibfnamefont {H.~A.}\
			\bibnamefont {Weiss}},\ }\href@noop {} {\bibfield  {journal} {\bibinfo
			{journal} {BMC infectious diseases}\ }\textbf {\bibinfo {volume} {19}},\
		\bibinfo {pages} {1} (\bibinfo {year} {2019})}\BibitemShut {NoStop}%
	\bibitem [{\citenamefont {Ferretti}\ \emph {et~al.}(2020)\citenamefont
		{Ferretti}, \citenamefont {Wymant}, \citenamefont {Kendall}, \citenamefont
		{Zhao}, \citenamefont {Nurtay}, \citenamefont {Abeler-D{\"o}rner},
		\citenamefont {Parker}, \citenamefont {Bonsall},\ and\ \citenamefont
		{Fraser}}]{ferretti2020quantifying}%
	\BibitemOpen
	\bibfield  {author} {\bibinfo {author} {\bibfnamefont {L.}~\bibnamefont
			{Ferretti}}, \bibinfo {author} {\bibfnamefont {C.}~\bibnamefont {Wymant}},
		\bibinfo {author} {\bibfnamefont {M.}~\bibnamefont {Kendall}}, \bibinfo
		{author} {\bibfnamefont {L.}~\bibnamefont {Zhao}}, \bibinfo {author}
		{\bibfnamefont {A.}~\bibnamefont {Nurtay}}, \bibinfo {author} {\bibfnamefont
			{L.}~\bibnamefont {Abeler-D{\"o}rner}}, \bibinfo {author} {\bibfnamefont
			{M.}~\bibnamefont {Parker}}, \bibinfo {author} {\bibfnamefont
			{D.}~\bibnamefont {Bonsall}}, \ and\ \bibinfo {author} {\bibfnamefont
			{C.}~\bibnamefont {Fraser}},\ }\href@noop {} {\bibfield  {journal} {\bibinfo
			{journal} {Science}\ }\textbf {\bibinfo {volume} {368}},\ \bibinfo {pages}
		{eabb6936} (\bibinfo {year} {2020})}\BibitemShut {NoStop}%
	\bibitem [{\citenamefont {Kucharski}\ \emph {et~al.}(2020)\citenamefont
		{Kucharski}, \citenamefont {Klepac}, \citenamefont {Conlan}, \citenamefont
		{Kissler}, \citenamefont {Tang}, \citenamefont {Fry}, \citenamefont {Gog},
		\citenamefont {Edmunds}, \citenamefont {Emery}, \citenamefont {Medley} \emph
		{et~al.}}]{kucharski2020effectiveness}%
	\BibitemOpen
	\bibfield  {author} {\bibinfo {author} {\bibfnamefont {A.~J.}\ \bibnamefont
			{Kucharski}}, \bibinfo {author} {\bibfnamefont {P.}~\bibnamefont {Klepac}},
		\bibinfo {author} {\bibfnamefont {A.~J.}\ \bibnamefont {Conlan}}, \bibinfo
		{author} {\bibfnamefont {S.~M.}\ \bibnamefont {Kissler}}, \bibinfo {author}
		{\bibfnamefont {M.~L.}\ \bibnamefont {Tang}}, \bibinfo {author}
		{\bibfnamefont {H.}~\bibnamefont {Fry}}, \bibinfo {author} {\bibfnamefont
			{J.~R.}\ \bibnamefont {Gog}}, \bibinfo {author} {\bibfnamefont {W.~J.}\
			\bibnamefont {Edmunds}}, \bibinfo {author} {\bibfnamefont {J.~C.}\
			\bibnamefont {Emery}}, \bibinfo {author} {\bibfnamefont {G.}~\bibnamefont
			{Medley}},  \emph {et~al.},\ }\href@noop {} {\bibfield  {journal} {\bibinfo
			{journal} {The Lancet Infectious Diseases}\ }\textbf {\bibinfo {volume}
			{20}},\ \bibinfo {pages} {1151} (\bibinfo {year} {2020})}\BibitemShut
	{NoStop}%
	\bibitem [{\citenamefont {Braithwaite}\ \emph {et~al.}(2020)\citenamefont
		{Braithwaite}, \citenamefont {Callender}, \citenamefont {Bullock},\ and\
		\citenamefont {Aldridge}}]{braithwaite2020automated}%
	\BibitemOpen
	\bibfield  {author} {\bibinfo {author} {\bibfnamefont {I.}~\bibnamefont
			{Braithwaite}}, \bibinfo {author} {\bibfnamefont {T.}~\bibnamefont
			{Callender}}, \bibinfo {author} {\bibfnamefont {M.}~\bibnamefont {Bullock}},
		\ and\ \bibinfo {author} {\bibfnamefont {R.~W.}\ \bibnamefont {Aldridge}},\
	}\href@noop {} {\bibfield  {journal} {\bibinfo  {journal} {The Lancet Digital
				Health}\ }\textbf {\bibinfo {volume} {2}},\ \bibinfo {pages} {e607} (\bibinfo
		{year} {2020})}\BibitemShut {NoStop}%
	\bibitem [{\citenamefont {Chien}\ \emph {et~al.}(2022)\citenamefont {Chien},
		\citenamefont {Be{\"y}},\ and\ \citenamefont {Koenig}}]{chien2022taiwan}%
	\BibitemOpen
	\bibfield  {author} {\bibinfo {author} {\bibfnamefont {L.-C.}\ \bibnamefont
			{Chien}}, \bibinfo {author} {\bibfnamefont {C.~K.}\ \bibnamefont {Be{\"y}}},
		\ and\ \bibinfo {author} {\bibfnamefont {K.~L.}\ \bibnamefont {Koenig}},\
	}\href@noop {} {\bibfield  {journal} {\bibinfo  {journal} {Disaster Medicine
				and Public Health Preparedness}\ }\textbf {\bibinfo {volume} {16}},\ \bibinfo
		{pages} {434} (\bibinfo {year} {2022})}\BibitemShut {NoStop}%
	\bibitem [{\citenamefont {Oh}\ \emph {et~al.}(2020)\citenamefont {Oh},
		\citenamefont {Lee}, \citenamefont {Schwarz}, \citenamefont {Ratcliffe},
		\citenamefont {Markuns},\ and\ \citenamefont {Hirschhorn}}]{oh2020national}%
	\BibitemOpen
	\bibfield  {author} {\bibinfo {author} {\bibfnamefont {J.}~\bibnamefont
			{Oh}}, \bibinfo {author} {\bibfnamefont {J.-K.}\ \bibnamefont {Lee}},
		\bibinfo {author} {\bibfnamefont {D.}~\bibnamefont {Schwarz}}, \bibinfo
		{author} {\bibfnamefont {H.~L.}\ \bibnamefont {Ratcliffe}}, \bibinfo {author}
		{\bibfnamefont {J.~F.}\ \bibnamefont {Markuns}}, \ and\ \bibinfo {author}
		{\bibfnamefont {L.~R.}\ \bibnamefont {Hirschhorn}},\ }\href@noop {}
	{\bibfield  {journal} {\bibinfo  {journal} {Health Systems \& Reform}\
		}\textbf {\bibinfo {volume} {6}},\ \bibinfo {pages} {e1753464} (\bibinfo
		{year} {2020})}\BibitemShut {NoStop}%
	\bibitem [{\citenamefont {Aslam}\ and\ \citenamefont
		{Hussain}(2020)}]{aslam2020fighting}%
	\BibitemOpen
	\bibfield  {author} {\bibinfo {author} {\bibfnamefont {H.}~\bibnamefont
			{Aslam}}\ and\ \bibinfo {author} {\bibfnamefont {R.}~\bibnamefont
			{Hussain}},\ }\href@noop {} {\enquote {\bibinfo {title} {Fighting covid-19:
				Lessons from china, south korea and japan},}\ } (\bibinfo {year}
	{2020})\BibitemShut {NoStop}%
	\bibitem [{\citenamefont {Huang}\ \emph {et~al.}(2020)\citenamefont {Huang},
		\citenamefont {Guo}, \citenamefont {Lee}, \citenamefont {Ho}, \citenamefont
		{Ang}, \citenamefont {Chow} \emph {et~al.}}]{huang2020performance}%
	\BibitemOpen
	\bibfield  {author} {\bibinfo {author} {\bibfnamefont {Z.}~\bibnamefont
			{Huang}}, \bibinfo {author} {\bibfnamefont {H.}~\bibnamefont {Guo}}, \bibinfo
		{author} {\bibfnamefont {Y.-M.}\ \bibnamefont {Lee}}, \bibinfo {author}
		{\bibfnamefont {E.~C.}\ \bibnamefont {Ho}}, \bibinfo {author} {\bibfnamefont
			{H.}~\bibnamefont {Ang}}, \bibinfo {author} {\bibfnamefont {A.}~\bibnamefont
			{Chow}},  \emph {et~al.},\ }\href@noop {} {\bibfield  {journal} {\bibinfo
			{journal} {JMIR mHealth and uHealth}\ }\textbf {\bibinfo {volume} {8}},\
		\bibinfo {pages} {e23148} (\bibinfo {year} {2020})}\BibitemShut {NoStop}%
	\bibitem [{\citenamefont {Mello}\ and\ \citenamefont
		{Wang}(2020)}]{mello2020ethics}%
	\BibitemOpen
	\bibfield  {author} {\bibinfo {author} {\bibfnamefont {M.~M.}\ \bibnamefont
			{Mello}}\ and\ \bibinfo {author} {\bibfnamefont {C.~J.}\ \bibnamefont
			{Wang}},\ }\href@noop {} {\bibfield  {journal} {\bibinfo  {journal}
			{Science}\ }\textbf {\bibinfo {volume} {368}},\ \bibinfo {pages} {951}
		(\bibinfo {year} {2020})}\BibitemShut {NoStop}%
	\bibitem [{\citenamefont {Bengio}\ \emph {et~al.}(2020)\citenamefont {Bengio},
		\citenamefont {Janda}, \citenamefont {Yu}, \citenamefont {Ippolito},
		\citenamefont {Jarvie}, \citenamefont {Pilat}, \citenamefont {Struck},
		\citenamefont {Krastev},\ and\ \citenamefont {Sharma}}]{bengio2020need}%
	\BibitemOpen
	\bibfield  {author} {\bibinfo {author} {\bibfnamefont {Y.}~\bibnamefont
			{Bengio}}, \bibinfo {author} {\bibfnamefont {R.}~\bibnamefont {Janda}},
		\bibinfo {author} {\bibfnamefont {Y.~W.}\ \bibnamefont {Yu}}, \bibinfo
		{author} {\bibfnamefont {D.}~\bibnamefont {Ippolito}}, \bibinfo {author}
		{\bibfnamefont {M.}~\bibnamefont {Jarvie}}, \bibinfo {author} {\bibfnamefont
			{D.}~\bibnamefont {Pilat}}, \bibinfo {author} {\bibfnamefont
			{B.}~\bibnamefont {Struck}}, \bibinfo {author} {\bibfnamefont
			{S.}~\bibnamefont {Krastev}}, \ and\ \bibinfo {author} {\bibfnamefont
			{A.}~\bibnamefont {Sharma}},\ }\href@noop {} {\bibfield  {journal} {\bibinfo
			{journal} {The Lancet Digital Health}\ }\textbf {\bibinfo {volume} {2}},\
		\bibinfo {pages} {e342} (\bibinfo {year} {2020})}\BibitemShut {NoStop}%
	\bibitem [{\citenamefont {Amann}\ \emph {et~al.}(2021)\citenamefont {Amann},
		\citenamefont {Sleigh},\ and\ \citenamefont {Vayena}}]{amann2021digital}%
	\BibitemOpen
	\bibfield  {author} {\bibinfo {author} {\bibfnamefont {J.}~\bibnamefont
			{Amann}}, \bibinfo {author} {\bibfnamefont {J.}~\bibnamefont {Sleigh}}, \
		and\ \bibinfo {author} {\bibfnamefont {E.}~\bibnamefont {Vayena}},\
	}\href@noop {} {\bibfield  {journal} {\bibinfo  {journal} {Plos one}\
		}\textbf {\bibinfo {volume} {16}},\ \bibinfo {pages} {e0246524} (\bibinfo
		{year} {2021})}\BibitemShut {NoStop}%
	\bibitem [{\citenamefont {Jacob}\ and\ \citenamefont
		{Lawar{\'e}e}(2021)}]{jacob2021adoption}%
	\BibitemOpen
	\bibfield  {author} {\bibinfo {author} {\bibfnamefont {S.}~\bibnamefont
			{Jacob}}\ and\ \bibinfo {author} {\bibfnamefont {J.}~\bibnamefont
			{Lawar{\'e}e}},\ }\href@noop {} {\bibfield  {journal} {\bibinfo  {journal}
			{Policy Design and Practice}\ }\textbf {\bibinfo {volume} {4}},\ \bibinfo
		{pages} {44} (\bibinfo {year} {2021})}\BibitemShut {NoStop}%
	\bibitem [{\citenamefont {Munzert}\ \emph {et~al.}(2021)\citenamefont
		{Munzert}, \citenamefont {Selb}, \citenamefont {Gohdes}, \citenamefont
		{Stoetzer},\ and\ \citenamefont {Lowe}}]{munzert2021tracking}%
	\BibitemOpen
	\bibfield  {author} {\bibinfo {author} {\bibfnamefont {S.}~\bibnamefont
			{Munzert}}, \bibinfo {author} {\bibfnamefont {P.}~\bibnamefont {Selb}},
		\bibinfo {author} {\bibfnamefont {A.}~\bibnamefont {Gohdes}}, \bibinfo
		{author} {\bibfnamefont {L.~F.}\ \bibnamefont {Stoetzer}}, \ and\ \bibinfo
		{author} {\bibfnamefont {W.}~\bibnamefont {Lowe}},\ }\href@noop {} {\bibfield
		{journal} {\bibinfo  {journal} {Nature Human Behaviour}\ }\textbf {\bibinfo
			{volume} {5}},\ \bibinfo {pages} {247} (\bibinfo {year} {2021})}\BibitemShut
	{NoStop}%
	\bibitem [{\citenamefont {Bay}\ \emph {et~al.}(2020)\citenamefont {Bay},
		\citenamefont {Kek}, \citenamefont {Tan}, \citenamefont {Hau}, \citenamefont
		{Yongquan}, \citenamefont {Tan},\ and\ \citenamefont
		{Quy}}]{bay2020bluetrace}%
	\BibitemOpen
	\bibfield  {author} {\bibinfo {author} {\bibfnamefont {J.}~\bibnamefont
			{Bay}}, \bibinfo {author} {\bibfnamefont {J.}~\bibnamefont {Kek}}, \bibinfo
		{author} {\bibfnamefont {A.}~\bibnamefont {Tan}}, \bibinfo {author}
		{\bibfnamefont {C.~S.}\ \bibnamefont {Hau}}, \bibinfo {author} {\bibfnamefont
			{L.}~\bibnamefont {Yongquan}}, \bibinfo {author} {\bibfnamefont
			{J.}~\bibnamefont {Tan}}, \ and\ \bibinfo {author} {\bibfnamefont {T.~A.}\
			\bibnamefont {Quy}},\ }\href@noop {} {\bibfield  {journal} {\bibinfo
			{journal} {Government Technology Agency-Singapore, Tech. Rep}\ }\textbf
		{\bibinfo {volume} {18}} (\bibinfo {year} {2020})}\BibitemShut {NoStop}%
	\bibitem [{NHS(2020)}]{NHSapp}%
	\BibitemOpen
	\href@noop {} {\enquote {\bibinfo {title} {Nhs covid-19 app},}\ }\bibinfo
	{howpublished} {\url{https://covid19.nhs.uk/}} (\bibinfo {year}
	{2020})\BibitemShut {NoStop}%
	\bibitem [{Aar(2020)}]{Aarogyasetu}%
	\BibitemOpen
	\href@noop {} {\enquote {\bibinfo {title} {Aarogya setu app},}\ }\bibinfo
	{howpublished} {\url{https://www.aarogyasetu.gov.in/}} (\bibinfo {year}
	{2020})\BibitemShut {NoStop}%
	\bibitem [{\citenamefont {Troncoso}\ \emph {et~al.}(2020)\citenamefont
		{Troncoso}, \citenamefont {Payer}, \citenamefont {Hubaux}, \citenamefont
		{Salath{\'e}}, \citenamefont {Larus}, \citenamefont {Bugnion}, \citenamefont
		{Lueks}, \citenamefont {Stadler}, \citenamefont {Pyrgelis}, \citenamefont
		{Antonioli} \emph {et~al.}}]{troncoso2020decentralized}%
	\BibitemOpen
	\bibfield  {author} {\bibinfo {author} {\bibfnamefont {C.}~\bibnamefont
			{Troncoso}}, \bibinfo {author} {\bibfnamefont {M.}~\bibnamefont {Payer}},
		\bibinfo {author} {\bibfnamefont {J.-P.}\ \bibnamefont {Hubaux}}, \bibinfo
		{author} {\bibfnamefont {M.}~\bibnamefont {Salath{\'e}}}, \bibinfo {author}
		{\bibfnamefont {J.}~\bibnamefont {Larus}}, \bibinfo {author} {\bibfnamefont
			{E.}~\bibnamefont {Bugnion}}, \bibinfo {author} {\bibfnamefont
			{W.}~\bibnamefont {Lueks}}, \bibinfo {author} {\bibfnamefont
			{T.}~\bibnamefont {Stadler}}, \bibinfo {author} {\bibfnamefont
			{A.}~\bibnamefont {Pyrgelis}}, \bibinfo {author} {\bibfnamefont
			{D.}~\bibnamefont {Antonioli}},  \emph {et~al.},\ }\href@noop {} {\enquote
		{\bibinfo {title} {Decentralized privacy-preserving proximity tracing},}\ }
	(\bibinfo {year} {2020}),\ \bibinfo {note} {arXiv:2005.12273}\BibitemShut
	{NoStop}%
	\bibitem [{\citenamefont {Chan}\ \emph {et~al.}(2020)\citenamefont {Chan},
		\citenamefont {Gollakota}, \citenamefont {Horvitz}, \citenamefont {Jaeger},
		\citenamefont {Kakade}, \citenamefont {Kohno}, \citenamefont {Langford},
		\citenamefont {Larson}, \citenamefont {Singanamalla}, \citenamefont
		{Sunshine} \emph {et~al.}}]{chan2020pact}%
	\BibitemOpen
	\bibfield  {author} {\bibinfo {author} {\bibfnamefont {J.}~\bibnamefont
			{Chan}}, \bibinfo {author} {\bibfnamefont {S.}~\bibnamefont {Gollakota}},
		\bibinfo {author} {\bibfnamefont {E.}~\bibnamefont {Horvitz}}, \bibinfo
		{author} {\bibfnamefont {J.}~\bibnamefont {Jaeger}}, \bibinfo {author}
		{\bibfnamefont {S.}~\bibnamefont {Kakade}}, \bibinfo {author} {\bibfnamefont
			{T.}~\bibnamefont {Kohno}}, \bibinfo {author} {\bibfnamefont
			{J.}~\bibnamefont {Langford}}, \bibinfo {author} {\bibfnamefont
			{J.}~\bibnamefont {Larson}}, \bibinfo {author} {\bibfnamefont
			{S.}~\bibnamefont {Singanamalla}}, \bibinfo {author} {\bibfnamefont
			{J.}~\bibnamefont {Sunshine}},  \emph {et~al.},\ }\href@noop {} {\enquote
		{\bibinfo {title} {Pact: Privacy sensitive protocols and mechanisms for
				mobile contact tracing},}\ } (\bibinfo {year} {2020}),\ \bibinfo {note}
	{arXiv:2004.03544}\BibitemShut {NoStop}%
	\bibitem [{\citenamefont {Apple}\ and\ \citenamefont
		{Google}(2020)}]{AppleGoogle}%
	\BibitemOpen
	\bibfield  {author} {\bibinfo {author} {\bibnamefont {Apple}}\ and\ \bibinfo
		{author} {\bibnamefont {Google}},\ }\href@noop {} {\enquote {\bibinfo {title}
			{Privacy-preserving contact tracing},}\ } (\bibinfo {year} {2020}),\ \bibinfo
	{note} {https://covid19.apple.com/contacttracing}\BibitemShut {NoStop}%
	\bibitem [{\citenamefont {Kendall}\ \emph {et~al.}(2020)\citenamefont
		{Kendall}, \citenamefont {Milsom}, \citenamefont {Abeler-D{\"o}rner},
		\citenamefont {Wymant}, \citenamefont {Ferretti}, \citenamefont {Briers},
		\citenamefont {Holmes}, \citenamefont {Bonsall}, \citenamefont {Abeler},\
		and\ \citenamefont {Fraser}}]{kendall2020epidemiological}%
	\BibitemOpen
	\bibfield  {author} {\bibinfo {author} {\bibfnamefont {M.}~\bibnamefont
			{Kendall}}, \bibinfo {author} {\bibfnamefont {L.}~\bibnamefont {Milsom}},
		\bibinfo {author} {\bibfnamefont {L.}~\bibnamefont {Abeler-D{\"o}rner}},
		\bibinfo {author} {\bibfnamefont {C.}~\bibnamefont {Wymant}}, \bibinfo
		{author} {\bibfnamefont {L.}~\bibnamefont {Ferretti}}, \bibinfo {author}
		{\bibfnamefont {M.}~\bibnamefont {Briers}}, \bibinfo {author} {\bibfnamefont
			{C.}~\bibnamefont {Holmes}}, \bibinfo {author} {\bibfnamefont
			{D.}~\bibnamefont {Bonsall}}, \bibinfo {author} {\bibfnamefont
			{J.}~\bibnamefont {Abeler}}, \ and\ \bibinfo {author} {\bibfnamefont
			{C.}~\bibnamefont {Fraser}},\ }\href@noop {} {\bibfield  {journal} {\bibinfo
			{journal} {The Lancet Digital Health}\ }\textbf {\bibinfo {volume} {2}},\
		\bibinfo {pages} {e658} (\bibinfo {year} {2020})}\BibitemShut {NoStop}%
	\bibitem [{\citenamefont {Salath{\'e}}\ \emph {et~al.}(2020)\citenamefont
		{Salath{\'e}}, \citenamefont {Althaus}, \citenamefont {Anderegg},
		\citenamefont {Antonioli}, \citenamefont {Ballouz}, \citenamefont {Bugnion},
		\citenamefont {{\v{C}}apkun}, \citenamefont {Jackson}, \citenamefont {Kim},
		\citenamefont {Larus} \emph {et~al.}}]{salathe2020early}%
	\BibitemOpen
	\bibfield  {author} {\bibinfo {author} {\bibfnamefont {M.}~\bibnamefont
			{Salath{\'e}}}, \bibinfo {author} {\bibfnamefont {C.~L.}\ \bibnamefont
			{Althaus}}, \bibinfo {author} {\bibfnamefont {N.}~\bibnamefont {Anderegg}},
		\bibinfo {author} {\bibfnamefont {D.}~\bibnamefont {Antonioli}}, \bibinfo
		{author} {\bibfnamefont {T.}~\bibnamefont {Ballouz}}, \bibinfo {author}
		{\bibfnamefont {E.}~\bibnamefont {Bugnion}}, \bibinfo {author} {\bibfnamefont
			{S.}~\bibnamefont {{\v{C}}apkun}}, \bibinfo {author} {\bibfnamefont
			{D.}~\bibnamefont {Jackson}}, \bibinfo {author} {\bibfnamefont {S.-I.}\
			\bibnamefont {Kim}}, \bibinfo {author} {\bibfnamefont {J.~R.}\ \bibnamefont
			{Larus}},  \emph {et~al.},\ }\href@noop {} {\bibfield  {journal} {\bibinfo
			{journal} {medRxiv}\ } (\bibinfo {year} {2020})}\BibitemShut {NoStop}%
	\bibitem [{\citenamefont {Wymant}\ \emph {et~al.}(2021)\citenamefont {Wymant},
		\citenamefont {Ferretti}, \citenamefont {Tsallis}, \citenamefont
		{Charalambides}, \citenamefont {Abeler-D{\"o}rner}, \citenamefont {Bonsall},
		\citenamefont {Hinch}, \citenamefont {Kendall}, \citenamefont {Milsom},
		\citenamefont {Ayres} \emph {et~al.}}]{wymant2021epidemiological}%
	\BibitemOpen
	\bibfield  {author} {\bibinfo {author} {\bibfnamefont {C.}~\bibnamefont
			{Wymant}}, \bibinfo {author} {\bibfnamefont {L.}~\bibnamefont {Ferretti}},
		\bibinfo {author} {\bibfnamefont {D.}~\bibnamefont {Tsallis}}, \bibinfo
		{author} {\bibfnamefont {M.}~\bibnamefont {Charalambides}}, \bibinfo {author}
		{\bibfnamefont {L.}~\bibnamefont {Abeler-D{\"o}rner}}, \bibinfo {author}
		{\bibfnamefont {D.}~\bibnamefont {Bonsall}}, \bibinfo {author} {\bibfnamefont
			{R.}~\bibnamefont {Hinch}}, \bibinfo {author} {\bibfnamefont
			{M.}~\bibnamefont {Kendall}}, \bibinfo {author} {\bibfnamefont
			{L.}~\bibnamefont {Milsom}}, \bibinfo {author} {\bibfnamefont
			{M.}~\bibnamefont {Ayres}},  \emph {et~al.},\ }\href@noop {} {\bibfield
		{journal} {\bibinfo  {journal} {Nature}\ }\textbf {\bibinfo {volume} {594}},\
		\bibinfo {pages} {408} (\bibinfo {year} {2021})}\BibitemShut {NoStop}%
	\bibitem [{\citenamefont {Rodr{\'\i}guez}\ \emph {et~al.}(2021)\citenamefont
		{Rodr{\'\i}guez}, \citenamefont {Gra{\~n}a}, \citenamefont
		{Alvarez-Le{\'o}n}, \citenamefont {Battaglini}, \citenamefont {Darias},
		\citenamefont {Hern{\'a}n}, \citenamefont {L{\'o}pez}, \citenamefont
		{Llaneza}, \citenamefont {Mart{\'\i}n}, \citenamefont {Ramirez-Rubio} \emph
		{et~al.}}]{rodriguez2021population}%
	\BibitemOpen
	\bibfield  {author} {\bibinfo {author} {\bibfnamefont {P.}~\bibnamefont
			{Rodr{\'\i}guez}}, \bibinfo {author} {\bibfnamefont {S.}~\bibnamefont
			{Gra{\~n}a}}, \bibinfo {author} {\bibfnamefont {E.~E.}\ \bibnamefont
			{Alvarez-Le{\'o}n}}, \bibinfo {author} {\bibfnamefont {M.}~\bibnamefont
			{Battaglini}}, \bibinfo {author} {\bibfnamefont {F.~J.}\ \bibnamefont
			{Darias}}, \bibinfo {author} {\bibfnamefont {M.~A.}\ \bibnamefont
			{Hern{\'a}n}}, \bibinfo {author} {\bibfnamefont {R.}~\bibnamefont
			{L{\'o}pez}}, \bibinfo {author} {\bibfnamefont {P.}~\bibnamefont {Llaneza}},
		\bibinfo {author} {\bibfnamefont {M.~C.}\ \bibnamefont {Mart{\'\i}n}},
		\bibinfo {author} {\bibfnamefont {O.}~\bibnamefont {Ramirez-Rubio}},  \emph
		{et~al.},\ }\href@noop {} {\bibfield  {journal} {\bibinfo  {journal} {Nature
				communications}\ }\textbf {\bibinfo {volume} {12}},\ \bibinfo {pages} {1}
		(\bibinfo {year} {2021})}\BibitemShut {NoStop}%
	\bibitem [{\citenamefont {Barrat}\ \emph {et~al.}(2021)\citenamefont {Barrat},
		\citenamefont {Cattuto}, \citenamefont {Kivel{\"a}}, \citenamefont
		{Lehmann},\ and\ \citenamefont {Saram{\"a}ki}}]{barrat2021effect}%
	\BibitemOpen
	\bibfield  {author} {\bibinfo {author} {\bibfnamefont {A.}~\bibnamefont
			{Barrat}}, \bibinfo {author} {\bibfnamefont {C.}~\bibnamefont {Cattuto}},
		\bibinfo {author} {\bibfnamefont {M.}~\bibnamefont {Kivel{\"a}}}, \bibinfo
		{author} {\bibfnamefont {S.}~\bibnamefont {Lehmann}}, \ and\ \bibinfo
		{author} {\bibfnamefont {J.}~\bibnamefont {Saram{\"a}ki}},\ }\href@noop {}
	{\bibfield  {journal} {\bibinfo  {journal} {Journal of the Royal Society
				Interface}\ }\textbf {\bibinfo {volume} {18}},\ \bibinfo {pages} {20201000}
		(\bibinfo {year} {2021})}\BibitemShut {NoStop}%
	\bibitem [{\citenamefont {Cencetti}\ \emph {et~al.}(2021)\citenamefont
		{Cencetti}, \citenamefont {Santin}, \citenamefont {Longa}, \citenamefont
		{Pigani}, \citenamefont {Barrat}, \citenamefont {Cattuto}, \citenamefont
		{Lehmann}, \citenamefont {Salathe},\ and\ \citenamefont
		{Lepri}}]{cencetti2021digital}%
	\BibitemOpen
	\bibfield  {author} {\bibinfo {author} {\bibfnamefont {G.}~\bibnamefont
			{Cencetti}}, \bibinfo {author} {\bibfnamefont {G.}~\bibnamefont {Santin}},
		\bibinfo {author} {\bibfnamefont {A.}~\bibnamefont {Longa}}, \bibinfo
		{author} {\bibfnamefont {E.}~\bibnamefont {Pigani}}, \bibinfo {author}
		{\bibfnamefont {A.}~\bibnamefont {Barrat}}, \bibinfo {author} {\bibfnamefont
			{C.}~\bibnamefont {Cattuto}}, \bibinfo {author} {\bibfnamefont
			{S.}~\bibnamefont {Lehmann}}, \bibinfo {author} {\bibfnamefont
			{M.}~\bibnamefont {Salathe}}, \ and\ \bibinfo {author} {\bibfnamefont
			{B.}~\bibnamefont {Lepri}},\ }\href@noop {} {\bibfield  {journal} {\bibinfo
			{journal} {Nature communications}\ }\textbf {\bibinfo {volume} {12}},\
		\bibinfo {pages} {1} (\bibinfo {year} {2021})}\BibitemShut {NoStop}%
	\bibitem [{\citenamefont {Contreras}\ \emph {et~al.}(2021)\citenamefont
		{Contreras}, \citenamefont {Dehning}, \citenamefont {Loidolt}, \citenamefont
		{Zierenberg}, \citenamefont {Spitzner}, \citenamefont {Urrea-Quintero},
		\citenamefont {Mohr}, \citenamefont {Wilczek}, \citenamefont {Wibral},\ and\
		\citenamefont {Priesemann}}]{contreras2021challenges}%
	\BibitemOpen
	\bibfield  {author} {\bibinfo {author} {\bibfnamefont {S.}~\bibnamefont
			{Contreras}}, \bibinfo {author} {\bibfnamefont {J.}~\bibnamefont {Dehning}},
		\bibinfo {author} {\bibfnamefont {M.}~\bibnamefont {Loidolt}}, \bibinfo
		{author} {\bibfnamefont {J.}~\bibnamefont {Zierenberg}}, \bibinfo {author}
		{\bibfnamefont {F.~P.}\ \bibnamefont {Spitzner}}, \bibinfo {author}
		{\bibfnamefont {J.~H.}\ \bibnamefont {Urrea-Quintero}}, \bibinfo {author}
		{\bibfnamefont {S.~B.}\ \bibnamefont {Mohr}}, \bibinfo {author}
		{\bibfnamefont {M.}~\bibnamefont {Wilczek}}, \bibinfo {author} {\bibfnamefont
			{M.}~\bibnamefont {Wibral}}, \ and\ \bibinfo {author} {\bibfnamefont
			{V.}~\bibnamefont {Priesemann}},\ }\href@noop {} {\bibfield  {journal}
		{\bibinfo  {journal} {Nature communications}\ }\textbf {\bibinfo {volume}
			{12}},\ \bibinfo {pages} {1} (\bibinfo {year} {2021})}\BibitemShut {NoStop}%
	\bibitem [{\citenamefont {Alsdurf}\ \emph {et~al.}(2020)\citenamefont
		{Alsdurf}, \citenamefont {Belliveau}, \citenamefont {Bengio}, \citenamefont
		{Deleu}, \citenamefont {Gupta}, \citenamefont {Ippolito}, \citenamefont
		{Janda}, \citenamefont {Jarvie}, \citenamefont {Kolody}, \citenamefont
		{Krastev} \emph {et~al.}}]{alsdurf2020covi}%
	\BibitemOpen
	\bibfield  {author} {\bibinfo {author} {\bibfnamefont {H.}~\bibnamefont
			{Alsdurf}}, \bibinfo {author} {\bibfnamefont {E.}~\bibnamefont {Belliveau}},
		\bibinfo {author} {\bibfnamefont {Y.}~\bibnamefont {Bengio}}, \bibinfo
		{author} {\bibfnamefont {T.}~\bibnamefont {Deleu}}, \bibinfo {author}
		{\bibfnamefont {P.}~\bibnamefont {Gupta}}, \bibinfo {author} {\bibfnamefont
			{D.}~\bibnamefont {Ippolito}}, \bibinfo {author} {\bibfnamefont
			{R.}~\bibnamefont {Janda}}, \bibinfo {author} {\bibfnamefont
			{M.}~\bibnamefont {Jarvie}}, \bibinfo {author} {\bibfnamefont
			{T.}~\bibnamefont {Kolody}}, \bibinfo {author} {\bibfnamefont
			{S.}~\bibnamefont {Krastev}},  \emph {et~al.},\ }\href@noop {} {\bibfield
		{journal} {\bibinfo  {journal} {arXiv preprint arXiv:2005.08502}\ } (\bibinfo
		{year} {2020})}\BibitemShut {NoStop}%
	\bibitem [{\citenamefont {Fenton}\ \emph {et~al.}(2020)\citenamefont {Fenton},
		\citenamefont {McLachlan}, \citenamefont {Lucas}, \citenamefont {Dube},
		\citenamefont {Hitman}, \citenamefont {Osman}, \citenamefont {Kyrimi},\ and\
		\citenamefont {Neil}}]{fenton2020privacy}%
	\BibitemOpen
	\bibfield  {author} {\bibinfo {author} {\bibfnamefont {N.}~\bibnamefont
			{Fenton}}, \bibinfo {author} {\bibfnamefont {S.}~\bibnamefont {McLachlan}},
		\bibinfo {author} {\bibfnamefont {P.}~\bibnamefont {Lucas}}, \bibinfo
		{author} {\bibfnamefont {K.}~\bibnamefont {Dube}}, \bibinfo {author}
		{\bibfnamefont {G.}~\bibnamefont {Hitman}}, \bibinfo {author} {\bibfnamefont
			{M.}~\bibnamefont {Osman}}, \bibinfo {author} {\bibfnamefont
			{E.}~\bibnamefont {Kyrimi}}, \ and\ \bibinfo {author} {\bibfnamefont
			{M.}~\bibnamefont {Neil}},\ }\href@noop {} {\bibfield  {journal} {\bibinfo
			{journal} {medRxiv}\ }\textbf {\bibinfo {volume} {" "}} (\bibinfo {year}
		{2020})}\BibitemShut {NoStop}%
	\bibitem [{\citenamefont {Baker}\ \emph {et~al.}(2021)\citenamefont {Baker},
		\citenamefont {Biazzo}, \citenamefont {Braunstein}, \citenamefont {Catania},
		\citenamefont {Dall’Asta}, \citenamefont {Ingrosso}, \citenamefont
		{Krzakala}, \citenamefont {Mazza}, \citenamefont {M{\'e}zard}, \citenamefont
		{Muntoni} \emph {et~al.}}]{baker2021epidemic}%
	\BibitemOpen
	\bibfield  {author} {\bibinfo {author} {\bibfnamefont {A.}~\bibnamefont
			{Baker}}, \bibinfo {author} {\bibfnamefont {I.}~\bibnamefont {Biazzo}},
		\bibinfo {author} {\bibfnamefont {A.}~\bibnamefont {Braunstein}}, \bibinfo
		{author} {\bibfnamefont {G.}~\bibnamefont {Catania}}, \bibinfo {author}
		{\bibfnamefont {L.}~\bibnamefont {Dall’Asta}}, \bibinfo {author}
		{\bibfnamefont {A.}~\bibnamefont {Ingrosso}}, \bibinfo {author}
		{\bibfnamefont {F.}~\bibnamefont {Krzakala}}, \bibinfo {author}
		{\bibfnamefont {F.}~\bibnamefont {Mazza}}, \bibinfo {author} {\bibfnamefont
			{M.}~\bibnamefont {M{\'e}zard}}, \bibinfo {author} {\bibfnamefont {A.~P.}\
			\bibnamefont {Muntoni}},  \emph {et~al.},\ }\href@noop {} {\bibfield
		{journal} {\bibinfo  {journal} {Proceedings of the National Academy of
				Sciences}\ }\textbf {\bibinfo {volume} {118}} (\bibinfo {year}
		{2021})}\BibitemShut {NoStop}%
	\bibitem [{\citenamefont {Murphy}\ \emph {et~al.}(2021)\citenamefont {Murphy},
		\citenamefont {Kumar},\ and\ \citenamefont {Serghiou}}]{murphy2021risk}%
	\BibitemOpen
	\bibfield  {author} {\bibinfo {author} {\bibfnamefont {K.}~\bibnamefont
			{Murphy}}, \bibinfo {author} {\bibfnamefont {A.}~\bibnamefont {Kumar}}, \
		and\ \bibinfo {author} {\bibfnamefont {S.}~\bibnamefont {Serghiou}},\ }in\
	\href@noop {} {\emph {\bibinfo {booktitle} {Machine Learning for Healthcare
				Conference}}}\ (\bibinfo {organization} {PMLR},\ \bibinfo {year} {2021})\
	pp.\ \bibinfo {pages} {373--390}\BibitemShut {NoStop}%
	\bibitem [{\citenamefont {Braunstein}\ and\ \citenamefont
		{Ingrosso}(2016)}]{braunstein_inference_2016}%
	\BibitemOpen
	\bibfield  {author} {\bibinfo {author} {\bibfnamefont {A.}~\bibnamefont
			{Braunstein}}\ and\ \bibinfo {author} {\bibfnamefont {A.}~\bibnamefont
			{Ingrosso}},\ }\href {\doibase 10.1038/srep27538} {\bibfield  {journal}
		{\bibinfo  {journal} {Sci Rep}\ }\textbf {\bibinfo {volume} {6}},\ \bibinfo
		{pages} {27538} (\bibinfo {year} {2016})}\BibitemShut {NoStop}%
	\bibitem [{\citenamefont {Altarelli}\ \emph {et~al.}(2014)\citenamefont
		{Altarelli}, \citenamefont {Braunstein}, \citenamefont {Dall’Asta},
		\citenamefont {Lage-Castellanos},\ and\ \citenamefont
		{Zecchina}}]{altarelli2014bayesian}%
	\BibitemOpen
	\bibfield  {author} {\bibinfo {author} {\bibfnamefont {F.}~\bibnamefont
			{Altarelli}}, \bibinfo {author} {\bibfnamefont {A.}~\bibnamefont
			{Braunstein}}, \bibinfo {author} {\bibfnamefont {L.}~\bibnamefont
			{Dall’Asta}}, \bibinfo {author} {\bibfnamefont {A.}~\bibnamefont
			{Lage-Castellanos}}, \ and\ \bibinfo {author} {\bibfnamefont
			{R.}~\bibnamefont {Zecchina}},\ }\href@noop {} {\bibfield  {journal}
		{\bibinfo  {journal} {Physical review letters}\ }\textbf {\bibinfo {volume}
			{112}},\ \bibinfo {pages} {118701} (\bibinfo {year} {2014})}\BibitemShut
	{NoStop}%
	\bibitem [{\citenamefont {Hinch}\ \emph {et~al.}(2021)\citenamefont {Hinch},
		\citenamefont {Probert}, \citenamefont {Nurtay}, \citenamefont {Kendall},
		\citenamefont {Wymant}, \citenamefont {Hall}, \citenamefont {Lythgoe},
		\citenamefont {Bulas~Cruz}, \citenamefont {Zhao}, \citenamefont {Stewart}
		\emph {et~al.}}]{hinch2021openabm}%
	\BibitemOpen
	\bibfield  {author} {\bibinfo {author} {\bibfnamefont {R.}~\bibnamefont
			{Hinch}}, \bibinfo {author} {\bibfnamefont {W.~J.}\ \bibnamefont {Probert}},
		\bibinfo {author} {\bibfnamefont {A.}~\bibnamefont {Nurtay}}, \bibinfo
		{author} {\bibfnamefont {M.}~\bibnamefont {Kendall}}, \bibinfo {author}
		{\bibfnamefont {C.}~\bibnamefont {Wymant}}, \bibinfo {author} {\bibfnamefont
			{M.}~\bibnamefont {Hall}}, \bibinfo {author} {\bibfnamefont {K.}~\bibnamefont
			{Lythgoe}}, \bibinfo {author} {\bibfnamefont {A.}~\bibnamefont {Bulas~Cruz}},
		\bibinfo {author} {\bibfnamefont {L.}~\bibnamefont {Zhao}}, \bibinfo {author}
		{\bibfnamefont {A.}~\bibnamefont {Stewart}},  \emph {et~al.},\ }\href@noop {}
	{\bibfield  {journal} {\bibinfo  {journal} {PLoS computational biology}\
		}\textbf {\bibinfo {volume} {17}},\ \bibinfo {pages} {e1009146} (\bibinfo
		{year} {2021})}\BibitemShut {NoStop}%
	\bibitem [{\citenamefont {Kerr}\ \emph
		{et~al.}(2021{\natexlab{a}})\citenamefont {Kerr}, \citenamefont {Stuart},
		\citenamefont {Mistry}, \citenamefont {Abeysuriya}, \citenamefont
		{Rosenfeld}, \citenamefont {Hart}, \citenamefont {N{\'u}{\~n}ez},
		\citenamefont {Cohen}, \citenamefont {Selvaraj}, \citenamefont {Hagedorn}
		\emph {et~al.}}]{kerr2021covasim}%
	\BibitemOpen
	\bibfield  {author} {\bibinfo {author} {\bibfnamefont {C.~C.}\ \bibnamefont
			{Kerr}}, \bibinfo {author} {\bibfnamefont {R.~M.}\ \bibnamefont {Stuart}},
		\bibinfo {author} {\bibfnamefont {D.}~\bibnamefont {Mistry}}, \bibinfo
		{author} {\bibfnamefont {R.~G.}\ \bibnamefont {Abeysuriya}}, \bibinfo
		{author} {\bibfnamefont {K.}~\bibnamefont {Rosenfeld}}, \bibinfo {author}
		{\bibfnamefont {G.~R.}\ \bibnamefont {Hart}}, \bibinfo {author}
		{\bibfnamefont {R.~C.}\ \bibnamefont {N{\'u}{\~n}ez}}, \bibinfo {author}
		{\bibfnamefont {J.~A.}\ \bibnamefont {Cohen}}, \bibinfo {author}
		{\bibfnamefont {P.}~\bibnamefont {Selvaraj}}, \bibinfo {author}
		{\bibfnamefont {B.}~\bibnamefont {Hagedorn}},  \emph {et~al.},\ }\href@noop
	{} {\bibfield  {journal} {\bibinfo  {journal} {PLOS Computational Biology}\
		}\textbf {\bibinfo {volume} {17}},\ \bibinfo {pages} {e1009149} (\bibinfo
		{year} {2021}{\natexlab{a}})}\BibitemShut {NoStop}%
	\bibitem [{\citenamefont {Lorch}\ \emph {et~al.}(2022)\citenamefont {Lorch},
		\citenamefont {Kremer}, \citenamefont {Trouleau}, \citenamefont {Tsirtsis},
		\citenamefont {Szanto}, \citenamefont {Sch{\"o}lkopf},\ and\ \citenamefont
		{Gomez-Rodriguez}}]{lorch2022quantifying}%
	\BibitemOpen
	\bibfield  {author} {\bibinfo {author} {\bibfnamefont {L.}~\bibnamefont
			{Lorch}}, \bibinfo {author} {\bibfnamefont {H.}~\bibnamefont {Kremer}},
		\bibinfo {author} {\bibfnamefont {W.}~\bibnamefont {Trouleau}}, \bibinfo
		{author} {\bibfnamefont {S.}~\bibnamefont {Tsirtsis}}, \bibinfo {author}
		{\bibfnamefont {A.}~\bibnamefont {Szanto}}, \bibinfo {author} {\bibfnamefont
			{B.}~\bibnamefont {Sch{\"o}lkopf}}, \ and\ \bibinfo {author} {\bibfnamefont
			{M.}~\bibnamefont {Gomez-Rodriguez}},\ }\href@noop {} {\bibfield  {journal}
		{\bibinfo  {journal} {ACM Transactions on Spatial Algorithms and Systems}\
		}\textbf {\bibinfo {volume} {8}},\ \bibinfo {pages} {1} (\bibinfo {year}
		{2022})}\BibitemShut {NoStop}%
	\bibitem [{\citenamefont {Kojaku}\ \emph {et~al.}(2021)\citenamefont {Kojaku},
		\citenamefont {H{\'e}bert-Dufresne}, \citenamefont {Mones}, \citenamefont
		{Lehmann},\ and\ \citenamefont {Ahn}}]{kojaku2021effectiveness}%
	\BibitemOpen
	\bibfield  {author} {\bibinfo {author} {\bibfnamefont {S.}~\bibnamefont
			{Kojaku}}, \bibinfo {author} {\bibfnamefont {L.}~\bibnamefont
			{H{\'e}bert-Dufresne}}, \bibinfo {author} {\bibfnamefont {E.}~\bibnamefont
			{Mones}}, \bibinfo {author} {\bibfnamefont {S.}~\bibnamefont {Lehmann}}, \
		and\ \bibinfo {author} {\bibfnamefont {Y.-Y.}\ \bibnamefont {Ahn}},\
	}\href@noop {} {\bibfield  {journal} {\bibinfo  {journal} {Nature physics}\
		}\textbf {\bibinfo {volume} {17}},\ \bibinfo {pages} {652} (\bibinfo {year}
		{2021})}\BibitemShut {NoStop}%
	\bibitem [{\citenamefont {Tufekci}(2020)}]{tufekci_this_2020}%
	\BibitemOpen
	\bibfield  {author} {\bibinfo {author} {\bibfnamefont {Z.}~\bibnamefont
			{Tufekci}},\ }\href
	{https://www.theatlantic.com/health/archive/2020/09/k-overlooked-variable-driving-pandemic/616548/}
	{\enquote {\bibinfo {title} {This {Overlooked} {Variable} {Is} the {Key} to
				the {Pandemic}},}\ } (\bibinfo {year} {2020}),\ \bibinfo {note} {section:
		Health}\BibitemShut {NoStop}%
	\bibitem [{\citenamefont {Bradshaw}\ \emph {et~al.}(2021)\citenamefont
		{Bradshaw}, \citenamefont {Alley}, \citenamefont {Huggins}, \citenamefont
		{Lloyd},\ and\ \citenamefont {Esvelt}}]{bradshaw2021bidirectional}%
	\BibitemOpen
	\bibfield  {author} {\bibinfo {author} {\bibfnamefont {W.~J.}\ \bibnamefont
			{Bradshaw}}, \bibinfo {author} {\bibfnamefont {E.~C.}\ \bibnamefont {Alley}},
		\bibinfo {author} {\bibfnamefont {J.~H.}\ \bibnamefont {Huggins}}, \bibinfo
		{author} {\bibfnamefont {A.~L.}\ \bibnamefont {Lloyd}}, \ and\ \bibinfo
		{author} {\bibfnamefont {K.~M.}\ \bibnamefont {Esvelt}},\ }\href@noop {}
	{\bibfield  {journal} {\bibinfo  {journal} {Nature communications}\ }\textbf
		{\bibinfo {volume} {12}},\ \bibinfo {pages} {1} (\bibinfo {year}
		{2021})}\BibitemShut {NoStop}%
	\bibitem [{\citenamefont {Stein}(2011)}]{stein2011super}%
	\BibitemOpen
	\bibfield  {author} {\bibinfo {author} {\bibfnamefont {R.~A.}\ \bibnamefont
			{Stein}},\ }\href@noop {} {\bibfield  {journal} {\bibinfo  {journal}
			{International Journal of Infectious Diseases}\ }\textbf {\bibinfo {volume}
			{15}},\ \bibinfo {pages} {e510} (\bibinfo {year} {2011})}\BibitemShut
	{NoStop}%
	\bibitem [{\citenamefont {Endo}\ \emph
		{et~al.}(2020{\natexlab{a}})\citenamefont {Endo}, \citenamefont {Abbott},
		\citenamefont {Kucharski}, \citenamefont {Funk} \emph
		{et~al.}}]{endo2020estimating}%
	\BibitemOpen
	\bibfield  {author} {\bibinfo {author} {\bibfnamefont {A.}~\bibnamefont
			{Endo}}, \bibinfo {author} {\bibfnamefont {S.}~\bibnamefont {Abbott}},
		\bibinfo {author} {\bibfnamefont {A.~J.}\ \bibnamefont {Kucharski}}, \bibinfo
		{author} {\bibfnamefont {S.}~\bibnamefont {Funk}},  \emph {et~al.},\
	}\href@noop {} {\bibfield  {journal} {\bibinfo  {journal} {Wellcome open
				research}\ }\textbf {\bibinfo {volume} {5}} (\bibinfo {year}
		{2020}{\natexlab{a}})}\BibitemShut {NoStop}%
	\bibitem [{\citenamefont {Lau}\ \emph {et~al.}(2020)\citenamefont {Lau},
		\citenamefont {Grenfell}, \citenamefont {Thomas}, \citenamefont {Bryan},
		\citenamefont {Nelson},\ and\ \citenamefont
		{Lopman}}]{lau2020characterizing}%
	\BibitemOpen
	\bibfield  {author} {\bibinfo {author} {\bibfnamefont {M.~S.}\ \bibnamefont
			{Lau}}, \bibinfo {author} {\bibfnamefont {B.}~\bibnamefont {Grenfell}},
		\bibinfo {author} {\bibfnamefont {M.}~\bibnamefont {Thomas}}, \bibinfo
		{author} {\bibfnamefont {M.}~\bibnamefont {Bryan}}, \bibinfo {author}
		{\bibfnamefont {K.}~\bibnamefont {Nelson}}, \ and\ \bibinfo {author}
		{\bibfnamefont {B.}~\bibnamefont {Lopman}},\ }\href@noop {} {\bibfield
		{journal} {\bibinfo  {journal} {Proceedings of the National Academy of
				Sciences}\ }\textbf {\bibinfo {volume} {117}},\ \bibinfo {pages} {22430}
		(\bibinfo {year} {2020})}\BibitemShut {NoStop}%
	\bibitem [{\citenamefont {Althouse}\ \emph {et~al.}(2020)\citenamefont
		{Althouse}, \citenamefont {Wenger}, \citenamefont {Miller}, \citenamefont
		{Scarpino}, \citenamefont {Allard}, \citenamefont {H{\'e}bert-Dufresne},\
		and\ \citenamefont {Hu}}]{althouse2020superspreading}%
	\BibitemOpen
	\bibfield  {author} {\bibinfo {author} {\bibfnamefont {B.~M.}\ \bibnamefont
			{Althouse}}, \bibinfo {author} {\bibfnamefont {E.~A.}\ \bibnamefont
			{Wenger}}, \bibinfo {author} {\bibfnamefont {J.~C.}\ \bibnamefont {Miller}},
		\bibinfo {author} {\bibfnamefont {S.~V.}\ \bibnamefont {Scarpino}}, \bibinfo
		{author} {\bibfnamefont {A.}~\bibnamefont {Allard}}, \bibinfo {author}
		{\bibfnamefont {L.}~\bibnamefont {H{\'e}bert-Dufresne}}, \ and\ \bibinfo
		{author} {\bibfnamefont {H.}~\bibnamefont {Hu}},\ }\href@noop {} {\bibfield
		{journal} {\bibinfo  {journal} {PLoS biology}\ }\textbf {\bibinfo {volume}
			{18}},\ \bibinfo {pages} {e3000897} (\bibinfo {year} {2020})}\BibitemShut
	{NoStop}%
	\bibitem [{\citenamefont {Sun}\ \emph {et~al.}(2021)\citenamefont {Sun},
		\citenamefont {Wang}, \citenamefont {Gao}, \citenamefont {Wang},
		\citenamefont {Luo}, \citenamefont {Ren}, \citenamefont {Zhan}, \citenamefont
		{Chen}, \citenamefont {Zhao}, \citenamefont {Huang} \emph
		{et~al.}}]{sun2021transmission}%
	\BibitemOpen
	\bibfield  {author} {\bibinfo {author} {\bibfnamefont {K.}~\bibnamefont
			{Sun}}, \bibinfo {author} {\bibfnamefont {W.}~\bibnamefont {Wang}}, \bibinfo
		{author} {\bibfnamefont {L.}~\bibnamefont {Gao}}, \bibinfo {author}
		{\bibfnamefont {Y.}~\bibnamefont {Wang}}, \bibinfo {author} {\bibfnamefont
			{K.}~\bibnamefont {Luo}}, \bibinfo {author} {\bibfnamefont {L.}~\bibnamefont
			{Ren}}, \bibinfo {author} {\bibfnamefont {Z.}~\bibnamefont {Zhan}}, \bibinfo
		{author} {\bibfnamefont {X.}~\bibnamefont {Chen}}, \bibinfo {author}
		{\bibfnamefont {S.}~\bibnamefont {Zhao}}, \bibinfo {author} {\bibfnamefont
			{Y.}~\bibnamefont {Huang}},  \emph {et~al.},\ }\href@noop {} {\bibfield
		{journal} {\bibinfo  {journal} {Science}\ }\textbf {\bibinfo {volume}
			{371}},\ \bibinfo {pages} {eabe2424} (\bibinfo {year} {2021})}\BibitemShut
	{NoStop}%
	\bibitem [{\citenamefont {Lemieux}\ \emph {et~al.}(2021)\citenamefont
		{Lemieux}, \citenamefont {Siddle}, \citenamefont {Shaw}, \citenamefont
		{Loreth}, \citenamefont {Schaffner}, \citenamefont {Gladden-Young},
		\citenamefont {Adams}, \citenamefont {Fink}, \citenamefont {Tomkins-Tinch},
		\citenamefont {Krasilnikova} \emph {et~al.}}]{lemieux2021phylogenetic}%
	\BibitemOpen
	\bibfield  {author} {\bibinfo {author} {\bibfnamefont {J.~E.}\ \bibnamefont
			{Lemieux}}, \bibinfo {author} {\bibfnamefont {K.~J.}\ \bibnamefont {Siddle}},
		\bibinfo {author} {\bibfnamefont {B.~M.}\ \bibnamefont {Shaw}}, \bibinfo
		{author} {\bibfnamefont {C.}~\bibnamefont {Loreth}}, \bibinfo {author}
		{\bibfnamefont {S.~F.}\ \bibnamefont {Schaffner}}, \bibinfo {author}
		{\bibfnamefont {A.}~\bibnamefont {Gladden-Young}}, \bibinfo {author}
		{\bibfnamefont {G.}~\bibnamefont {Adams}}, \bibinfo {author} {\bibfnamefont
			{T.}~\bibnamefont {Fink}}, \bibinfo {author} {\bibfnamefont {C.~H.}\
			\bibnamefont {Tomkins-Tinch}}, \bibinfo {author} {\bibfnamefont {L.~A.}\
			\bibnamefont {Krasilnikova}},  \emph {et~al.},\ }\href@noop {} {\bibfield
		{journal} {\bibinfo  {journal} {Science}\ }\textbf {\bibinfo {volume}
			{371}},\ \bibinfo {pages} {eabe3261} (\bibinfo {year} {2021})}\BibitemShut
	{NoStop}%
	\bibitem [{\citenamefont {Smith}(2022)}]{smith2022yes}%
	\BibitemOpen
	\bibfield  {author} {\bibinfo {author} {\bibfnamefont {J.}~\bibnamefont
			{Smith}},\ }\href@noop {} {\bibfield  {journal} {\bibinfo  {journal} {The
				Lancet Infectious Diseases}\ }\textbf {\bibinfo {volume} {22}},\ \bibinfo
		{pages} {1416} (\bibinfo {year} {2022})}\BibitemShut {NoStop}%
	\bibitem [{\citenamefont {Paredes}\ \emph {et~al.}(2024)\citenamefont
		{Paredes}, \citenamefont {Ahmed}, \citenamefont {Figgins}, \citenamefont
		{Colizza}, \citenamefont {Lemey}, \citenamefont {McCrone}, \citenamefont
		{M{\"u}ller}, \citenamefont {Tran-Kiem},\ and\ \citenamefont
		{Bedford}}]{paredes2024underdetected}%
	\BibitemOpen
	\bibfield  {author} {\bibinfo {author} {\bibfnamefont {M.~I.}\ \bibnamefont
			{Paredes}}, \bibinfo {author} {\bibfnamefont {N.}~\bibnamefont {Ahmed}},
		\bibinfo {author} {\bibfnamefont {M.}~\bibnamefont {Figgins}}, \bibinfo
		{author} {\bibfnamefont {V.}~\bibnamefont {Colizza}}, \bibinfo {author}
		{\bibfnamefont {P.}~\bibnamefont {Lemey}}, \bibinfo {author} {\bibfnamefont
			{J.~T.}\ \bibnamefont {McCrone}}, \bibinfo {author} {\bibfnamefont
			{N.}~\bibnamefont {M{\"u}ller}}, \bibinfo {author} {\bibfnamefont
			{C.}~\bibnamefont {Tran-Kiem}}, \ and\ \bibinfo {author} {\bibfnamefont
			{T.}~\bibnamefont {Bedford}},\ }\href@noop {} {\bibfield  {journal} {\bibinfo
			{journal} {Cell}\ }\textbf {\bibinfo {volume} {187}},\ \bibinfo {pages}
		{1374} (\bibinfo {year} {2024})}\BibitemShut {NoStop}%
	\bibitem [{\citenamefont {Ward}\ \emph {et~al.}(2024)\citenamefont {Ward},
		\citenamefont {Overton}, \citenamefont {Paton}, \citenamefont {Christie},
		\citenamefont {Cumming},\ and\ \citenamefont
		{Fyles}}]{ward2024understanding}%
	\BibitemOpen
	\bibfield  {author} {\bibinfo {author} {\bibfnamefont {T.}~\bibnamefont
			{Ward}}, \bibinfo {author} {\bibfnamefont {C.~E.}\ \bibnamefont {Overton}},
		\bibinfo {author} {\bibfnamefont {R.~S.}\ \bibnamefont {Paton}}, \bibinfo
		{author} {\bibfnamefont {R.}~\bibnamefont {Christie}}, \bibinfo {author}
		{\bibfnamefont {F.}~\bibnamefont {Cumming}}, \ and\ \bibinfo {author}
		{\bibfnamefont {M.}~\bibnamefont {Fyles}},\ }\href@noop {} {\bibfield
		{journal} {\bibinfo  {journal} {Nature Communications}\ }\textbf {\bibinfo
			{volume} {15}},\ \bibinfo {pages} {2199} (\bibinfo {year}
		{2024})}\BibitemShut {NoStop}%
	\bibitem [{\citenamefont {Oshitani}\ \emph {et~al.}(2020)\citenamefont
		{Oshitani} \emph {et~al.}}]{oshitani2020cluster}%
	\BibitemOpen
	\bibfield  {author} {\bibinfo {author} {\bibfnamefont {H.}~\bibnamefont
			{Oshitani}} \emph {et~al.},\ }\href@noop {} {\bibfield  {journal} {\bibinfo
			{journal} {Japanese Journal of Infectious Diseases}\ ,\ \bibinfo {pages}
			{JJID}} (\bibinfo {year} {2020})}\BibitemShut {NoStop}%
	\bibitem [{\citenamefont {Lee}\ \emph {et~al.}(2020)\citenamefont {Lee},
		\citenamefont {Yuh}, \citenamefont {Yang}, \citenamefont {Cho}, \citenamefont
		{Yoo}, \citenamefont {Koh}, \citenamefont {Marshall}, \citenamefont {Oh},
		\citenamefont {Ha}, \citenamefont {Han} \emph {et~al.}}]{lee2020nationwide}%
	\BibitemOpen
	\bibfield  {author} {\bibinfo {author} {\bibfnamefont {S.~W.}\ \bibnamefont
			{Lee}}, \bibinfo {author} {\bibfnamefont {W.~T.}\ \bibnamefont {Yuh}},
		\bibinfo {author} {\bibfnamefont {J.~M.}\ \bibnamefont {Yang}}, \bibinfo
		{author} {\bibfnamefont {Y.-S.}\ \bibnamefont {Cho}}, \bibinfo {author}
		{\bibfnamefont {I.~K.}\ \bibnamefont {Yoo}}, \bibinfo {author} {\bibfnamefont
			{H.~Y.}\ \bibnamefont {Koh}}, \bibinfo {author} {\bibfnamefont
			{D.}~\bibnamefont {Marshall}}, \bibinfo {author} {\bibfnamefont
			{D.}~\bibnamefont {Oh}}, \bibinfo {author} {\bibfnamefont {E.~K.}\
			\bibnamefont {Ha}}, \bibinfo {author} {\bibfnamefont {M.~Y.}\ \bibnamefont
			{Han}},  \emph {et~al.},\ }\href@noop {} {\bibfield  {journal} {\bibinfo
			{journal} {JMIR medical informatics}\ }\textbf {\bibinfo {volume} {8}},\
		\bibinfo {pages} {e20992} (\bibinfo {year} {2020})}\BibitemShut {NoStop}%
	\bibitem [{\citenamefont {Taylor}(2020)}]{taylor2020uruguay}%
	\BibitemOpen
	\bibfield  {author} {\bibinfo {author} {\bibfnamefont {L.}~\bibnamefont
			{Taylor}},\ }\href@noop {} {\bibfield  {journal} {\bibinfo  {journal} {bmj}\
		}\textbf {\bibinfo {volume} {370}} (\bibinfo {year} {2020})}\BibitemShut
	{NoStop}%
	\bibitem [{\citenamefont {Endo}\ \emph
		{et~al.}(2020{\natexlab{b}})\citenamefont {Endo}, \citenamefont {Leclerc},
		\citenamefont {Knight}, \citenamefont {Medley}, \citenamefont {Atkins},
		\citenamefont {Funk}, \citenamefont {Kucharski} \emph
		{et~al.}}]{endo2020implication}%
	\BibitemOpen
	\bibfield  {author} {\bibinfo {author} {\bibfnamefont {A.}~\bibnamefont
			{Endo}}, \bibinfo {author} {\bibfnamefont {Q.~J.}\ \bibnamefont {Leclerc}},
		\bibinfo {author} {\bibfnamefont {G.~M.}\ \bibnamefont {Knight}}, \bibinfo
		{author} {\bibfnamefont {G.~F.}\ \bibnamefont {Medley}}, \bibinfo {author}
		{\bibfnamefont {K.~E.}\ \bibnamefont {Atkins}}, \bibinfo {author}
		{\bibfnamefont {S.}~\bibnamefont {Funk}}, \bibinfo {author} {\bibfnamefont
			{A.~J.}\ \bibnamefont {Kucharski}},  \emph {et~al.},\ }\href@noop {}
	{\bibfield  {journal} {\bibinfo  {journal} {Wellcome open research}\ }\textbf
		{\bibinfo {volume} {5}} (\bibinfo {year} {2020}{\natexlab{b}})}\BibitemShut
	{NoStop}%
	\bibitem [{\citenamefont {Flaxman}\ \emph {et~al.}(2020)\citenamefont
		{Flaxman}, \citenamefont {Mishra}, \citenamefont {Gandy}, \citenamefont
		{Unwin}, \citenamefont {Mellan}, \citenamefont {Coupland}, \citenamefont
		{Whittaker}, \citenamefont {Zhu}, \citenamefont {Berah}, \citenamefont
		{Eaton} \emph {et~al.}}]{flaxman2020estimating}%
	\BibitemOpen
	\bibfield  {author} {\bibinfo {author} {\bibfnamefont {S.}~\bibnamefont
			{Flaxman}}, \bibinfo {author} {\bibfnamefont {S.}~\bibnamefont {Mishra}},
		\bibinfo {author} {\bibfnamefont {A.}~\bibnamefont {Gandy}}, \bibinfo
		{author} {\bibfnamefont {H.~J.~T.}\ \bibnamefont {Unwin}}, \bibinfo {author}
		{\bibfnamefont {T.~A.}\ \bibnamefont {Mellan}}, \bibinfo {author}
		{\bibfnamefont {H.}~\bibnamefont {Coupland}}, \bibinfo {author}
		{\bibfnamefont {C.}~\bibnamefont {Whittaker}}, \bibinfo {author}
		{\bibfnamefont {H.}~\bibnamefont {Zhu}}, \bibinfo {author} {\bibfnamefont
			{T.}~\bibnamefont {Berah}}, \bibinfo {author} {\bibfnamefont {J.~W.}\
			\bibnamefont {Eaton}},  \emph {et~al.},\ }\href@noop {} {\bibfield  {journal}
		{\bibinfo  {journal} {Nature}\ }\textbf {\bibinfo {volume} {584}},\ \bibinfo
		{pages} {257} (\bibinfo {year} {2020})}\BibitemShut {NoStop}%
	\bibitem [{\citenamefont {Chang}\ \emph {et~al.}(2021)\citenamefont {Chang},
		\citenamefont {Pierson}, \citenamefont {Koh}, \citenamefont {Gerardin},
		\citenamefont {Redbird}, \citenamefont {Grusky},\ and\ \citenamefont
		{Leskovec}}]{chang2021mobility}%
	\BibitemOpen
	\bibfield  {author} {\bibinfo {author} {\bibfnamefont {S.}~\bibnamefont
			{Chang}}, \bibinfo {author} {\bibfnamefont {E.}~\bibnamefont {Pierson}},
		\bibinfo {author} {\bibfnamefont {P.~W.}\ \bibnamefont {Koh}}, \bibinfo
		{author} {\bibfnamefont {J.}~\bibnamefont {Gerardin}}, \bibinfo {author}
		{\bibfnamefont {B.}~\bibnamefont {Redbird}}, \bibinfo {author} {\bibfnamefont
			{D.}~\bibnamefont {Grusky}}, \ and\ \bibinfo {author} {\bibfnamefont
			{J.}~\bibnamefont {Leskovec}},\ }\href@noop {} {\bibfield  {journal}
		{\bibinfo  {journal} {Nature}\ }\textbf {\bibinfo {volume} {589}},\ \bibinfo
		{pages} {82} (\bibinfo {year} {2021})}\BibitemShut {NoStop}%
	\bibitem [{\citenamefont {Pullano}\ \emph {et~al.}(2021)\citenamefont
		{Pullano}, \citenamefont {Di~Domenico}, \citenamefont {Sabbatini},
		\citenamefont {Valdano}, \citenamefont {Turbelin}, \citenamefont {Debin},
		\citenamefont {Guerrisi}, \citenamefont {Kengne-Kuetche}, \citenamefont
		{Souty}, \citenamefont {Hanslik} \emph {et~al.}}]{pullano2021underdetection}%
	\BibitemOpen
	\bibfield  {author} {\bibinfo {author} {\bibfnamefont {G.}~\bibnamefont
			{Pullano}}, \bibinfo {author} {\bibfnamefont {L.}~\bibnamefont
			{Di~Domenico}}, \bibinfo {author} {\bibfnamefont {C.~E.}\ \bibnamefont
			{Sabbatini}}, \bibinfo {author} {\bibfnamefont {E.}~\bibnamefont {Valdano}},
		\bibinfo {author} {\bibfnamefont {C.}~\bibnamefont {Turbelin}}, \bibinfo
		{author} {\bibfnamefont {M.}~\bibnamefont {Debin}}, \bibinfo {author}
		{\bibfnamefont {C.}~\bibnamefont {Guerrisi}}, \bibinfo {author}
		{\bibfnamefont {C.}~\bibnamefont {Kengne-Kuetche}}, \bibinfo {author}
		{\bibfnamefont {C.}~\bibnamefont {Souty}}, \bibinfo {author} {\bibfnamefont
			{T.}~\bibnamefont {Hanslik}},  \emph {et~al.},\ }\href@noop {} {\bibfield
		{journal} {\bibinfo  {journal} {Nature}\ }\textbf {\bibinfo {volume} {590}},\
		\bibinfo {pages} {134} (\bibinfo {year} {2021})}\BibitemShut {NoStop}%
	\bibitem [{\citenamefont {Aleta}\ \emph {et~al.}(2020)\citenamefont {Aleta},
		\citenamefont {Martin-Corral}, \citenamefont {Pastore~y Piontti},
		\citenamefont {Ajelli}, \citenamefont {Litvinova}, \citenamefont {Chinazzi},
		\citenamefont {Dean}, \citenamefont {Halloran}, \citenamefont {Longini~Jr},
		\citenamefont {Merler} \emph {et~al.}}]{aleta2020modelling}%
	\BibitemOpen
	\bibfield  {author} {\bibinfo {author} {\bibfnamefont {A.}~\bibnamefont
			{Aleta}}, \bibinfo {author} {\bibfnamefont {D.}~\bibnamefont
			{Martin-Corral}}, \bibinfo {author} {\bibfnamefont {A.}~\bibnamefont
			{Pastore~y Piontti}}, \bibinfo {author} {\bibfnamefont {M.}~\bibnamefont
			{Ajelli}}, \bibinfo {author} {\bibfnamefont {M.}~\bibnamefont {Litvinova}},
		\bibinfo {author} {\bibfnamefont {M.}~\bibnamefont {Chinazzi}}, \bibinfo
		{author} {\bibfnamefont {N.~E.}\ \bibnamefont {Dean}}, \bibinfo {author}
		{\bibfnamefont {M.~E.}\ \bibnamefont {Halloran}}, \bibinfo {author}
		{\bibfnamefont {I.~M.}\ \bibnamefont {Longini~Jr}}, \bibinfo {author}
		{\bibfnamefont {S.}~\bibnamefont {Merler}},  \emph {et~al.},\ }\href@noop {}
	{\bibfield  {journal} {\bibinfo  {journal} {Nature Human Behaviour}\ }\textbf
		{\bibinfo {volume} {4}},\ \bibinfo {pages} {964} (\bibinfo {year}
		{2020})}\BibitemShut {NoStop}%
	\bibitem [{\citenamefont {Gupta}\ \emph {et~al.}(2020)\citenamefont {Gupta},
		\citenamefont {Maharaj}, \citenamefont {Weiss}, \citenamefont {Rahaman},
		\citenamefont {Alsdurf}, \citenamefont {Sharma}, \citenamefont {Minoyan},
		\citenamefont {Harnois-Leblanc}, \citenamefont {Schmidt}, \citenamefont
		{Charles} \emph {et~al.}}]{gupta2020covi}%
	\BibitemOpen
	\bibfield  {author} {\bibinfo {author} {\bibfnamefont {P.}~\bibnamefont
			{Gupta}}, \bibinfo {author} {\bibfnamefont {T.}~\bibnamefont {Maharaj}},
		\bibinfo {author} {\bibfnamefont {M.}~\bibnamefont {Weiss}}, \bibinfo
		{author} {\bibfnamefont {N.}~\bibnamefont {Rahaman}}, \bibinfo {author}
		{\bibfnamefont {H.}~\bibnamefont {Alsdurf}}, \bibinfo {author} {\bibfnamefont
			{A.}~\bibnamefont {Sharma}}, \bibinfo {author} {\bibfnamefont
			{N.}~\bibnamefont {Minoyan}}, \bibinfo {author} {\bibfnamefont
			{S.}~\bibnamefont {Harnois-Leblanc}}, \bibinfo {author} {\bibfnamefont
			{V.}~\bibnamefont {Schmidt}}, \bibinfo {author} {\bibfnamefont {P.-L.~S.}\
			\bibnamefont {Charles}},  \emph {et~al.},\ }\href@noop {} {\bibfield
		{journal} {\bibinfo  {journal} {arXiv preprint arXiv:2010.16004}\ } (\bibinfo
		{year} {2020})}\BibitemShut {NoStop}%
	\bibitem [{\citenamefont {Lasser}\ \emph {et~al.}(2022)\citenamefont {Lasser},
		\citenamefont {Sorger}, \citenamefont {Richter}, \citenamefont {Thurner},
		\citenamefont {Schmid},\ and\ \citenamefont {Klimek}}]{lasser2022assessing}%
	\BibitemOpen
	\bibfield  {author} {\bibinfo {author} {\bibfnamefont {J.}~\bibnamefont
			{Lasser}}, \bibinfo {author} {\bibfnamefont {J.}~\bibnamefont {Sorger}},
		\bibinfo {author} {\bibfnamefont {L.}~\bibnamefont {Richter}}, \bibinfo
		{author} {\bibfnamefont {S.}~\bibnamefont {Thurner}}, \bibinfo {author}
		{\bibfnamefont {D.}~\bibnamefont {Schmid}}, \ and\ \bibinfo {author}
		{\bibfnamefont {P.}~\bibnamefont {Klimek}},\ }\href@noop {} {\bibfield
		{journal} {\bibinfo  {journal} {Nature communications}\ }\textbf {\bibinfo
			{volume} {13}},\ \bibinfo {pages} {1} (\bibinfo {year} {2022})}\BibitemShut
	{NoStop}%
	\bibitem [{\citenamefont {Kerr}\ \emph
		{et~al.}(2021{\natexlab{b}})\citenamefont {Kerr}, \citenamefont {Mistry},
		\citenamefont {Stuart}, \citenamefont {Rosenfeld}, \citenamefont {Hart},
		\citenamefont {N{\'u}{\~n}ez}, \citenamefont {Cohen}, \citenamefont
		{Selvaraj}, \citenamefont {Abeysuriya}, \citenamefont {Jastrz{\k{e}}bski}
		\emph {et~al.}}]{kerr2021controlling}%
	\BibitemOpen
	\bibfield  {author} {\bibinfo {author} {\bibfnamefont {C.~C.}\ \bibnamefont
			{Kerr}}, \bibinfo {author} {\bibfnamefont {D.}~\bibnamefont {Mistry}},
		\bibinfo {author} {\bibfnamefont {R.~M.}\ \bibnamefont {Stuart}}, \bibinfo
		{author} {\bibfnamefont {K.}~\bibnamefont {Rosenfeld}}, \bibinfo {author}
		{\bibfnamefont {G.~R.}\ \bibnamefont {Hart}}, \bibinfo {author}
		{\bibfnamefont {R.~C.}\ \bibnamefont {N{\'u}{\~n}ez}}, \bibinfo {author}
		{\bibfnamefont {J.~A.}\ \bibnamefont {Cohen}}, \bibinfo {author}
		{\bibfnamefont {P.}~\bibnamefont {Selvaraj}}, \bibinfo {author}
		{\bibfnamefont {R.~G.}\ \bibnamefont {Abeysuriya}}, \bibinfo {author}
		{\bibfnamefont {M.}~\bibnamefont {Jastrz{\k{e}}bski}},  \emph {et~al.},\
	}\href@noop {} {\bibfield  {journal} {\bibinfo  {journal} {Nature
				communications}\ }\textbf {\bibinfo {volume} {12}},\ \bibinfo {pages} {2993}
		(\bibinfo {year} {2021}{\natexlab{b}})}\BibitemShut {NoStop}%
	\bibitem [{\citenamefont {Dinnes}\ \emph {et~al.}(2021)\citenamefont {Dinnes},
		\citenamefont {Deeks}, \citenamefont {Berhane}, \citenamefont {Taylor},
		\citenamefont {Adriano}, \citenamefont {Davenport}, \citenamefont {Dittrich},
		\citenamefont {Emperador}, \citenamefont {Takwoingi}, \citenamefont
		{Cunningham}, \citenamefont {Beese}, \citenamefont {Domen}, \citenamefont
		{Dretzke}, \citenamefont {Ruffano}, \citenamefont {Harris}, \citenamefont
		{Price}, \citenamefont {Taylor-Phillips}, \citenamefont {Hooft},
		\citenamefont {Leeflang}, \citenamefont {McInnes}, \citenamefont {Spijker},
		\citenamefont {Bruel},\ and\ \citenamefont {Group}}]{dinnes2021rapid}%
	\BibitemOpen
	\bibfield  {author} {\bibinfo {author} {\bibfnamefont {J.}~\bibnamefont
			{Dinnes}}, \bibinfo {author} {\bibfnamefont {J.~J.}\ \bibnamefont {Deeks}},
		\bibinfo {author} {\bibfnamefont {S.}~\bibnamefont {Berhane}}, \bibinfo
		{author} {\bibfnamefont {M.}~\bibnamefont {Taylor}}, \bibinfo {author}
		{\bibfnamefont {A.}~\bibnamefont {Adriano}}, \bibinfo {author} {\bibfnamefont
			{C.}~\bibnamefont {Davenport}}, \bibinfo {author} {\bibfnamefont
			{S.}~\bibnamefont {Dittrich}}, \bibinfo {author} {\bibfnamefont
			{D.}~\bibnamefont {Emperador}}, \bibinfo {author} {\bibfnamefont
			{Y.}~\bibnamefont {Takwoingi}}, \bibinfo {author} {\bibfnamefont
			{J.}~\bibnamefont {Cunningham}}, \bibinfo {author} {\bibfnamefont
			{S.}~\bibnamefont {Beese}}, \bibinfo {author} {\bibfnamefont
			{J.}~\bibnamefont {Domen}}, \bibinfo {author} {\bibfnamefont
			{J.}~\bibnamefont {Dretzke}}, \bibinfo {author} {\bibfnamefont {L.~F.~d.}\
			\bibnamefont {Ruffano}}, \bibinfo {author} {\bibfnamefont {I.~M.}\
			\bibnamefont {Harris}}, \bibinfo {author} {\bibfnamefont {M.~J.}\
			\bibnamefont {Price}}, \bibinfo {author} {\bibfnamefont {S.}~\bibnamefont
			{Taylor-Phillips}}, \bibinfo {author} {\bibfnamefont {L.}~\bibnamefont
			{Hooft}}, \bibinfo {author} {\bibfnamefont {M.~M.}\ \bibnamefont {Leeflang}},
		\bibinfo {author} {\bibfnamefont {M.~D.}\ \bibnamefont {McInnes}}, \bibinfo
		{author} {\bibfnamefont {R.}~\bibnamefont {Spijker}}, \bibinfo {author}
		{\bibfnamefont {A.~V.~d.}\ \bibnamefont {Bruel}}, \ and\ \bibinfo {author}
		{\bibfnamefont {C.~C.-. D. T.~A.}\ \bibnamefont {Group}},\ }\href {\doibase
		10.1002/14651858.CD013705.pub2} {\bibfield  {journal} {\bibinfo  {journal}
			{Cochrane Database of Systematic Reviews}\ } (\bibinfo {year} {2021}),\
		10.1002/14651858.CD013705.pub2},\ \bibinfo {note} {publisher: John Wiley \&
		Sons, Ltd}\BibitemShut {NoStop}%
	\bibitem [{\citenamefont {Wong}\ and\ \citenamefont
		{Collins}(2020)}]{wong2020evidence}%
	\BibitemOpen
	\bibfield  {author} {\bibinfo {author} {\bibfnamefont {F.}~\bibnamefont
			{Wong}}\ and\ \bibinfo {author} {\bibfnamefont {J.~J.}\ \bibnamefont
			{Collins}},\ }\href {\doibase 10.1073/pnas.2018490117} {\bibfield  {journal}
		{\bibinfo  {journal} {Proceedings of the National Academy of Sciences}\
		}\textbf {\bibinfo {volume} {117}},\ \bibinfo {pages} {29416} (\bibinfo
		{year} {2020})},\ \Eprint
	{http://arxiv.org/abs/https://www.pnas.org/doi/pdf/10.1073/pnas.2018490117}
	{https://www.pnas.org/doi/pdf/10.1073/pnas.2018490117} \BibitemShut {NoStop}%
	\bibitem [{\citenamefont {Kennedy-Shaffer}\ \emph {et~al.}(2021)\citenamefont
		{Kennedy-Shaffer}, \citenamefont {Baym},\ and\ \citenamefont
		{Hanage}}]{kennedy2021perfect}%
	\BibitemOpen
	\bibfield  {author} {\bibinfo {author} {\bibfnamefont {L.}~\bibnamefont
			{Kennedy-Shaffer}}, \bibinfo {author} {\bibfnamefont {M.}~\bibnamefont
			{Baym}}, \ and\ \bibinfo {author} {\bibfnamefont {W.~P.}\ \bibnamefont
			{Hanage}},\ }\href@noop {} {\bibfield  {journal} {\bibinfo  {journal} {The
				Lancet Microbe}\ }\textbf {\bibinfo {volume} {2}},\ \bibinfo {pages} {e219}
		(\bibinfo {year} {2021})}\BibitemShut {NoStop}%
	\bibitem [{\citenamefont {Team}(2020)}]{sibGitHub}%
	\BibitemOpen
	\bibfield  {author} {\bibinfo {author} {\bibfnamefont {S.}~\bibnamefont
			{Team}},\ }\href@noop {} {\enquote {\bibinfo {title} {Sib},}\ }\bibinfo
	{howpublished} {\url{https://github.com/sibyl-team/}} (\bibinfo {year}
	{2020})\BibitemShut {NoStop}%
	\bibitem [{\citenamefont {Lauer}\ \emph {et~al.}(2020)\citenamefont {Lauer},
		\citenamefont {Grantz}, \citenamefont {Bi}, \citenamefont {Jones},
		\citenamefont {Zheng}, \citenamefont {Meredith}, \citenamefont {Azman},
		\citenamefont {Reich},\ and\ \citenamefont
		{Lessler}}]{lauer_incubation_2020}%
	\BibitemOpen
	\bibfield  {author} {\bibinfo {author} {\bibfnamefont {S.~A.}\ \bibnamefont
			{Lauer}}, \bibinfo {author} {\bibfnamefont {K.~H.}\ \bibnamefont {Grantz}},
		\bibinfo {author} {\bibfnamefont {Q.}~\bibnamefont {Bi}}, \bibinfo {author}
		{\bibfnamefont {F.~K.}\ \bibnamefont {Jones}}, \bibinfo {author}
		{\bibfnamefont {Q.}~\bibnamefont {Zheng}}, \bibinfo {author} {\bibfnamefont
			{H.~R.}\ \bibnamefont {Meredith}}, \bibinfo {author} {\bibfnamefont {A.~S.}\
			\bibnamefont {Azman}}, \bibinfo {author} {\bibfnamefont {N.~G.}\ \bibnamefont
			{Reich}}, \ and\ \bibinfo {author} {\bibfnamefont {J.}~\bibnamefont
			{Lessler}},\ }\href {\doibase 10.7326/M20-0504} {\bibfield  {journal}
		{\bibinfo  {journal} {Annals of Internal Medicine}\ } (\bibinfo {year}
		{2020}),\ 10.7326/M20-0504}\BibitemShut {NoStop}%
	\bibitem [{\citenamefont {Panovska-Griffiths}\ \emph
		{et~al.}(2020)\citenamefont {Panovska-Griffiths}, \citenamefont {Kerr},
		\citenamefont {Stuart}, \citenamefont {Mistry}, \citenamefont {Klein},
		\citenamefont {Viner},\ and\ \citenamefont
		{Bonell}}]{panovska2020determining}%
	\BibitemOpen
	\bibfield  {author} {\bibinfo {author} {\bibfnamefont {J.}~\bibnamefont
			{Panovska-Griffiths}}, \bibinfo {author} {\bibfnamefont {C.~C.}\ \bibnamefont
			{Kerr}}, \bibinfo {author} {\bibfnamefont {R.~M.}\ \bibnamefont {Stuart}},
		\bibinfo {author} {\bibfnamefont {D.}~\bibnamefont {Mistry}}, \bibinfo
		{author} {\bibfnamefont {D.~J.}\ \bibnamefont {Klein}}, \bibinfo {author}
		{\bibfnamefont {R.~M.}\ \bibnamefont {Viner}}, \ and\ \bibinfo {author}
		{\bibfnamefont {C.}~\bibnamefont {Bonell}},\ }\href@noop {} {\bibfield
		{journal} {\bibinfo  {journal} {The Lancet Child \& Adolescent Health}\
		}\textbf {\bibinfo {volume} {4}},\ \bibinfo {pages} {817} (\bibinfo {year}
		{2020})}\BibitemShut {NoStop}%
	\bibitem [{\citenamefont {Pham}\ \emph {et~al.}(2021)\citenamefont {Pham},
		\citenamefont {Stuart}, \citenamefont {Nguyen}, \citenamefont {Luong},
		\citenamefont {Tran}, \citenamefont {Pham}, \citenamefont {Phan},
		\citenamefont {Dang}, \citenamefont {Tran}, \citenamefont {Do} \emph
		{et~al.}}]{pham2021estimating}%
	\BibitemOpen
	\bibfield  {author} {\bibinfo {author} {\bibfnamefont {Q.~D.}\ \bibnamefont
			{Pham}}, \bibinfo {author} {\bibfnamefont {R.~M.}\ \bibnamefont {Stuart}},
		\bibinfo {author} {\bibfnamefont {T.~V.}\ \bibnamefont {Nguyen}}, \bibinfo
		{author} {\bibfnamefont {Q.~C.}\ \bibnamefont {Luong}}, \bibinfo {author}
		{\bibfnamefont {Q.~D.}\ \bibnamefont {Tran}}, \bibinfo {author}
		{\bibfnamefont {T.~Q.}\ \bibnamefont {Pham}}, \bibinfo {author}
		{\bibfnamefont {L.~T.}\ \bibnamefont {Phan}}, \bibinfo {author}
		{\bibfnamefont {T.~Q.}\ \bibnamefont {Dang}}, \bibinfo {author}
		{\bibfnamefont {D.~N.}\ \bibnamefont {Tran}}, \bibinfo {author}
		{\bibfnamefont {H.~T.}\ \bibnamefont {Do}},  \emph {et~al.},\ }\href@noop {}
	{\bibfield  {journal} {\bibinfo  {journal} {The Lancet Global Health}\
		}\textbf {\bibinfo {volume} {9}},\ \bibinfo {pages} {e916} (\bibinfo {year}
		{2021})}\BibitemShut {NoStop}%
	\bibitem [{cov(2022{\natexlab{a}})}]{covasibyl}%
	\BibitemOpen
	\href@noop {} {\enquote {\bibinfo {title} {Covasibyl - interventions code for
				the covasim model},}\ }\bibinfo {howpublished}
	{\url{https://github.com/sibyl-team/covasibyl}} (\bibinfo {year}
	{2022}{\natexlab{a}})\BibitemShut {NoStop}%
	\bibitem [{cov(2022{\natexlab{b}})}]{covasim-experiments}%
	\BibitemOpen
	\href@noop {} {\enquote {\bibinfo {title} {Code for the simulations on the
				covasim model},}\ }\bibinfo {howpublished}
	{\url{https://github.com/sibyl-team/covasim-experiments}} (\bibinfo {year}
	{2022}{\natexlab{b}})\BibitemShut {NoStop}%
	\bibitem [{sim(2022)}]{simulator}%
	\BibitemOpen
	\href@noop {} {\enquote {\bibinfo {title} {Github fork of the spatiotemporal
				model used in this work},}\ }\bibinfo {howpublished}
	{\url{https://github.com/sibyl-team/simulator}} (\bibinfo {year}
	{2022})\BibitemShut {NoStop}%
\end{thebibliography}
%

\pagebreak

\appendix

\setcounter{figure}{0}
\renewcommand{\figurename}{Figure}
\renewcommand{\thefigure}{S\arabic{figure}}

\section{Epidemic models for COVID-19 disease diffusion\label{sec:allmodels}}
This section provides a concise overview of three distinct agent-based models that have been specifically developed to simulate and forecast COVID-19 epidemic trends within populations characterized by contact structures and age stratification. These advanced models also serve as invaluable tools in evaluating the efficacy of diverse non-pharmaceutical containment measures implemented to effectively curb the transmission of the SARS-CoV-2 virus.

\subsection{OpenABM-Covid19\label{sec:ABM}}

The Open Agent-Based Model (OpenABM) is a pioneering COVID-19 epidemic simulator that was introduced during the first European outbreak in early 2020 \cite{ferretti2020quantifying,hinch2021openabm}. This agent-based model represents a population of individuals, with their demographic characteristics such as household size and age distribution derived from the 2011 census data of the United Kingdom.

In the OpenABM, individuals interact on a daily basis through a contact network that combines three synthetic graphs representing different social layers: households, occupations, and casual interactions. The household layer consists of static complete graphs that connect individuals within the same household, with these interactions occurring identically every day. Additionally, each individual is also part of an occupation network, which models interactions within schools (for children), workplaces (for adults), and recurrent social activities (for elderly individuals): the occupation networks are modeled as static Watts-Strogatz small-world networks, from which different subsets of interactions are randomly sampled on a daily basis. Furthermore, the model includes casual interactions that are randomly drawn each day, independent of previous connections. The number of connections an individual has in this layer follows a negative binomial distribution, which explicitly considers the presence of potential super-spreaders in the model.

The epidemic model used in the OpenABM is a discrete-time generalized non-Markovian Susceptible-Infected-Recovered (SIR) model that incorporates three distinct infection routes. These routes differentiate between asymptomatic individuals, two types of pre-symptomatic individuals, and two classes of symptomatic individuals with varying levels of symptom severity (mild and severe). Infected individuals can undergo transitions toward recovery, hospitalization, or death. The model's structure, including all possible transitions between epidemic states, is depicted schematically in Figure~\ref{fig:dynamics_allmodels}\figpanel{A}. 
The transition times between states are drawn from Gamma distributions with phenomenological parameter values. A comprehensive description of the model's parameters, primarily extrapolated from epidemiological and medical analyses conducted during the first outbreak in Hubei, can be found in Ref.\cite{hinch2021openabm}. 

The spread of the virus occurs when infected individuals come into contact with susceptible individuals. Daily contacts are assumed to be instantaneous and carry an infection probability that depends on several factors. The infection probability exhibits a non-trivial time dependence, increasing from zero and peaking around six days after the infector's own infection; it then gradually decreases toward zero. This time-dependent pattern implicitly considers the incubation period of the virus  \cite{lauer_incubation_2020}, while the other epidemic models under study explicitly include an Exposed state to account for this period. The magnitude of the infection probability is influenced by the infector's state (symptomatic or asymptomatic) and the susceptibility of the potentially infected individual (with higher susceptibility among elderly individuals).

\subsection{Covasim\label{sec:Covasim}}

Covasim is an agent-based simulator introduced in \cite{kerr2021covasim} and has been utilized in several studies to develop and evaluate country-specific containment policies for COVID-19 \cite{panovska2020determining, kerr2021controlling, pham2021estimating}. Similar to OpenABM, Covasim is a discrete-time agent-based model that operates on a multi-layer contact network. The population structure is based on country-specific demographic information, such as age, sex, and comorbidity data.  The contact network consists of various social layers, including households, workplaces, schools, long-term care facilities, and community contacts encompassing shared public spaces and public transportation. Within Covasim, these networks can be generated using SynthPops, an open-source data-driven model capable of creating realistic synthetic contact networks for populations.
All contact networks considered in Covasim are static, except for community contacts, which are randomly redrawn over time. To generate realistic contact networks, we employ the Synthpops package along with population data specific to King County, Washington,  following the methodology outlined in \cite{kerr2021covasim}. To ensure computational feasibility, the population size is limited to 70,000 individuals. In Covasim, it is also possible to model a large population while working with a smaller effective population size by using a ``dynamical rescaling'' technique that assigns multiple individuals to a single infected agent and dynamically adjusts the population size based on the number of infected agents \cite{kerr2021covasim}. However, in the present study, we deliberately disable this feature as it is not suitable for implementing any contact tracing methodology.

The epidemic process in Covasim is a generalized version of a non-Markovian Susceptible-Exposed-Infectious-Recovered (SEIR) model in discrete time. It introduces additional states to differentiate infectious individuals into asymptomatic and symptomatic categories. The symptomatic category is further subdivided into pre-symptomatic, mildly symptomatic, severely symptomatic, and critically symptomatic. The model also includes states to account for the recovery and death of individuals. Figure~\ref{fig:dynamics_allmodels}\figpanel{B} reports a schematic description of the model's structure, including all possible transitions between epidemic states.  Transition times between these states are drawn from log-normal distributions with different parameters. The daily transmission probability in a contact between an infectious individual and a susceptible one depends on various factors, including individual parameters (such as relative transmissibility and susceptibility), the symptomaticity level of the infectious individual, and the social layer to which the contact belongs. Following the observation that viral load is highest around or slightly before the onset of symptoms and decreases afterward, the transmission probability is assumed to decrease over time, reaching a plateau at half its initial value a few days after the infector's own infection.
Covasim includes a predefined intervention based on manual contact tracing, which is used to trace the source of newly infected individuals. In this implementation, the contacts of recently infected individuals, excluding community contacts, are randomly traced with a given probability. Individuals identified through contact tracing are preemptively quarantined, regardless of their state. For individuals in preemptive quarantine, a 40\% effective reduction in transmissibility due to mobility and interactions with others is assumed. This intervention is applied in all simulations in Figure 1 of the main text.
To implement contact tracing strategies, including Test-Trace-Quarantine (TTQ), we leverage the modular nature of the Covasim model, which easily enables the definition of new intervention strategies. The code for these interventions is available at \cite{covasibyl}, while the scripts used for running the interventions can be found at \cite{covasim-experiments}. All parameter values in the model are set to those used in \cite{kerr2021covasim}.

\subsection{StEM\label{sec:StEM}}

The Spatiotemporal Epidemic Model (StEM), introduced in Ref.~\cite{lorch2022quantifying}, is an advanced epidemic simulator that encompasses two interconnected continuous-time discrete-space processes. The first process is a mobility simulation, wherein individuals can travel to multiple locations and get in contact with others who are present in the same place simultaneously. The second process is a proper epidemic simulation, which accounts for the spread of the virus through the population initiated by one or a few initially infected individuals and facilitated by the aforementioned contacts. 

The StEM model places significant emphasis on generating realistic mobility data. Once an urban area is selected for simulation, the mobility generator requires a set of population data, including population density, age group distributions, and household composition. These data are used to generate and locate the set of possible households, along with their inhabitants, on a map. The various sites that individuals can visit during the mobility simulation are instead generated leveraging geolocation data. These locations fall into the following categories: education (schools, universities), social activities (bars, restaurants, cafes), business (offices, shops), supermarkets, and bus stops. Due to the unavailability of public data on the actual mobility patterns of individuals, the StEM model adopts an approach where visits to specific places are simulated under the assumption that people tend to visit only a limited subset of the possible venues. The probability of visiting a particular site decreases when the distance between the site and the individual's household increases. 

\begin{figure}[h]
	\centering
	\begin{overpic}[width=\textwidth]{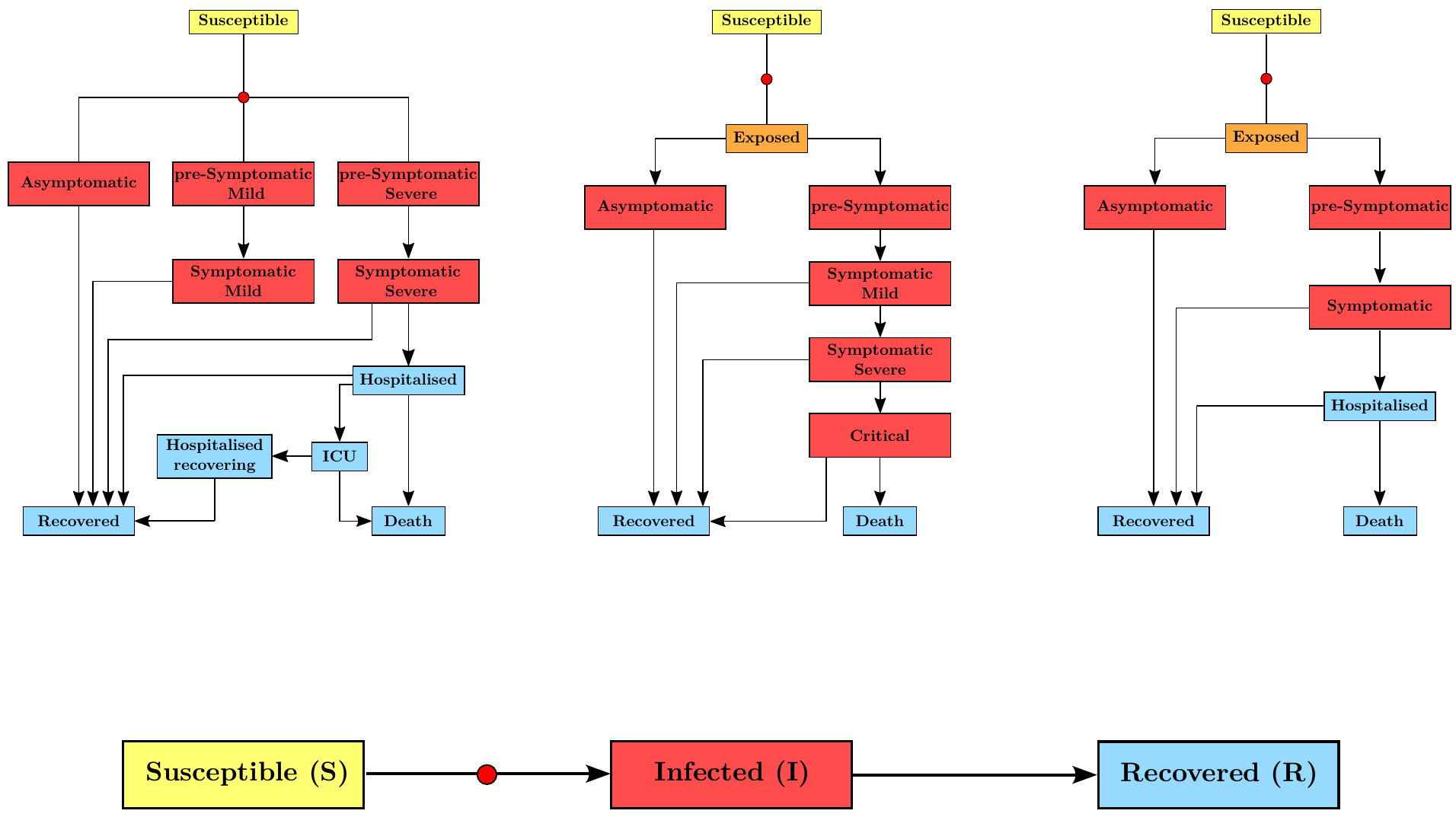}
		\put(7,54.5){{\figpanel{A}}}
		\put(44,54.45){{\figpanel{B}}}
		\put(78,54.5){{\figpanel{C}}}
		\put(6,8){{\figpanel{D}}}
	\end{overpic}
	\caption{Schematic representation of the epidemic dynamics in the three models considered in this work. \figpanel{A}: OpenABM; \figpanel{B}: Covasim.  \figpanel{C}: StEM. Color codes identify how these states are considered within the reduction to an effective SIR model \figpanel{D}. Infection events are graphically represented by red dots; each arrow represents a transition between two states, whose waiting time is drawn from certain probability distributions; the probability to undergo one of the different infection pathways is always age-dependent in the three models.
		\label{fig:dynamics_allmodels}}
\end{figure}

The epidemic model employed by StEM extends the Susceptible-Exposed-Infectious-Recovered (SEIR) model in continuous time, incorporating multiple infected states to accommodate the differentiation of pre-symptomatic, symptomatic, and asymptomatic individuals. Recovery, hospitalization, and death are possible evolutions of the infected states. Figure~\ref{fig:dynamics_allmodels}\figpanel{C} schematically shows the different individual epidemic states in the StEM model and all possible transitions between them. 

The mobility of individuals and the evolution of their epidemic states are modeled by a collection of counting processes that are mathematically represented by stochastic differential equations (SDE) with jumps (as the dynamics require discrete state transitions in continuous time). For practical convenience, all mobility data are first generated and then used to identify exposure events, which occur when a susceptible individual and an infector in a pre-symptomatic, symptomatic, or asymptomatic state, are simultaneously present in the same venue. The exposure counting process considers, together with a transmission rate that depends on the state of the infector, a venue-dependent exposure rate and a kernel term that quantifies possible environmental transmissions (due to the presence of the virus in the air or on the surfaces). The remaining events are individual-dependent, characterized by transition times with log-normal distributions, whose parameters are derived from the relevant data extracted from clinical COVID-19 literature. 

The sampling algorithm employed for the practical implementation of the epidemic dynamics is based on an event queue to track state transitions for each individual. Starting from the initial state of the population, the algorithm samples the next transition time for each individual and adds the corresponding event to the queue. The algorithm then iterates through the event queue, updating individual states and sampling the next transition times until the queue is empty. The approach described in Ref.~\cite{lorch2022quantifying} also includes a mitigation strategy that incorporates both manual and digital contact tracing. 

In the present work, interventions are implemented on a daily basis, and therefore, the epidemic inference algorithms are also designed using a discrete-time framework. For the sake of algorithmic efficiency, the very  large amount of contacts generated by the continuous-time process defined in the StEM model, are aggregated over a finite time window of one day: it means that all contacts between individuals $i$ and $j$ occurring within the same day 
are aggregated into a single effective coarse-grained daily contact. The duration of this aggregated contact is equal to the sum of the durations of the actual contacts. We conducted additional analyses with shorter time windows, ranging from $6$ hours, to verify if this would provide any additional advantages in terms of epidemic inference performance. However, it was found that smaller time windows did not yield any significant improvements. 
We also assume that the contact tracing process models the probability of exposure at day $t$ between the two individuals in contact for a total time $\Delta t$ as the integrated probability on a single contact of duration  $\Delta t$ (with the same instantaneous transmission rate defined for the underlying StEM model used in Ref.~\cite{lorch2022quantifying}). 
The simulation code used for these analyses is built upon the original model code presented in Ref.~\cite{lorch2022quantifying}. The code can be found in the GitHub fork \cite{simulator}.

\section{Overdispersion properties}
All three models discussed in the previous section account in different ways for the presence of super-spreaders in the simulation, meaning that the frequency of secondary infections is characterized by an overdispersed distribution with heavy tails. 

As explained in the main text, while in OpenABM and StEM, the source of overdispersion resides in the contact network structure (through an explicit contact graph with connectivity following a fat-tailed distribution in the first case and as a result of the continuous-time mobility simulation in the second), in Covasim overdispersion arises directly through the relative transmissibility intensity parameter, that is drawn from a heavy-tailed distribution for each individual in the population.

\begin{figure}[ht]
	\begin{center}
		\includegraphics[width=0.6\linewidth]{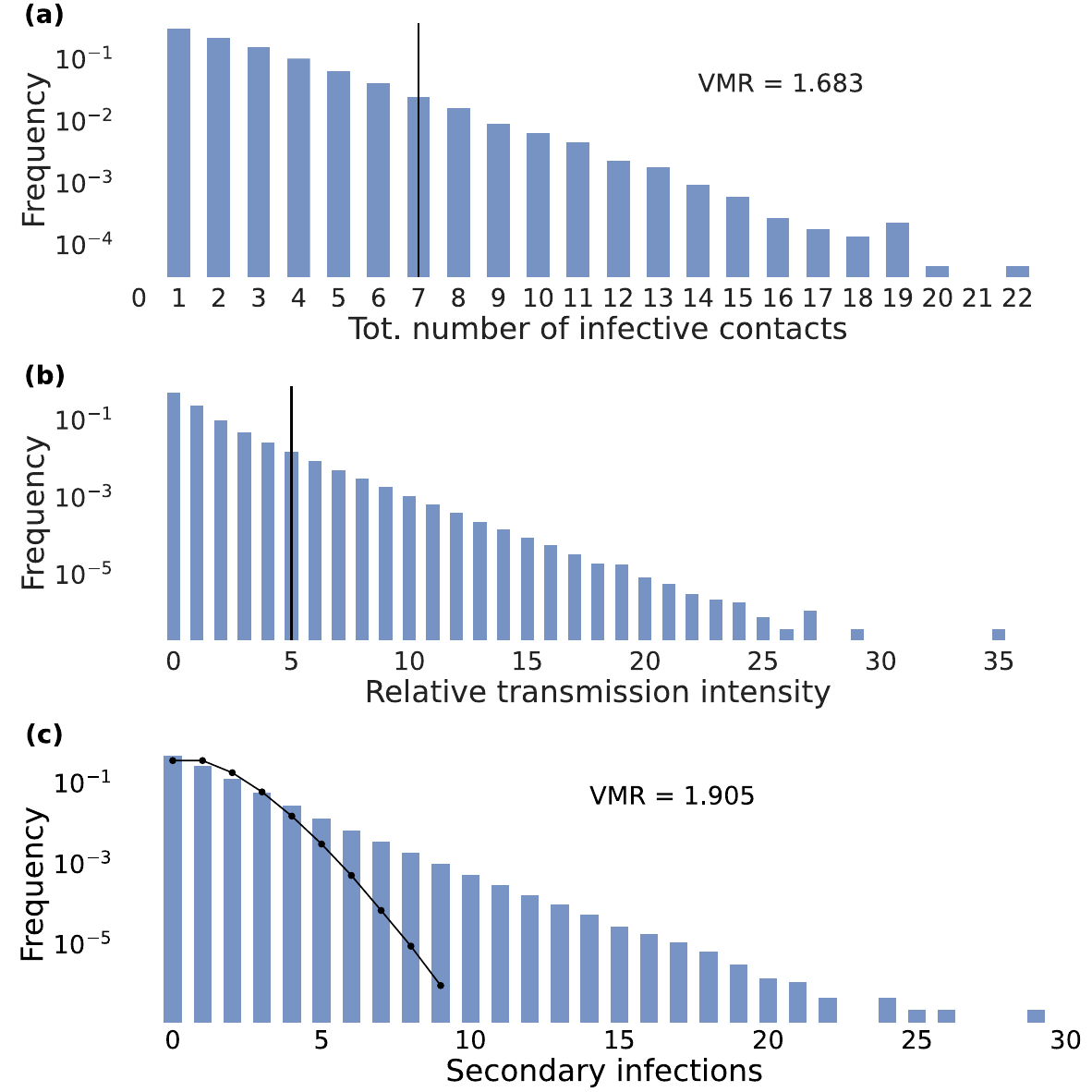}
		\caption{
			Empirical measures of overdispersion in the three epidemic models. \figpanel{a} empirical distribution of the number of infective contact for StEM; \figpanel{b} empirical
			density of the relative transmission intensity in Covasim; \figpanel{c} empirical distribution of secondary
			infections in OpenABM. 
			The threshold used to identify the super-spreaders is 7 infective contacts in StEM \figpanel{a} and OpenABM (c). For Covasim \figpanel{b} we used the model default threshold for superspreaders, i.e. those having a relative transmission intensity $>5$ The variance-to-mean ratio (VMR) is larger than one in the two cases (the precise values are reported in panels \figpanel{a} and \figpanel{c}. As a comparison, the dotted curve in \figpanel{c} represents the behavior of a Poisson distribution of infections with an expectation value equal to the empirical expectation value computed from the histogram. It is worth noting that two distributions (the empirical one and the Poisson one, the latter having $\rm VMR=1$ by construction) are remarkably different in the behavior of the tail, as the empirical one is significantly heavy-tailed.
			\label{fig:overdispersion}}
	\end{center}
\end{figure}

To highlight this behavior, we computed model-based measures characterizing the overdispersion of infections for a typical realization of an epidemic on a large population. Results are shown in Figure \ref{fig:overdispersion} for the three models: panel \figpanel{a} shows the empirical distribution of the number of infective contacts in StEM; panel \figpanel{b} displays the empirical density of the relative transmission intensity in Covasim, and \figpanel{c} represents the empirical distribution of secondary infections in OpenABM.
In all three cases, overdispersion is confirmed by a variance-to-mean ratio (VMR) larger than $1$ (as a comparison, a Poisson distribution with thin tails would have $\text{VMR}=1$). We show a comparison in panel (c) of Figure~\ref{fig:overdispersion}.

\section{Probabilistic inference through effective SIR models}
The epidemic dynamics of the three models discussed in the previous section are characterized by multiple compartmental states, various infection routes, and specific transition time distributions. These models offer a level of detail that surpasses the typical models used in statistical physics, such as the SIR model. Although viable in principle, utilizing these detailed models as prior distributions for the forward dynamics within the proposed Bayesian inference scheme would require implementing model-specific approximations, with more complex algorithmic implementations and a consequent increase in the computational cost. In particular, in the case of the approach based on Belief Propagation, this extension would involve generalizing existing message-passing approaches to effectively handle the unique characteristics and intricacies of these models. Not only would this approach 
be computationally cumbersome, but it would also rely heavily on the specific details of the considered model, making it less generic. On the other hand, there is compelling evidence that even using a simplified description of the dynamics in the inference procedure, it is still possible to detect the individuals with the highest risk of infection, even when accounting for complex dynamics that aim to mimic realistic epidemic spreading. This is confirmed by recent studies on epidemic mitigation through statistical inference techniques \citep{baker2021epidemic}. 
Consequently, the probabilistic inference methods employed here (based on BP and SMF approximations) assume that the epidemic propagation can be adequately described by a SIR model. For the sake of simplicity, and following \cite{baker2021epidemic}, we adopted discrete-time dynamics in both methods. The discrete-time framework is directly applicable to two of the three models considered here (OpenABM and Covasim); regarding  StEM (cf. Section \ref{sec:StEM}), an additional pre-processing step is necessary to aggregate contacts within a one-day window. 
In the next section, we will provide a brief description of the Bayesian formulation with a non-Markovian SIR model as a prior distribution. Subsequently, in Section \ref{subsec:mappingSIR} we will discuss how to leverage the information embedded in the three models to enhance the performance of the inference algorithms.

\subsection{Bayesian inference on non-Markovian SIR model}
Let us consider a graph $G=\left(V,E\right)$ which represents the time-evolving contact network of a set of $N=\left| V \right|$ individuals. Each node in the graph corresponds to an individual and is associated with a time-dependent variable, denoted as $x_{i}^{t}$, representing the individual's state at time $t$. The variable $x_i^t$ takes values from a finite set of epidemic states: in the SIR model, $x_{i}^{t}\in\mathcal{X}=\left\{ S,I,R\right\} $, representing  respectively the states Susceptible ($S$), Infected ($I$) and Recovered/Removed ($R$). The dynamic is assumed to be discretized in time, with $t$ ranging from $0$ to $T$, representing the time period under consideration (e.g. days). The Markovian SIR dynamics is usually fully specified by two sets of parameters: the transmission probabilities
$\left\{ \lambda^t_{i\to k}\right\}$, representing the probability that $i$ will infect another individual $k$ at time $t$, and the recovery rates $\{\mu_{i}\}$, representing the daily probability with which $i$ can recover. In a Markovian discrete-time process, the distribution of recovery times is geometric, however, this is not generally the case for real-world diseases such as COVID-19. For this reason, we consider here a non-Markovian version of the SIR model in which both the transmission probabilities $\left\{ \lambda^t_{i\to k}\right\}$ and the daily recovery probabilities also depend on the time elapsed since agent $i$'s infection. These time dependencies can be used to describe time-dependent infectiousness (for instance, the initial incubation period of the virus in the organism), as well as clinical interventions (recovery, treatments, the appearance of symptoms, and so on), that influence the time-dependency of the recovery probability. Recovery rates are replaced by possibly individual-dependent recovery time distributions $p_i(\tau_i)$, where $\tau_i$ is the number of days agent $i$ takes to recover from infection.

In the following discussion, we introduce the notations and equations used to describe the dynamics of the SIR model. We denote with $\boldsymbol{x}_{i}=\left(x_{i}^{0},\ldots,x_{i}^{T}\right)$
(resp. $\boldsymbol{x}^{t}=\left(x_{1}^{t},\ldots,x_{N}^{t}\right)$) the trajectory of node $i$ (resp. the state of all nodes at
time $t$). Let $t_{i}=\text{min}\left\{ t:x_{i}^{t}=I\right\} $ be the infection time of agent $i$, the transition probabilities
$W\left[x_{i}^{t+1}\mid\boldsymbol{x}^{0},\ldots,\boldsymbol{x}^{t}\right]$
for node $i$ occurring between time $t$ and time $t+1$ are then
\begin{subequations}
	\begin{align}
		W\left[x_{i}^{t+1}=S\mid\boldsymbol{x}^{0},\ldots,\boldsymbol{x}^{t}\right] & =\delta_{x_{i}^{t},S}\prod_{k\neq i}\left(1-\lambda^t_{k\to i}\left(t_{k}\right)\delta_{x_{k}^{t},I}\right);\label{eq:SIR_transition_s_tp1}\\
		W\left[x_{i}^{t+1}=I\mid\boldsymbol{x}^{0},\ldots,\boldsymbol{x}^{t}\right] & =\left[1-\mu_{i}\left(t-t_{i}\right)\right]\delta_{x_{i}^{t},I}+\delta_{x_{i}^{t},S}\left[1-\prod_{k\neq i}\left(1-\lambda^t_{k\to i}\left(t_{k}\right)\delta_{x_{k}^{t},I}\right)\right]; \label{eq:SIR_transition_I_tp1}\\
		W\left[x_{i}^{t+1}=R\mid\boldsymbol{x}^{0},\ldots,\boldsymbol{x}^{t}\right] & =\mu_{i}\left(t-t_{i}\right)\delta_{x_{i}^{t},I}+\delta_{x_{i}^{t},R},\label{eq:SIR_transition_R_tp1}
	\end{align}
	\label{subeq:SIR_transition_all}
\end{subequations}
where in all the formulas $\delta$ denotes the Kronecker symbol, and we set $\lambda^t_{i\to j}(t_i)=\lambda^t_{j\to i}(t_j)=0$ if node $i$ and $j$ are not in contact at time $t$. The recovery probability $\mu_i(t-t_i)$ after $t-t_i$ days since infection is defined from the p.d.f. $p(t)$ as the hazard function 
\begin{equation}
	\mu_i(t-t_i) = \frac{p(t-t_i)}{1 - \sum_{s=0}^{t-t_i} p(s) }.
\end{equation} 
Denoting with $\mathbb{X}=\left\{ x_{i}^{t}\right\} _{i=1,\dots,N}^{t=0,\dots,T}$
the full epidemic trajectory, its probability can be written as
\begin{equation}
	p\left(\mathbb{X}\right)=p\left(\boldsymbol{x}^{0}\right)\prod_{t=0}^{T-1}W\left[\boldsymbol{x}^{t+1}\mid\boldsymbol{x}^{0},\ldots,\boldsymbol{x}^{t}\right]\,,\label{eq:prior_SIR}
\end{equation}
where the first term accounts for the initial condition. It is typically  assumed the latter is factorized over nodes, namely $p\left(\boldsymbol{x}^{0}\right)=\prod_{i}p\left(x_{i}^{0}\right)$. 
The Bayesian approach offers a convenient framework for incorporating observations and leveraging information about an individual's state at a specific time. These observations can include various factors such as diagnostic test results or the manifestation of
symptoms. Denoting as $\boldsymbol{\mathcal{O}}=\left\{ \mathcal{O}_{r}\right\} $
the set of observations (positive or negative outcomes of the medical tests, accounting for the presence/absence of infection), the posterior probability of the trajectory $\mathbb{X}$ can be expressed
using Bayes theorem as follows
\begin{equation}
	p\left(\mathbb{X}\mid\boldsymbol{\mathcal{O}}\right) =\frac{1}{p\left(\boldsymbol{\mathcal{O}}\right)}p\left(\mathbb{X}\right)p\left(\boldsymbol{\mathcal{O}}\mid\mathbb{X}\right).\label{eq:posterior_SIR}
\end{equation}
Observations are assumed to be statistically independent, so that $p\left(\mathcal{\boldsymbol{O}\mid \mathbb{X}}\right) = \prod_r p\left(O_r \mid \mathbb{X}\right)$. 

\subsection{Risk estimate and testing strategies from probabilistic inference}
In this section, we briefly revise how to quantitatively estimate the risk of infection, on a daily basis, of each individual. The individuals that have been confined in the previous iterations of the mitigation strategy are not considered in this analysis; we assume that once an individual is quarantined, 
he/she can no longer be infectious for the time window considered in the simulations. SMF and BP provide an estimate for the individual marginal probabilities at any time $t$ of $p\left(\mathbb{X}\mid\boldsymbol{\mathcal{O}}^{t}\right)$, where the observation $\boldsymbol{\mathcal{O}}^{t}$ collects all available information up to the discrete-time $t$. Let us call this estimate as $q_{i}^{t}\left(x_{i}^{t} | \boldsymbol{\mathcal{O}}^{t} \right)$ for $i = 1,\ldots,N$ (see \citep{baker2021epidemic} for a detailed description of the SMF and BP approximations). For SMF we rank individuals according to the marginal probabilities and we perform a fixed number of tests starting from those showing the highest risk. BP allows one to estimate, together with the individual marginal probability of being in one of the three possible states at any time, the probability of the infection time of each individual. This information is exploited during the ranking procedure; we sort non-confined individuals according to the probability of their infection time to lie in a time window of $\tau = 7$ days before the intervention time $t$.
\\
Only when a probability threshold $p_{th}$ is set, we test individuals displaying $q_i^{t}\left( x_{i}^t = I | \boldsymbol{\mathcal{O}}^{t}\right) > p_{th}$ for both SMF and BP: as a consequence, the number of diagnostic tests performed during the simulation adaptively changes according to the probabilistic risk estimates. We show in Figure~ \ref{fig:thresholds}-\ref{fig:thresholdc} the performances of BP and SMF for several values of $p_{th}$. When $p_{th}$ is relatively small, i.e. $p_{th} = 10^{-5}$ for StEM and $p_{th} = 2\time 10^{-3}$ for Covasim, many tests are performed on the first days after the intervention time $t_{i} = 14$ with the effect of dramatically reducing the number of infectious and the effective reproduction number which, after about twenty days, is under $R_{t} = 1$ (Note that for SMF the containment performance is less remarked than that obtained by BP). The drawback of setting such a small threshold is that the fraction of the population tested in the first days of intervention is significantly large; for StEM about $1\%$ of the population needs to be tested the first day $t_{i}$.
If the threshold is too large, i.e. $p_{th} = 5 \times 10^{-3}$ for StEM and $p_{th} = 2 \times 10^{-2}$ for Covasim, a few tests are performed on the first days, but since the spreading is only slightly affected by the containment policy (as suggested by the cumulative number of the infected individuals and the $R_{t}$), the estimated individual probabilities of being infected generally grow. Therefore, the number of tests performed after one month of simulation significantly increases. In this regime, testing large fractions of the population does not carry the same benefit shown when adopting a small threshold. 
These results suggest that a probabilistic threshold must be set by looking for an optimal trade-off between the cost associated with the number of diagnostic tests and the containment efficacy of the isolation strategy. 

\subsection{Coarse-graining and computation of the effective epidemic parameters\label{subsec:mappingSIR}}
In order to be able to effectively exploit a simpler \textit{a priori} model of epidemic dynamics, such as the SIR model, in the study of COVID-19 inference problems, it is essential to establish a mapping between the original epidemic models used for forward dynamics and the aforementioned SIR model. This mapping serves two key purposes: firstly, it enables the alignment of observations from the original models with the observed states in the SIR parameterization, and secondly, it facilitates the definition of effective SIR parameters that accurately capture the dynamics of the original models.

However, when employing digital contact tracing implementations, individual-based information is often unavailable due to privacy concerns. For instance, factors like age, which influences infection probabilities, cannot be directly incorporated into effective SIR modeling. Nevertheless, it is still possible to utilize clinical-based information from each epidemic model, such as time-dependent transmissivity and recovery times, to determine the parameters of the effective SIR description employed in the inference algorithms based on Belief Propagation (BP) and Mean-Field (SMF).

It is important to note the distinction between the two inference algorithms considered in this context: BP has the capability to handle non-Markovian dynamics when there exists an explicit dependence on the time elapsed since infection in the parameters. In contrast, the SMF approach is limited to Markovian dynamics and requires additional approximations, which will be discussed subsequently.

The mapping process is illustrated in Figure \ref{fig:dynamics_allmodels}\figpanel{D} through a color map. In this representation, all intermediate and different infection states in the three models are naturally considered as infected states in the SIR description, while the remaining states (e.g., Hospitalized) until reaching the two absorbing states are considered as Removed ($R$).

Another consideration arises from the presence of an Exposed state in Covasim and StEM (Figure \ref{fig:dynamics_allmodels} \figpanel{B}-\figpanel{C}). In the standard SEIR model, the exposed state accounts for an initial latency period during which individuals are infected but not yet contagious. This can be effectively incorporated into the SIR description by assuming an initial characteristic time during which the probability of infection is negligible.

The following two subsections delve into the computation of effective SIR parameters, specifically the infection probabilities and the distribution of recovery times, from which the daily recovery probability can be derived. These obtained parameters are then used as part of the prior SIR model in the SMF-based and BP-based inference algorithms.

\subsubsection{Infection probability}

In the three models under investigation, the infection probability exhibits a complex structure that depends on various factors, including detailed individual information about the two individuals involved in the contact (such as their age), the social layer to which the contact belongs (e.g., households, workplaces), and the time elapsed since the infector node's infection. However, due to privacy concerns, contact tracing techniques do not have access to such detailed individual information.

To address this limitation, we make certain assumptions in the prior SIR model by considering effective infection probabilities derived from the average transmission properties of the population. Specifically, for the Belief Propagation (BP) and Mean-Field (SMF) schemes, we define the following effective infection probabilities 
\begin{align}
	\lambda^{t,{\rm BP}}_{i\to j} \left(t_i \right) & \equiv w^t_{i\to j} \lambda_{0}^{\rm BP} \Lambda \left( t - t_i\right)\\
	\lambda^{t,{\rm SMF}}_{i\to j}  &\equiv w^t_{i\to j} \lambda_{0}^{\rm SMF}.
\end{align}
The variable $w^t_{i \to j}$ is non-zero only if there is a contact between individuals $i$ and $j$ at time $t$ and takes value equal to 1 for all contacts except for those occurring inside the households which are assumed to have a strength $g>1$, whose precise value depends on the underlying model. Secondly, $\lambda_0^{\rm BP}$ is a population-averaged constant quantity containing the dependency of the infection probability on the individual features of the agents and on the different social contexts the contact can belong to. 
The time-dependent part of the infectivity is taken into account by the phenomenological function $\Lambda(t)$ whose functional form is obtained interpolating the results of the average temporal behavior of the infection probability over the whole population, and it is considered only within the BP framework. The functional form of $\Lambda(t)$ is shown for the three different epidemic models in Figure~\ref{fig:probiandr} (left panel). The initial slow growth from zero of the function $\Lambda(t)$ is a signature of the existence in the original model of an incubation period (in Covasim and StEM, this is explicitly represented by the existence of an exposed state). The different long-time behavior of the function $\Lambda(t)$ for the three models reflects the different properties of the underlying epidemic models. 

On the other hand, in the SMF scheme, the infection probability is modeled as a constant in time, so that no incubation or time-dependent viral charge is taken into account. The values $\lambda_0^{\rm SMF}$ are computed by performing a time-independent average of the transmission properties of infected individuals across the whole population.

As explained in Sec.~\ref{sec:ABM}, in the case of the StEM model, contacts are aggregated on a daily basis before performing the contact tracing analysis. Hence, in defining the mapping on the effective SIR model, we also made the assumption that the probability of exposure between the two individuals in contact for a total time $\Delta t$ on a certain day is obtained by integrating the instantaneous infection rate of the StEM model over a time equal to the duration $\Delta t$ of the contact. This approach utilizes the values of instantaneous infection rate as defined in the underlying StEM model presented in Ref.~\cite{lorch2022quantifying}.

\subsubsection{Recovery time}
In the effective SIR description, infected individuals (belonging to any subcategory in the three different epidemic models as discussed in the mapping of Figure~\ref{fig:dynamics_allmodels}-\figpanel{d}) are considered to be in the infected state $I$ until they move to one of the states mapped to the recovered state $R$. The specific transition pathways leading to recovery vary depending on the model and are parameterized by known probability distributions, typically Gamma or log-normal distributions. To define an effective recovery time distribution, we need to account for the intermediate states and calculate the total duration an individual remains infected along each pathway.

Let us denote with $\boldsymbol{\chi}$ as a generic infection route, defined by a sequence of states starting from the initial infected state (including the Exposed state, when present) and ending with the first non-infectious state (i.e. the blue states in Figure \ref{fig:dynamics_allmodels}). Each route consists of $n$ of intermediate states, namely $\boldsymbol{\chi} = \{\chi_1, \ldots, \chi_n\}$: for example, in the StEM model (Figure \ref{fig:dynamics_allmodels}-\figpanel{C}), a possible route is $\{$Exposed $\to$ pre-Symptomatic $\to$ Symptomatic $\to$ Hospitalized$\}$, with $n=4$. The time $\tau_{\boldsymbol{\chi}}$ an individual spends following the entire route $\boldsymbol{\chi}$ is the sum of the intermediate transition times:
\begin{equation}
	\tau_{\boldsymbol{\chi}} = \tau_{\chi_1 \to \chi_2, \ldots, \to \chi_n} = \sum_{i=1}^{n-1} \tau_{\chi_{i} \to \chi_{i+1}}.\label{eq:taupath_sum}
\end{equation}
Since each transition time is drawn from a known probability density function (specified in each model's settings), the sum represents a random variable whose probability density function (p.d.f.) can be easily obtained by the convolution of the individual transition time distributions. Upon infection, each individual is assigned to one of the possible pathways with a probability that depends on their age. Thus, an additional average over the age distribution is required. Let $a$ represent an individual's age and $p(a)$ denote the empirical age distribution (available for each model). The overall distribution of the recovery time $\tau_R$ in the effective SIR model within the BP framework can be expressed as
\begin{equation}
	p\left( \tau_R \right) = \left \langle \sum_{\boldsymbol{\chi}} \phi\left(\boldsymbol{\chi},a \right) p\left( \tau_{\boldsymbol{\chi} } \right) \right \rangle_{p\left(a\right)}.\label{eq:tauRBP_sumoverpaths}
\end{equation}

\begin{figure}
	\begin{center}
		\includegraphics[width=0.9\linewidth]{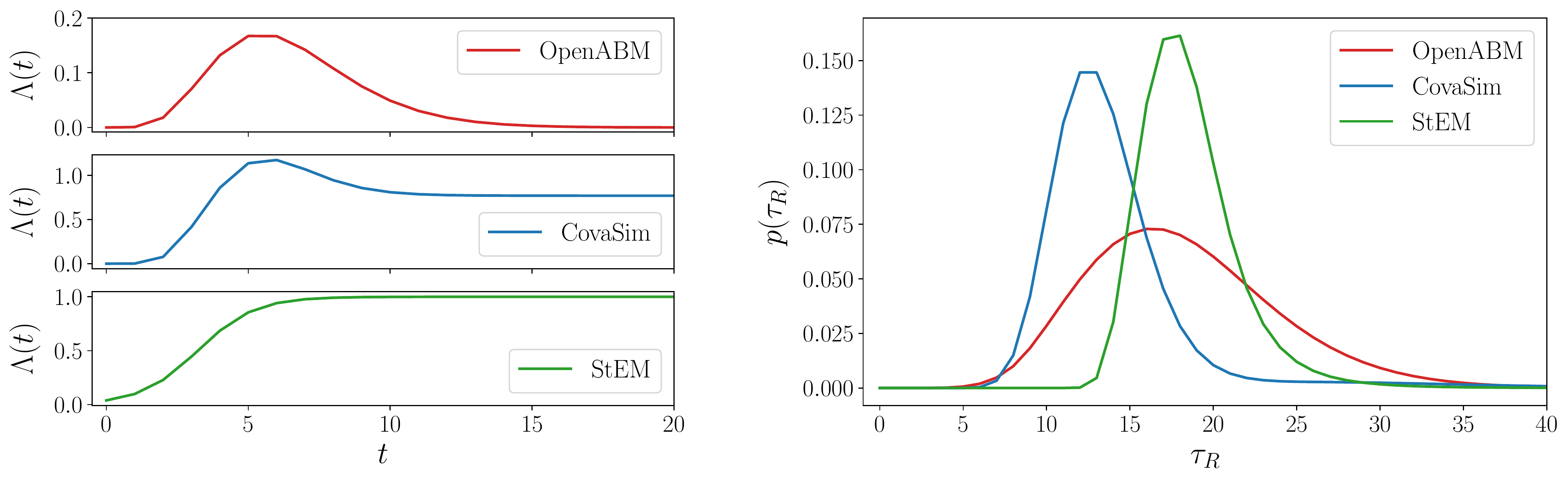}
		\caption{Left: Time-dependent contribution to the infection probability $\Lambda \left(t \right)$ for the three epidemic models, where $t$ denotes the time elapsed since infection. Right: probability distribution of the recovery time in the effective SIR reduction for the three models. Both timescales are in day units.}
		\label{fig:probiandr}
	\end{center}
\end{figure}

\begin{figure}[h!]
	\begin{center}
		\includegraphics[width=\linewidth]{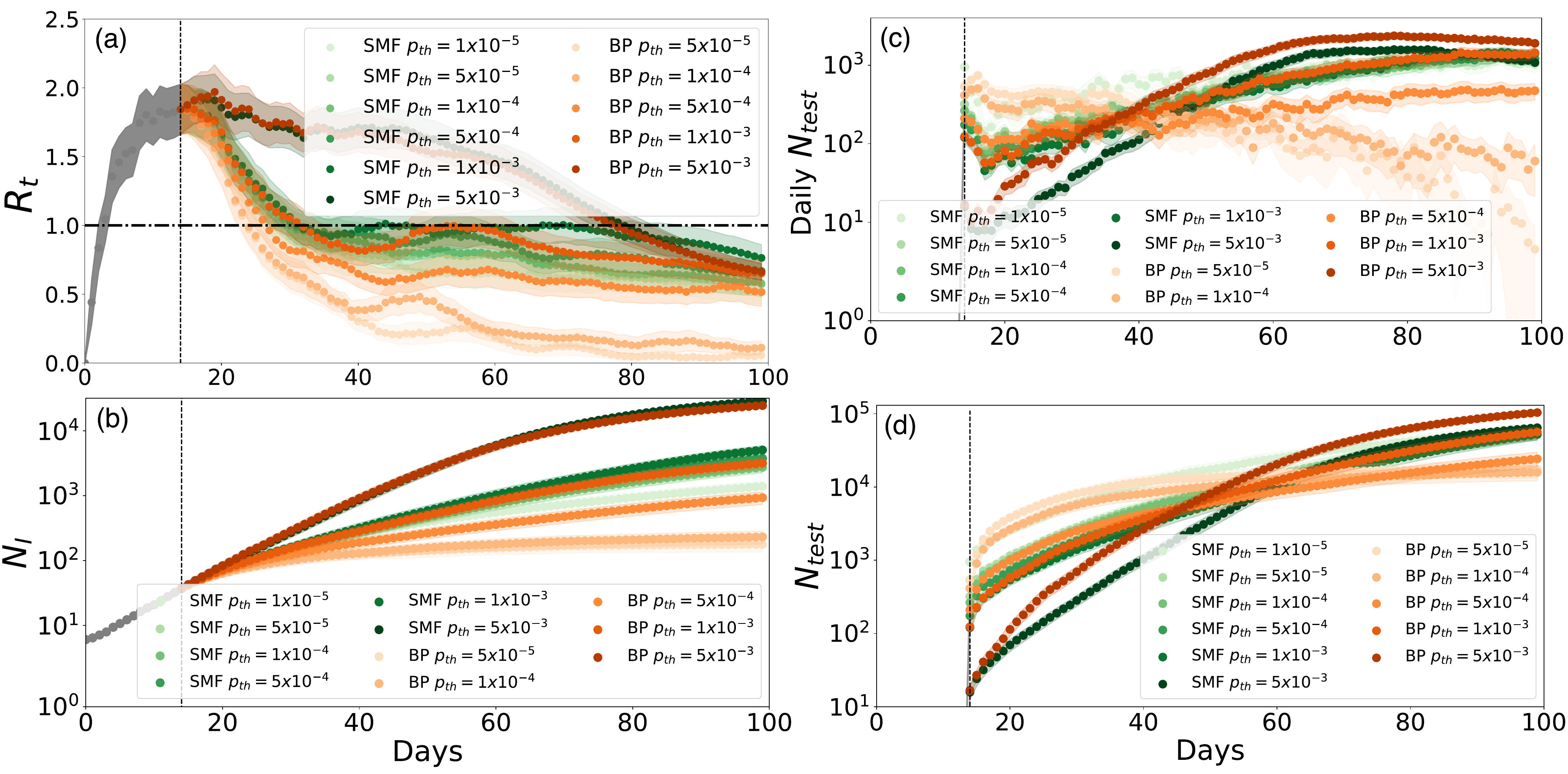}
		\caption{Performances of the adaptive strategy for BP and SMF on the StEM model, when only individuals exceeding the probability $p_{th}$ of being infected are tested. Color gradient reflects the values of the threshold probability: the lighter the color, the smaller the threshold. In panels (a), (b), (c), and (d) we plot the value of the effective reproduction number, the cumulative number of infected, the tests performed daily, and the cumulative number of tests as a function of the time, respectively.
			\label{fig:thresholds}}
	\end{center}
\end{figure}

\begin{figure}
	\begin{center}
		\includegraphics[width=\linewidth]{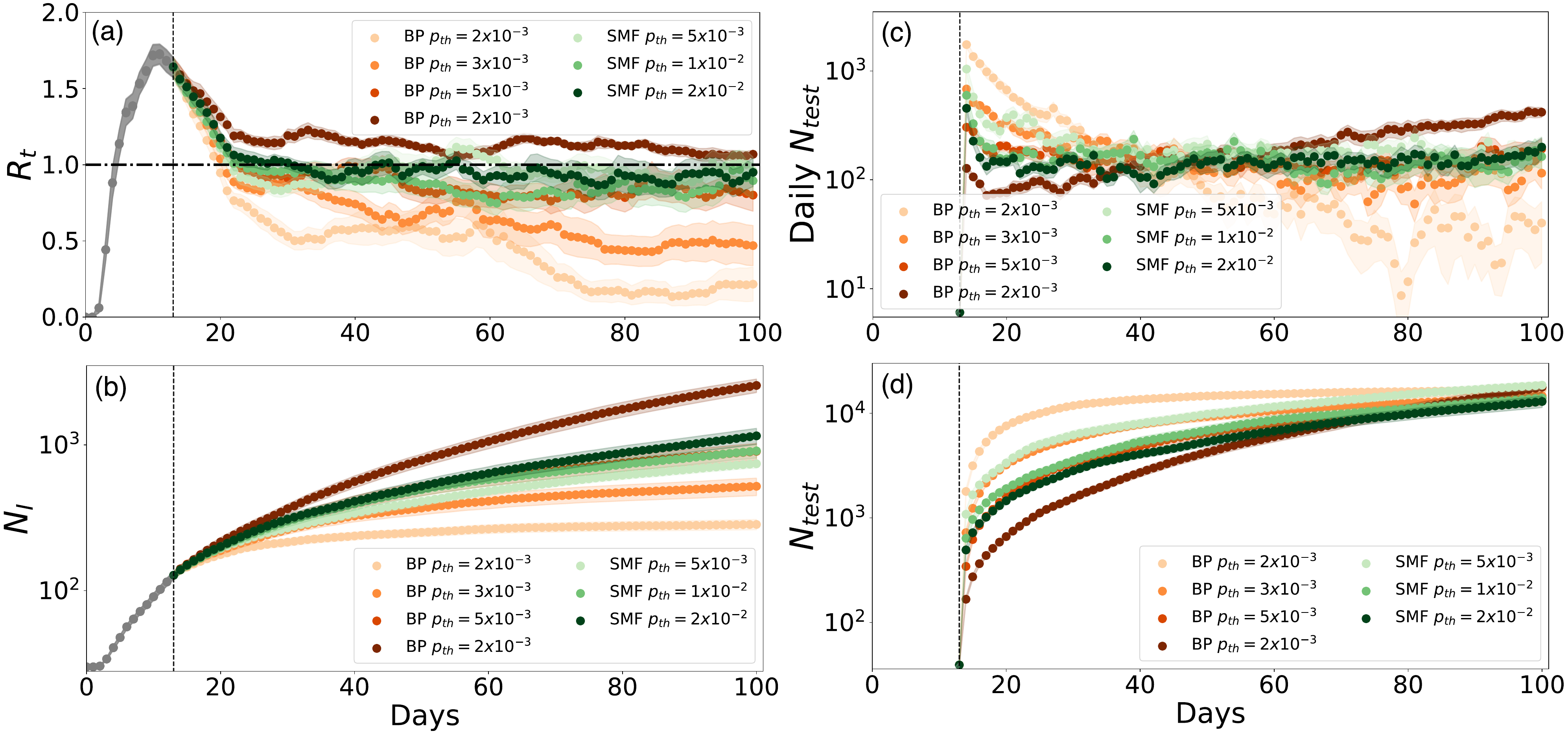}
		\caption{Performances of the adaptive strategy for BP and SMF on the Covasim model, when only individuals exceeding the probability $p_{th}$ of being infected are tested. Color gradient reflects the values of the threshold probability: the lighter the color, the smaller the threshold. In panels (a), (b), (c), and (d) we plot the value of the effective reproduction number, the cumulative number of infected, the tests performed daily, and the cumulative number of tests as a function of the time, respectively. All simulations are run over the Covasim model with the same parameters of Figure 1 of the main text. \label{fig:thresholdc}}
	\end{center}
\end{figure}

Here, the sum runs over all possible infection routes in the epidemic model, $\phi$ represents an age-dependent and route-dependent weight associated with each specific pathway, and each $\tau_{\boldsymbol{\chi}}$ follows the form given in Eq. \eqref{eq:taupath_sum} (including its associated probabilistic properties). The results of computing the p.d.f. in Eq. \eqref{eq:tauRBP_sumoverpaths} for the three different epidemic models are presented in Figure \ref{fig:probiandr} (right panel). These empirical recovery time distributions can be directly used in the BP algorithm for posterior inference (Eq. \eqref{eq:posterior_SIR}). We stress that the implementation of BP discussed in \citep{baker2021epidemic} and available at \citep{sibGitHub} does not require a specific functional form. It is worth noting that BP also allows for parameter inference of the effective SIR model by gradient descent on its free energy, although we have left this task for future investigations \citep{altarelli2014bayesian}.

When using the SMF algorithm, a further approximation is employed to simplify the recovery process to a Markovian one. Consequently, we approximate Eq. \eqref{eq:tauRBP_sumoverpaths} with a Geometric distribution, resulting in $\tau_R^{\text{MF}} \sim \text{Geometric}(1/\mu_R)$, where $\mu_R$ is the mean of Eq. \eqref{eq:tauRBP_sumoverpaths}. 
\section{Definition of the effective reproduction number.}
The effective reproduction number used in Figure 1 of the main text is computed at each iteration $t$ by considering the effective SIR dynamics assumed by the containment methods. It corresponds, from the point of view of the generative models in Covasim and StEM, to the ratio between the new exposures and the infected (exposed or infectious) individuals at that time, multiplied by the average duration of the infection (including the time spent in the exposed state) that has been computed empirically from simulations. This definition is similar to that used in Ref.~\citep{kerr2021covasim}. For the sake of clarity, we plot in Figure 1 of the main text at day $t$ the average of this metric computed on a moving time window of the seven days foregoing day $t$.

\newpage

\section{Supplementary Figures}

\begin{figure}
	\begin{center}
		\includegraphics[width=\linewidth]{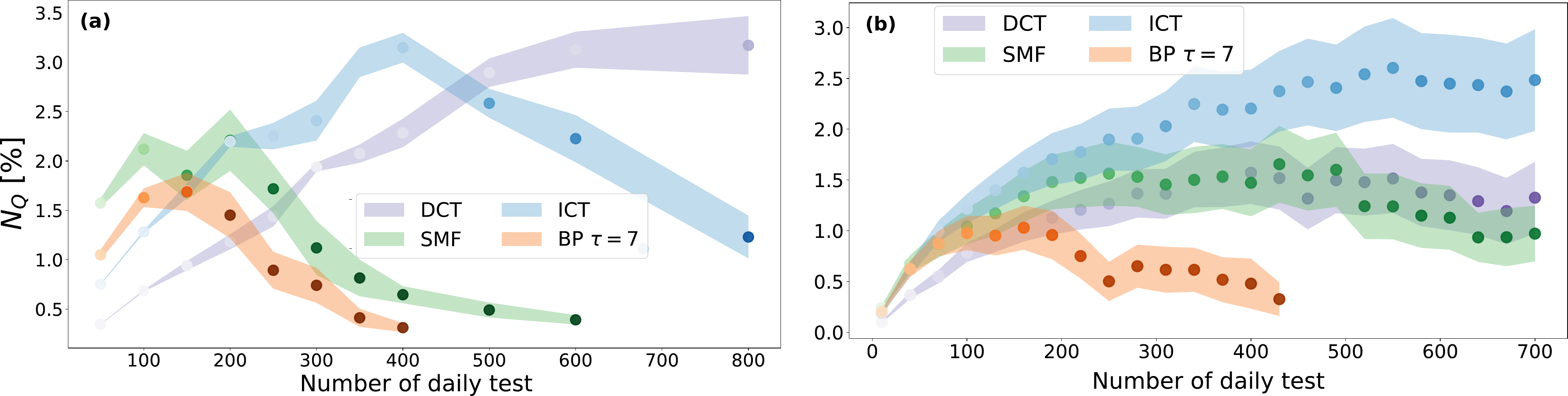}
		\caption{Percentage of confined individuals as a function of the diagnostic tests performed on a daily basis for (a) Covasim and (b) StEM. The setup is the one investigated in Figure 2 of the main text. The color gradient mirrors the reduction of the spreading due to the containment policies: the darkest (lightest) dots are associated with a reduction equal to one (zero). 
			\label{fig:Fig4SI}}
	\end{center}
\end{figure}

\end{document}